\documentclass[aps,twocolumn,epsf,floats,pre,superscriptaddress,nofootinbib]{revtex4-2}

\usepackage{graphicx}
\usepackage{amsmath}
\usepackage{microtype}
\usepackage{tikz}
\usepackage[compat=1.1.0]{tikz-feynman}
\usepackage{commath}
\usepackage{enumitem}
\usepackage{amsfonts}
\usepackage{hyperref}
\usepackage{mathtools}
\usepackage{scrtime}
\usepackage[dont-mess-around]{fnpct}
\usepackage{import}
\usepackage{float}
\usepackage{etoolbox}
\usepackage[font=small,labelfont=bf,
   justification=justified,
   format=plain]{caption}
\usepackage[tiny, center, uppercase]{titlesec}
\usepackage{subcaption} 
\usepackage{bibunits}

\defaultbibliography{399-bibliography.bib}
\defaultbibliographystyle{apsrev4-2}

\newcommand{\Ham}{\mathcal{H}}
\newcommand{\PP}{\mathbb{P}}
\newcommand{\II}{\mathbb{I}}
\newcommand{\tdelta}{\hat{\delta}}
\newcommand{\tSigma}{\tilde{\Sigma}}
\newcommand{\tPi}{\tilde{\Pi}}
\newcommand{\tGamma}{\tilde{\Gamma}}
\newcommand{\tlambda}{\tilde{\lambda}}
\newcommand{\tLambda}{\tilde{\Lambda}}

\newcommand{\iu}{\text{i}}
\newcommand{\eu}{\text{e}}

\newcommand{\pder}[2]{\frac{\partial #1}{\partial #2}}
\newcommand{\dder}[2]{\frac{\delta #1}{\delta #2}}
\newcommand{\vecc}[1]{\boldsymbol{#1}}
\newcommand{\tildehat}[1]{\breve{#1}}
\newcommand{\di}{\text{d}}

\newcommand{\vv}{{\boldsymbol{v}}}
\newcommand{\hpsi}{\hat{\psi}}
\newcommand{\vpsi}{{\boldsymbol{\psi}}}
\newcommand{\hvpsi}{\hat{\vpsi}}
\newcommand{\hphi}{{\hat{\phi}}}
\newcommand{\vphi}{{\vecc{\phi}}}
\newcommand{\hvphi}{{\vecc{\hphi}}}
\newcommand{\hs}{\hat{s}}
\newcommand{\vs}{\boldsymbol{s}}
\newcommand{\hvs}{\hat{\vs}}

\newcommand{\vx}{\boldsymbol{x}}

\newcommand{\vk}{\boldsymbol{k}}
\newcommand{\tvk}{\tilde{\vk}}
\newcommand{\vq}{\boldsymbol{q}}
\newcommand{\tvq}{\tilde{\vq}}
\newcommand{\vh}{\boldsymbol{h}}
\newcommand{\tvh}{\tilde{\vh}}
\newcommand{\vp}{\boldsymbol{p}}
\newcommand{\tvp}{\tilde{\vp}}
\newcommand{\vnabla}{{\boldsymbol{\nabla}}}
\newcommand{\E}[1]{\mathbb{E}\left[#1\right]}

\begin{document}

\title{Natural Swarms in $\bf 3.99$ Dimensions}

\author{Andrea Cavagna}
\affiliation{Istituto Sistemi Complessi, Consiglio Nazionale delle Ricerche, 00185 Rome, Italy}
\affiliation{Dipartimento di Fisica, Universit\`a\ Sapienza, 00185 Rome, Italy}
 
\author{Luca Di Carlo}
\affiliation{Dipartimento di Fisica, Universit\`a\ Sapienza, 00185 Rome, Italy}
\affiliation{Istituto Sistemi Complessi, Consiglio Nazionale delle Ricerche, 00185 Rome, Italy}

\author{Irene Giardina}
\affiliation{Dipartimento di Fisica, Universit\`a\ Sapienza, 00185 Rome, Italy}
\affiliation{Istituto Sistemi Complessi, Consiglio Nazionale delle Ricerche, 00185 Rome, Italy}
\affiliation{INFN, Unit\`a di Roma 1, 00185 Rome, Italy}

\author{Tom\'as S. Grigera}
\affiliation{Instituto de F\'\i{}sica de L\'\i{}quidos y Sistemas Biol\'ogicos CONICET -  Universidad Nacional de La Plata,  La Plata, Argentina}
\affiliation{CCT CONICET La Plata, Consejo Nacional de Investigaciones Cient\'\i{}ficas y T\'ecnicas, Argentina}
\affiliation{Departamento de F\'\i{}sica, Facultad de Ciencias Exactas, Universidad Nacional de La Plata, Argentina}

\author{Stefania Melillo}
\affiliation{Istituto Sistemi Complessi, Consiglio Nazionale delle Ricerche, 00185 Rome, Italy}
\affiliation{Dipartimento di Fisica, Universit\`a\ Sapienza, 00185 Rome, Italy}

\author{Leonardo Parisi}
\affiliation{Istituto Sistemi Complessi, Consiglio Nazionale delle Ricerche, 00185 Rome, Italy}
\affiliation{Dipartimento di Fisica, Universit\`a\ Sapienza, 00185 Rome, Italy}

\author{Giulia Pisegna}
\affiliation{Dipartimento di Fisica, Universit\`a\ Sapienza, 00185 Rome, Italy}
\affiliation{Istituto Sistemi Complessi, Consiglio Nazionale delle Ricerche, 00185 Rome, Italy}

\author{Mattia Scandolo}
\affiliation{Dipartimento di Fisica, Universit\`a\ Sapienza, 00185 Rome, Italy}
\affiliation{Istituto Sistemi Complessi, Consiglio Nazionale delle Ricerche, 00185 Rome, Italy}

\begin{abstract}
The dynamical critical exponent $z$ of natural swarms of insects is calculated using the renorma\-li\-zation group to order $\epsilon = 4-d$.  A novel fixed point emerges, where both \textcolor{black}{activity and inertia} are relevant. In three dimensions the critical exponent at the new fixed point is \textcolor{black}{$z = 1.35$, in agreement with both experiments ($1.37 \pm 0.11$) and numerical simulations ($1.35 \pm 0.04$).}
\end{abstract}

\maketitle

\begin{bibunit}
Collective behaviour is found in a great variety of biological systems, from bacterial clusters and cell colonies, up to insect swarms, bird flocks, and vertebrate groups. A unifying ingredient, which also provides an insightful connection with statistical physics, is the presence of strong correlations: the correlation length, $\xi$, is often significantly larger than the microscopic scales \cite{cavagna2010scalefree, strandburg2013visual, ginelli2015herds, attanasi2014collective, dombrowski2004self, zhang2010collective, tang2017critical}; in some instances $\xi$ grows with the system's size, giving rise to scale-free correlations \cite{cavagna2010scalefree, zhang2010collective, attanasi2014finite}. In the case of natural swarms of insects, a second key hallmark of statistical physics has been verified, namely dynamic scaling \cite{cavagna2017swarm}; this is noteworthy, as dynamical scaling entangles spatial and temporal relaxation into one law, known as {\it critical slowing down} \cite{HH1969scaling}: the collective relaxation time grows as a power of the correlation length, $\tau \sim \xi^z$, thus defining the dynamical critical exponent, $z$.
Strong correlations and scaling laws are the two essential prerequisites of the Renormalization Group (RG) \cite{wilson1972critical,wilson1974renormalizarion}: by coarse-graining short-wavelength fluctuations, the parameters of different systems flow towards few fixed points ruling their large-scale behaviour; RG fixed points therefore organize into universality classes the macroscopic behaviour of strongly correlated systems, thus providing parameter-free predictions of the critical exponents.
\textcolor{black}{The emergence of scale-free correlations and scaling laws calls for an exploration of the RG path also in collective biological systems.}

\textcolor{black}{In the broader field of active matter \cite{marchetti2013hydro}, RG is already a key tool.}  The pioneering hydrodynamic theory of Toner and Tu \cite{tonertu1995, tonertu1998} has been studied through the RG both in the polarised \cite{chen2018incompressible, toner2019gnf} and near ordering phase \cite{chen2015critical, skultety2020universality}, with applications in systems with nematic or polar order \cite{mishra2010dynamic,ramaswamy2010mechanics, tonertu2005hydro}. RG has also been employed to study motility-induced phase separation \cite{caballero2018bulk, maggi2021critical}, active membranes \cite{cagnetta2021universal}, bacterial chemotaxis \cite{mahdisoltani2021nonequilibrium}, cellular growth \cite{gelimson2015collective}.
Direct comparisons with experiments are few, though: the exponent of giant number fluctuations in $d=2$ \cite{tonertu1998, toner2019gnf} was confirmed in experiments on vibrated polar disks \cite{deseigne2010collective}, while in \cite{mahault2019quantitative} the exponents of the Vicsek Model in the ordered phase were found to be incompatible with those conjectured by Toner and
Tu \cite{tonertu1995}. Other RG exponents have been checked in numerical simulations \cite{tu1998sound, chate2008collective, ginelli2010largescale, doostmohammadi2017onset}. Comparisons with biological experiments are scarcer. Experiments studying giant number fluctuations in swimming filamentous bacteria displaying long-range nematic order \cite{nishiguchi2017longrange} found an exponent in disagreement with RG predictions of active nematic \cite{ramaswamy2010mechanics} and polar \cite{toner2019gnf} systems. To the best of our knowledge, there is yet no successful test of an RG prediction against experiments on living active systems.

Here, we apply the RG to the dynamics of insect swarms. Swarms of midges in the field are near-critical, strongly correlated systems \cite{attanasi2014finite}, \textcolor{black}{with short-range interactions \cite{attanasi2014collective}}, obeying dynamic scaling \cite{cavagna2017swarm} with an experimental exponent \textcolor{black}{$z_\mathrm{exp}=1.37\pm 0.11$}, significantly smaller than the value $z \approx 2$ of standard ferromagnets \cite{HH1977}. This large gap indicates that fundamental new physics is required. Although the relation, $\tau \sim \xi^z$, is not merely a dispersion law, smaller $z$ nevertheless suggests that fluctuations are more swiftly transported across the system. Hence, the effort to match theory with experiments in natural swarms must look for more efficient mechanisms of information transfer.
Activity is the first obvious candidate, as it allows fluctuations to propagate not only thanks to the inter-individual interaction, but also through the self-propelled motion of the particles \cite{tonertu1995}. Incompressible, near-critical  active matter was first studied in \cite{chen2015critical}, where an RG analysis found that activity lowers $z$ from $2$ to $1.73$. This was an important step towards bridging the gap between experiments and theory in natural swarms, although the chasm with the experimental exponent remains significant.

\textcolor{black}{The second ingredient known to foster information propagation is inertia. Behavioural inertia in the rotations of the individual velocities was first introduced to explain the propagation of collective turns in bird flocks \cite{attanasi2014information, cavagna2015flocking}, but experiments found clear evidence of underdamped inertial relaxation also in natural swarms of midges \cite{cavagna2017swarm}. 
At the general level, inertial dynamics stems from the existence of a reversible coupling between the primary field (playing the role of the generalized coordinate) and the generator of the symmetry (playing the role of the generalized momentum); in the case at hand, the symmetry is the rotation of the primary field, hence we call `spin' its generator. In absence of explicit dissipation, this reversible coupling leads to global conservation of the spin, a conservation law which is known --  at equilibrium -- to significantly decrease the dynamical exponent, from $z\approx 2$ of standard ferromagnets (Model A in the classification of Halperin and Hohenberg \cite{HH1977}), to $z=1.5$ of superfluid helium and quantum antiferromagnets (Models E/F and G of \cite{HH1977}).\footnote{\textcolor{black}{In Model A, $z=2$ at one loop and two-loop corrections are very small \cite{HH1977}; on the other hand, in Models E/F and G, $z=d/2$ is an exact result \cite{dedominicis1978field}.}}}
Overall, this scenario suggests that the combined effect of activity {\it and} inertia may account for the experimental exponent of natural swarms. Here, we perform an RG study of such theory, and find $z=1.34(8)$ in $d=3$, \textcolor{black}{a value in agreement with both experiments on real swarms, $z_\mathrm{exp}=1.37\pm 0.11$, and numerical simulations, $z_\mathrm{sim}=1.35\pm 0.04$.} The RG result is a parameter-free prediction, with no input beyond the information that both activity and inertia must be part of the theory.

A hydrodynamic theory of active matter with reversible inertial couplings requires three fields: velocity, spin and density \cite{cavagna2015silent, yang2015hydrodynamics}, making the calculation technically unfeasible.
To make progress, following \cite{chen2015critical}, we eliminate the density field by imposing incompressibility, $\vnabla\cdot \vv\left(\vx,t\right)=0$.
\textcolor{black}{Beyond its technical inevitability, this is a reasonable physical assumption. 
In compressible active systems the transition is first-order \cite{gregoire2004}, a framework that would make RG pointless and would rule out scaling. However, dynamic scaling {\it is} observed in natural swarms, suggesting a more complex scenario:  in absence of density fluctuations, the transition becomes second-order and recent studies \cite{cavagna2020equilibrium, dicarlo2022evidence, qi2022finite} suggest the existence of a crossover from a finite-size regime, where density fluctuations are weak and second-order physics is observed, to a infinite-size regime, where density fluctuations dominate, rendering the transition first-order. Weak density fluctuations and scaling laws in natural swarms \cite{cavagna2017swarm}, suggest then that the dynamics of these finite-size systems can be studied through an incompressible second-order theory.}

The polarization field, $\vpsi$, is defined by the relation, $\vv\left(\vx,t\right)= v_0\, \vpsi\left(\vx,t\right)$, where $v_0$ is the microscopic speed; having $v_0$ as an explicit parameter is useful to take the zero-activity limit, $v_0\to 0$, and compare with  equilibrium calculations \cite{cavagna2019short, cavagna2019long, cavagna2021dynamical}. The generator of the rotations of $\vpsi\left(\vx,t\right)$ is the spin, $\vs\left(\vx,t\right)$; in $d=3$ the spin is a vector; however, incompressibility requires $\vpsi$ to have the same dimension as space, and because RG entails an expansion in powers of $\epsilon=4-d$, we need a generic spin in $d$ dimensions, which is a $d\times d$ anti-symmetric tensor. 
\textcolor{black}{
Reversible inertial dynamics arises from the Poisson brackets \cite{HH1977},
\begin{equation}
\{s_{\alpha\beta}\left(\vx,t\right),\psi_\gamma\left(\vx',t\right)\}=2 g \, \II_{\alpha\beta\gamma\rho} \psi_\rho (\vx,t)\delta^{(d)}(\vx-\vx') \ ,
\label{eq:poisson}
\end{equation}
stating that $s_{\alpha\beta}\left(\vx\right)$ is the generator of the rotational symmetry, thus leading to the conservation of the total spin, $S_{\alpha\beta}(t)=\int d^d x \ s_{\alpha\beta}\left(\vx,t\right)$; the crucial constant $g$ is the reversible coupling regulating this symplectic structure \cite{HH1977}.} Finally, $\II_{\alpha\beta\gamma\nu}=( \delta_{\alpha\gamma} \delta_{\beta\nu} - \delta_{\alpha\nu} \delta_{\beta\gamma})/2$ is the identity in the space of $s_{\alpha\beta}$.

The dynamical field theory we study combines \textcolor{black}{the irreversible off-equilibrium hydrodynamic approach of Toner and Tu \cite{tonertu1995, tonertu1998, chen2015critical}, with the reversible conservative structure} used to describe superfluid helium and quantum antiferromagnets (Models E/F and G of \cite{HH1977}). The equations of motion are,
\begin{eqnarray}
D_t\psi_\alpha 
&=&
 - \Gamma \dder{\Ham}{\psi_\alpha} + g\,   \psi_\beta \dder{\Ham}{s_{\alpha\beta}} - \partial_\alpha \mathcal{P} + \theta_\alpha \ ,
\label{eq:psi}
\\
D_t s_{\alpha\beta}
&=&
- \Lambda_{\alpha\beta\gamma\nu} 
 \dder{\Ham}{s_{\gamma\nu}} 
 + 2 g \, \II_{\alpha\beta\gamma\nu} \psi_\gamma \dder{\Ham}{\psi_\nu} 
 + \zeta_{\alpha\beta} 
 \ ,
\label{eq:s}
\end{eqnarray}
where the material derivatives are defined as,
\begin{eqnarray}
D_t  \psi_\alpha &=&\partial_t \psi_\alpha + \gamma_v v_0 \,  \psi_\nu \partial_\nu  \psi_\alpha \ \ , \ \ 
\nonumber
\\
D_t s_{\alpha\beta}&=&\partial_t s_{\alpha\beta}+ \gamma_s   v_0\, \psi_\nu \partial_\nu  s_{\alpha\beta} \ .
\label{material}
\end{eqnarray}
Activity breaks Galilean invariance \cite{tonertu1998}, so that the couplings $\gamma_v$ and $\gamma_s$ can be different from $1$ and from each other. $\Ham$ is the classic Landau-Ginzburg coarse-grained Hamiltonian \cite{tonertu1995},
\begin{equation}
\Ham = \int d^d x \left[\frac{1}{2} \partial_\beta \psi_\alpha\partial_\beta \psi_\alpha + \frac{r}{2} \psi^2 + \frac{u}{4} \psi^4 + \frac{1}{2} s^2 \right] .
\end{equation}
The $\psi$-dependent part of $\mathcal H$ is the standard alignment interaction, while $s^2/2$ is the `kinetic' term \cite{HH1977}. \textcolor{black}{Since natural swarms have scale-free correlations \cite{attanasi2014finite}, we will perform the calculation on the critical manifold, $r=0$.}

\textcolor{black}{The terms proportional to $g$ in \eqref{eq:psi}-\eqref{eq:s} are the reversible forces, characterizing inertial dynamics: instead of acting directly on the polarization, the alignment force $\delta\Ham/\delta\psi$ acts on the spin, which in turns rotates $\vpsi$ \cite{cavagna2015flocking}.
On the other hand, the terms proportional to the kinetic coefficients $\Gamma$ and $\Lambda$ represent the irreversible forces, responsible of relaxation}.
The pressure $\mathcal{P}$ enforces incompressibility, which in $k$-space translates into $k_\alpha\psi_\alpha=0$, or $P^\perp_{\alpha\beta}\psi_\beta=\psi_\alpha$, with the projector $P^\perp_{\alpha\beta} = \delta_{\alpha\beta}-k_\alpha k_\beta/k^2 $. To respect this constraint, one must project  the dynamic equation for $\vpsi$, eq.\eqref{eq:psi}, and its noise correlator \cite{chen2015critical}, 
\begin{equation}
\langle \theta_\alpha\left(\vk,t\right) \theta_\beta \left(\vk',t'\right) \rangle= (2\pi)^d2 \tGamma\, P^\perp_{\alpha\beta}\,\delta^{(d)}\left(\vk+\vk'\right)\delta\left(t-t'\right) 
\nonumber
\end{equation}
where $\tGamma\neq\Gamma$ out of equilibrium. 
\textcolor{black}{Notice that, if we write a general anisotropic form of this kinetic coefficient, $\Gamma_{\alpha\beta}=\Gamma^\perp P^\perp_{\alpha\beta} + \Gamma^\parallel (\II - P^\perp_{\alpha\beta})$, only $\Gamma^\perp$ survives the projection of eq.\eqref{eq:psi}, so that effectively $\Gamma=\Gamma^\perp$ in our notation (and similarly for $\tilde\Gamma$).\footnote{\textcolor{black}{Further non-diagonal forms of the kinetic coefficient may be conceivable, but we do not study them here, as they would make the calculation too intricate.}}
On the other hand, because eq.\eqref{eq:s} is {\it not} projected,} anisotropic corrections to $\Lambda$ in $k$-space are generated by the RG \cite{cavagna2021dynamical}, so it is convenient to assume from the outset a general anisotropic form,
\begin{equation}
\Lambda_{\alpha\beta\gamma\nu} =  \left( \lambda^\perp \PP_{\alpha\beta\gamma\nu}^\perp +
\lambda^\parallel (\II-\PP^\perp)_{\alpha\beta\gamma\nu} \right) k^2 \ ,
\label{bazooka}
\end{equation}
where $\PP_{\alpha\beta\gamma\nu}^\perp$ is the projector in the anti-symmetric space \cite{cavagna2021dynamical}. \textcolor{black}{The fact that this kinetic coefficient is proportional to $k^2$ grants that also the irreversible terms conserve the total spin; notably, thanks to the Poisson structure, this $k^2$ term gets {\it generated} by the RG if one tries neglecting it \cite{cavagna2019long}.} Finally, the spin noise has correlator,
\begin{equation}
\langle \zeta_{\alpha\beta}\left(\vk,t\right) \zeta_{\gamma\nu}\left(\vk',t'\right) \rangle = (2\pi)^d
4 \tLambda_{\alpha\beta\gamma\nu}
\delta^{(d)}\left(\vk+\vk'\right)\delta\left(t-t'\right)
\nonumber
\end{equation}
where $\tLambda$ has the same structure as $\Lambda$, although out of equilibrium we may have, $\lambda^{\perp,\parallel}\neq\tlambda^{\perp,\parallel}$.
\textcolor{black}{In addition to the terms discussed here, new interactions compatible with symmetries are generated by the RG transformation. In order to be self-consistent and closed, the RG calculation must take into account these new terms (see SI-IC4).}

\begin{figure*}[t]
	\centering
	\includegraphics[width=1.0\textwidth]{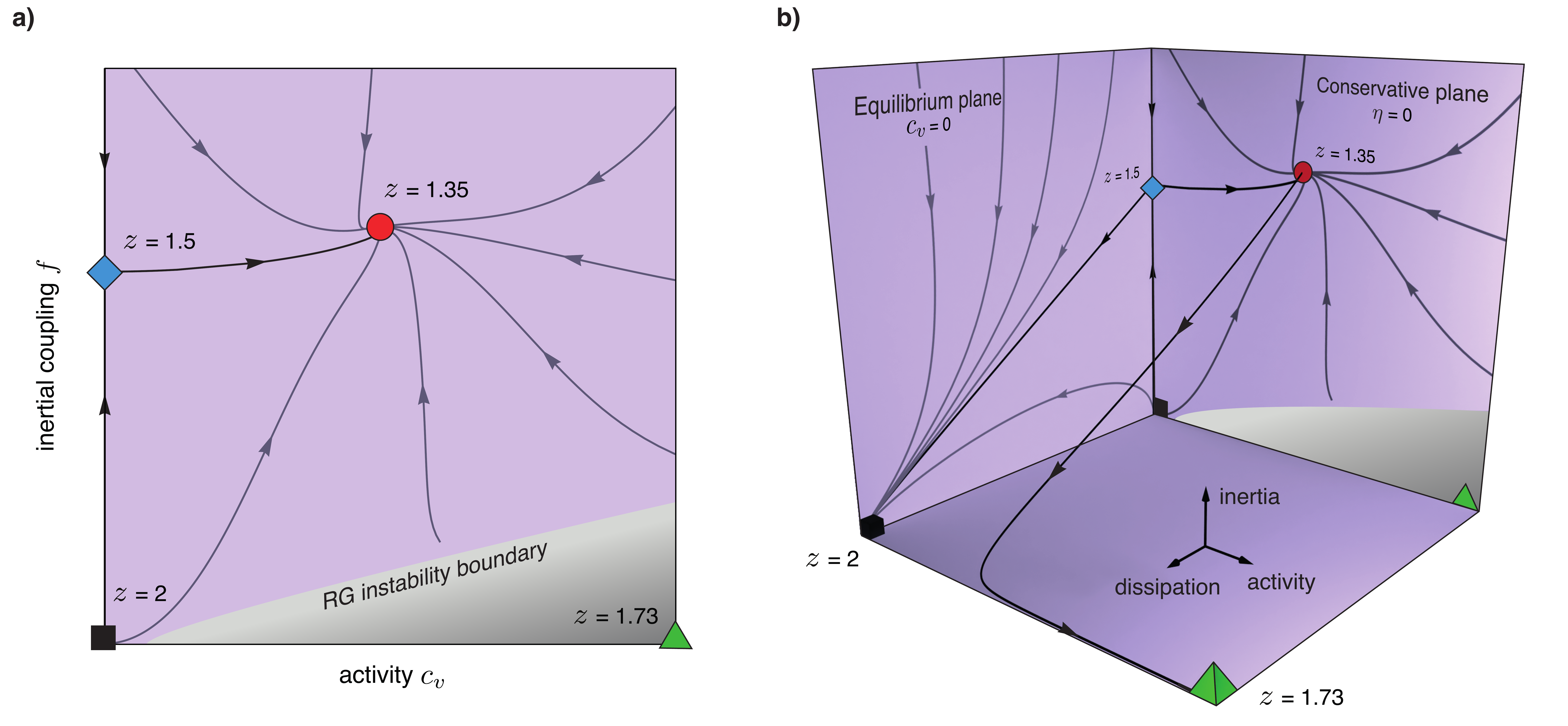}
	\caption{{\bf RG flow.} 
		\textcolor{black}{
			{\bf a)} {\it The flow in the conservative case}: The novel fixed point (red circle), with non-zero off-equilibrium activity and non-zero inertial coupling, is the only stable one, with a dynamical critical exponent $z=1.35$ in $d=3$. The equilibrium non-inertial fixed point (black square), $z=2.0$, corresponds to standard ferromagnets (Model A of \cite{HH1977}); the equilibrium inertial fixed point (blue diamond), $z=1.5$, corresponds to superfluids and quantum antiferromagnets (Models E and G of \cite{HH1977}); finally, the active non-inertial fixed point (green triangle), $z=1.73$, corresponds to active matter without reversible coupling between velocity and spin \cite{chen2015critical}; this last fixed point is not connected to the active inertial one onto this plane.
			{\bf b)} {\it The flow with spin dissipation}: Spin dissipation, $\eta$, is a relevant parameter that brings the flow out of the conservative plane; because $\eta$ grows up to infinity with the RG iterations, it is convenient to use the reduced dissipation $\hat\eta=\eta/(1+\eta)$ to represent the flow. If we perturb the active inertial fixed point, $z=1.35$, with some dissipation, the RG flow leaves the $\hat\eta=0$ plane, until it eventually reaches the active overdamped fixed point for $\hat\eta=1$ (green pyramid), where $z=1.73$. 
			When $\hat\eta\neq 0$ it is better to represent the flow through the reduced inertial coupling, $\hat f=(1-\hat\eta)f$, instead of $f$, so that in the overdamped limit, $\hat\eta=1$, we have one less parameter, as the inertial coupling drops out of the calculation. The overdamped fixed point, $z=1.73$, is best seen as belonging to the overdamped $\hat\eta=1$ line, rather than to the conservative but non-inertial line, $\eta=0, f=0$: even though the value of $\hat f$ is the same on the two lines, only the first one corresponds to the correct overdamped limit. All flow lines are actual numerical solutions of the RG equations.}
	}
	\label{fig:flow}
\end{figure*}

Within the RG analysis it is possible to define a set of effective parameters and couplings that are independent of the field dimensions and upon which physical predictions uniquely depend (see SI-IC5). \textcolor{black}{Because the most important factors are activity and inertia, we focus here on their effective coupling constants:}\footnote{We set the cutoff to $1$ to simplify the notation.}
\begin{equation}
c_v=v_0\;  \gamma_v\; \frac{\tGamma^{1/2}}{\Gamma^{3/2}} \;  \quad , \quad  f=  g^2 \; \frac{\tlambda^{\parallel}}{{\lambda^{\parallel}}^2 \Gamma}\   \ .
\end{equation}
$c_v$ is the effective coupling regulating activity, which vanishes for $v_0\to 0$; \textcolor{black}{on the other hand, $f$ quantifies the effective reversible coupling giving rise to  inertial dynamics, and we will therefore refer to it as the inertial coupling constant.}
The scaling dimension of all effective couplings is proportional to $4-d$, indicating that the upper critical dimension is $d_c=4$ and that an expansion in powers of $\epsilon=4-d$ is appropriate. A momentum shell RG calculation at one loop \cite{wilson1972critical, wilson1974renormalizarion} produces 65 Feynman diagrams
(full details are provided in SI-I and SI-II). A rich fixed-point structure emerges (Fig.\ref{fig:flow}a).

\textcolor{black}{The simplest fixed point corresponds to zero activity and zero inertial coupling, $c_v^*=f^*=0$ (black square in Fig.\ref{fig:flow}a). This {\it equilibrium non-inertial} fixed point describes non-active systems, as classical ferromagnets, where the polarization is not coupled to the spin; here $z=2$ at one loop (Model A of \cite{HH1977}). Incompressibility is merely a solenoidal constraint on $\vpsi$, leading to the universality class of dipolar ferromagnets \cite{bruce1974critical}. 
If we perturb this fixed point by adding an inertial coupling we reach the {\it equilibrium inertial} fixed point (blue diamond), which has still zero activity, but non-zero inertial coupling, $c_v^*=0, f^*\neq 0$.  This fixed point describes equilibrium superfluids and antiferromagnets (Models E/F and G of \cite{HH1977}) and it has $z=d/2$, hence $z=1.5$ in $d=3$; here too incompressibility is a solenoidal constraint on $\vpsi$, which changes the static universality class, but not the dynamical one \cite{cavagna2021dynamical}.
This fixed point is unstable against activity, which leads the RG flow towards a novel {\it active inertial} fixed point (red circle), where both $c_v^*\neq0$ and  $f^*\neq0$. The combined effect of activity and inertia lowers significantly the dynamical critical exponent; in $d=3$ we find, $\color{black}{z=1.34(8)}$.
This fixed point is stable against perturbations of all the parameters considered up to now.
As we shall discuss more thoroughly later on, we believe this to be the fixed point describing natural swarms.}

\textcolor{black}{Finally, there is a fourth fixed point (green triangle), which has non-zero activity, $c_v^*\neq 0$, but zero inertial coupling, $f^*=0$, corresponding to $\color{black}{z =1.73}$ in $d=3$; here the inertial reversible terms are absent from the dynamics, hence the polarization is decoupled from the spin \cite{chen2015critical}.
This {\it active non-inertial} fixed point  is stable against activity fluctuations, but as soon as we perturb it with an inertial coupling, $f\neq 0$, the RG flow diverges (shaded area). There is a sound reason for this: the correct way to attain non-inertial dynamics is not to kill the reversible coupling between coordinate and momentum, but to introduce  dissipation and let it take over in the overdamped limit. This is the consistency check our calculation must pass next.}

\begin{figure*}[t]
	\centering
	\includegraphics[width=0.99\textwidth]{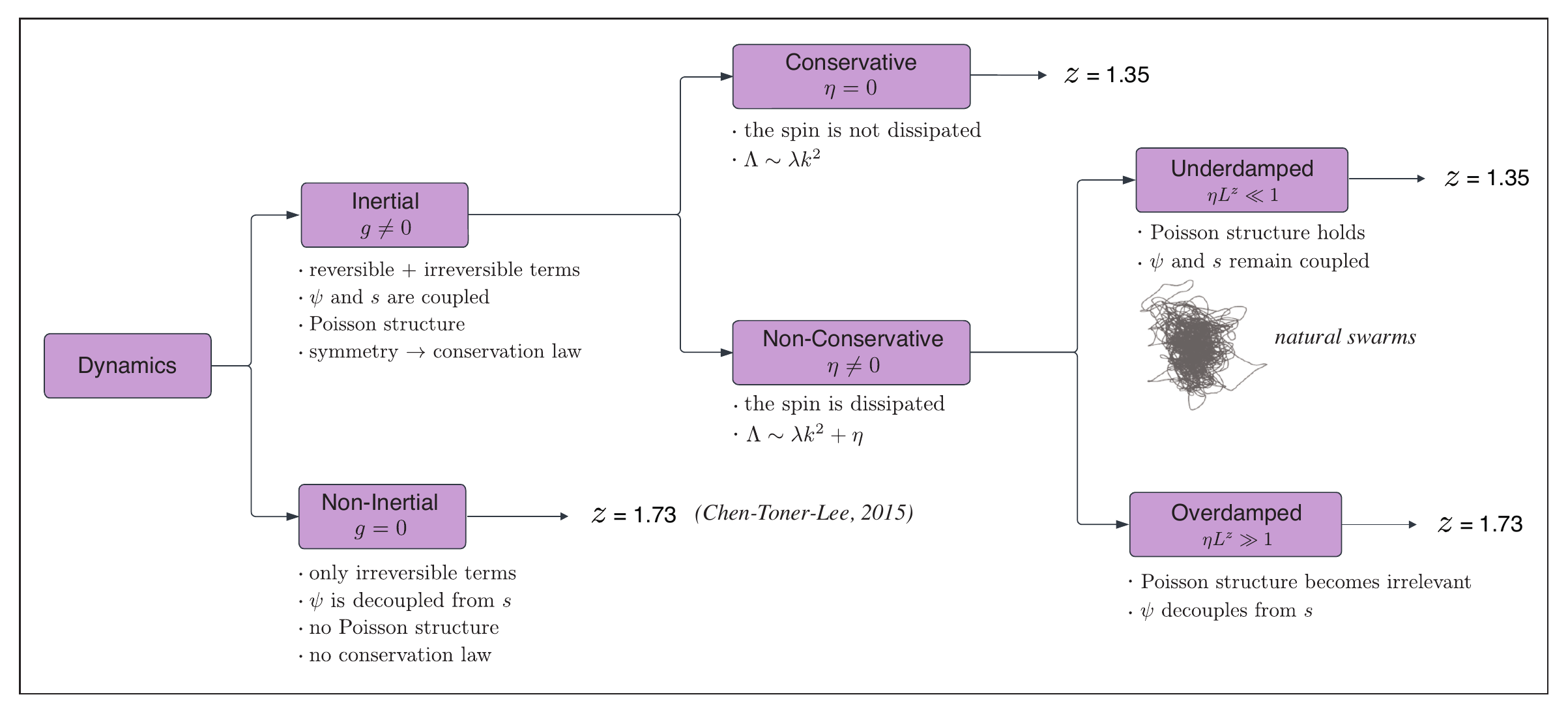}
	\caption{
		\textcolor{black}{
			{\bf Map of the incompressible active theory.} 
			The coarse-grained dynamical equations may either have or not have  reversible terms giving rise to the inertial coupling between the polarization $\psi$ (i.e. the 
			generalized coordinate) and the spin $s$ (i.e. the generalized momentum). 
			In the first case ($g\neq0$) we have an {\it inertial} theory, with a Poisson structure expressing the fact that $s$ is the generator of the rotational symmetry, thus leading to the conservation of the total spin. 
			In the second case ($g=0$) we recover the {\it non-inertial} theory of \cite{chen2015critical}, where polarization is decoupled from the spin and the symmetry does not entail any Poisson structure (the equation for $s$ becomes irrelevant); in this case $z=1.73$. 
			On the other hand, in the inertial theory the irreversible kinetic coefficient of the spin may be either conservative or non-conservative. In the {\it conservative} case there is no spin-dissipation ($\eta=0$), which produces the inertial-conservative fixed point with $z=1.35$. In the {\it non-conservative} case, the kinetic coefficient contains a 
			dissipative term ($\eta\neq 0$), although the impact of dissipation depends on how strong that is compared to system's size.
			In the {\it underdamped} regime, $\eta L^z\ll 1$, collective fluctuations are still ruled by the inertial-conservative fixed point, so that $z=1.35$; this is the regime of natural swarms. Conversely, in the {\it overdamped} regime, $\eta L^z\gg 1$, the Poisson structure is washed out, the spin drops out of the calculation and collective fluctuations are ruled by the fully non-conservative fixed point, hence giving $z=1.73$.}
	}
	\label{fig:tree}
\end{figure*}

\textcolor{black}{So far our theory conserved the total spin, thanks to the Poisson structure generating the reversible terms in the dynamics and to the fact that the irreversible kinetic coefficient $\Lambda$ is zero at $k=0$. Although the Poisson structure has no reasons to change, $\Lambda$ could: within natural swarms we cannot exclude that some spin dissipation exists, not as a result of a violation of the rotational symmetry, but because individual midges might exchange spin with the environment in a way that is unaccounted for in the equations of motion. Spin dissipation is produced by a $k$-independent term $\eta$ in the kinetic coefficient, $\Lambda \sim (\lambda^\parallel +\lambda^\perp)k^2 + \eta$, and similarly in  $\tilde \Lambda$.
The RG calculation (reported in SI-I.E) shows that the scaling dimension of $\eta$ is always positive, so that  if we perturb the active inertial fixed point with $\eta\neq 0$, the RG flow gets  out of the conservative plane and eventually reaches a fixed point at $\eta=\infty$, where polarization decouples from the spin and $z=1.73$ (see Fig. \ref{fig:flow}b). This is the correct way to obtain the overdamped limit in which inertia becomes irrelevant, hence we call this the {\it active overdamped} fixed point (a hopefully clarifying map of the theory is depicted in Fig.\ref{fig:tree}). Yet: if the overdamped fixed point is the only asymptotically stable one, why should we be interested in the inertial fixed point?}

\textcolor{black}{
The answer is that {\it finite-size} systems can be ruled by a partially-stable RG fixed point, if the physical parameters are close enough to it. Consider a ferromagnet slightly above its critical temperature; the stable fixed point is $T^*=\infty$, and yet, if the temperature is close enough to $T_c$, the critical fixed point governs the physics as long as the system's size is smaller than the correlation length, $L\ll\xi$. This is a general mechanism: when the physical couplings are close to a mixed-stability fixed point, the RG flow remains for many iterations in the vicinity of it, and because iterating RG corresponds to observing at larger and larger scales, the flow of a finite-size system may never get out of that basin of attraction. This balance is always regulated by a crossover length scale, $\cal R$, which is in general a more complicated quantity than $\xi$; but the upshot is the same: as long as $L\ll {\cal R}^\kappa$ (where $\kappa$ is the {\it crossover exponent}) the metastable fixed point rules the system \cite{cavagna2019long, cavagna2020equilibrium}. 
How is this relevant for natural swarms? 
The underdamped shape of the dynamic correlation functions in natural swarms (Fig.\ref{fig:crossover}a) is solid experimental evidence that spin dissipation is {\it weak}. On the other hand, the rewiring of the interaction network in swarms occurs over the same time scale as velocity relaxation (Fig.\ref{fig:crossover}b), i.e. activity is {\it strong}. 
Hence, the RG flow starts close to the conservative plane, $\eta=0$, but far from the equilibrium plane, $c_v=0$; as a result, RG rapidly leads the system in the vicinity of the active inertial fixed point, $z=1.35$, lingering there for many iterations, before flowing to the overdamped fixed point (Fig.\ref{fig:crossover}c and \ref{fig:crossover}d). We find $\mathcal R = \sqrt{\lambda^\parallel/\eta}$ and $\kappa=2/z$ (SI-IE2), so that for $L \ll (\lambda^\parallel/\eta)^{1/z}$ a finite-size system is ruled by the active inertial fixed point. Given that $\lambda^\parallel$ is finite along the flow, we conclude that as long as $\eta\, L^{z} \ll 1$, the underdamped inertial scenario must hold; because experimental relaxation {\it is} underdamped (Fig.\ref{fig:crossover}a), we conclude that the dynamical critical exponent in natural swarms is $z=1.35$.
}

\textcolor{black}{
Critical slowing down in natural swarms was first experimentally observed in \cite{cavagna2017swarm}; the spatio-temporal span of the events in that study, though, was somewhat too limited to have an accurate determination of $z$, as the largest swarm had $N=278$ individuals; here we added $8$ new swarming events to the experimental dataset, notably including a swarm of $780$ insects. The relaxation time $\tau$ vs correlation length $\xi$ is reported in Fig.\ref{fig:experiments}a. In \cite{cavagna2017swarm} the exponent was determined through Least Squares (LS) linear regression of $\log\tau$ vs $\log\xi$; however, LS works under the hypothesis that the independent variable is perfectly determined and that all experimental uncertainty is in the dependent variable; when this hypothesis is violated, LS systematically underestimate the slope \cite{sokal1995biometry}. In our experiments errors certainly impact on both $\tau$ and $\xi$, hence LS is not appropriate and this is why $z$ was unfortunately underestimated in \cite{cavagna2017swarm}. Reduced Major Axis (RMA) regression \cite{sokal1995biometry}, on the other hand, treats fluctuations over $x$ and $y$ on the same statistical footing (see Methods and SI-IV); applied to our dataset RMA gives $z_\mathrm{exp}=1.37\pm 0.11$ (Fig.\ref{fig:experiments}). The substantial error bar should make us cautious about the agreement between experiments and theory, also considering the rather uncontrolled approximations our calculation made, most notably incompressibility and the first-order perturbative expansion in powers of $\epsilon$, with $\epsilon=1$. For this reason, we make a final sanity check of our RG calculation through numerical simulations.
}

\begin{figure*}[t]
	\centering
	\includegraphics[width=0.9\textwidth]{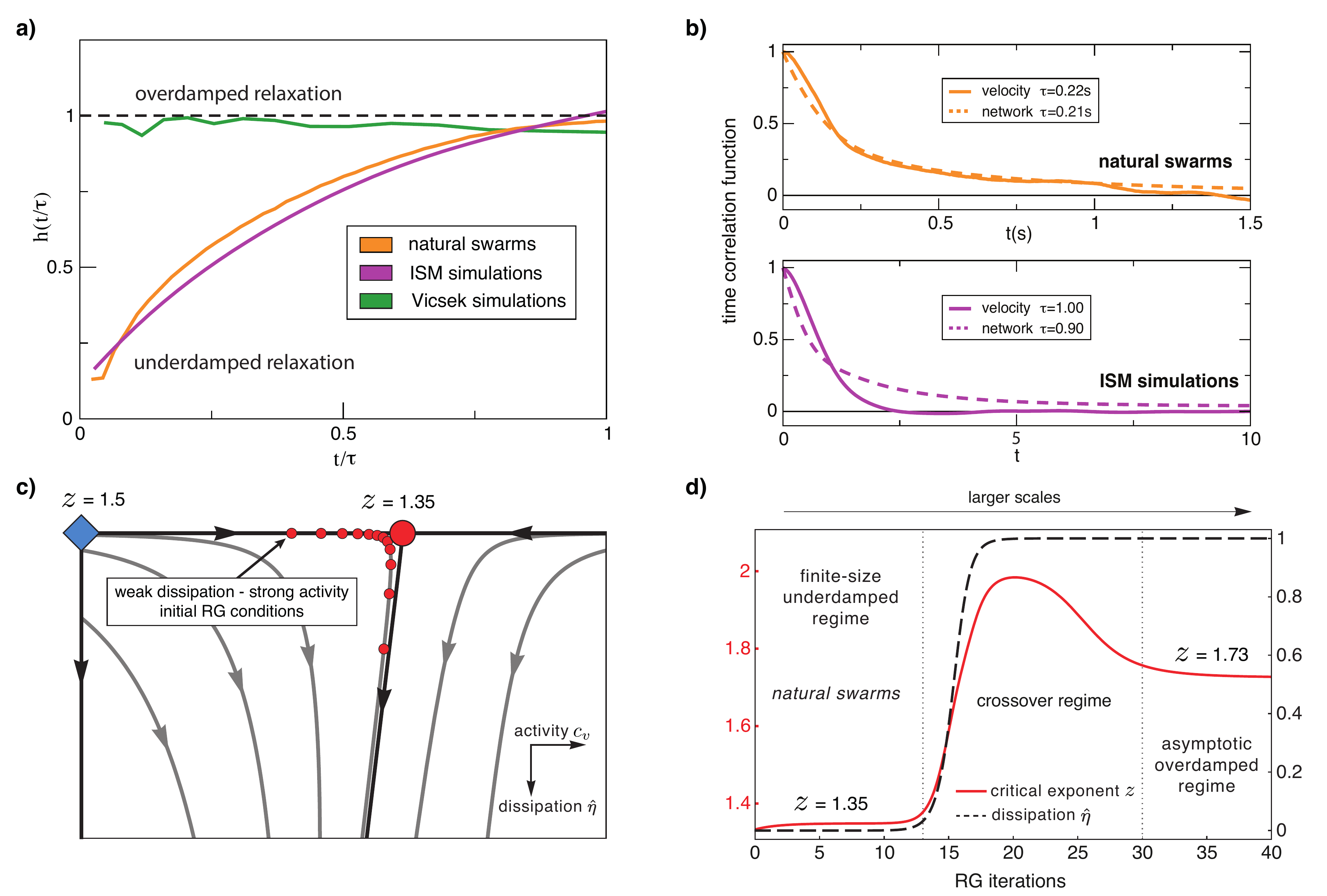}
	\caption{{\bf RG crossover.} 
		\textcolor{black}{
			{\bf a)} {\it Weak dissipation}: Given the velocity correlation function, $C(t)$, and its relaxation time, $\tau$, we can define the shape function as, $h(t/\tau) = -\log C(t/\tau)/(t/\tau)$; in the limit  $t/\tau\to 0$, $h(t/\tau)\to 1$ for overdamped exponential dynamics, while $h(t/\tau)\to 0$ for inertial underdamped dynamics \cite{cavagna2017swarm}. Experiments on natural swarms (orange line) and numerical simulations of near-critical ISM (purple line), both display underdamped inertial relaxation. The Vicsek model, on the other hand, belongs to the overdamped class (green line - data from \cite{cavagna2017swarm}).
			{\bf b)} {\it Strong activity}: Velocity (full line) vs network (dashed line) dynamical correlation functions, for natural swarms (orange) and near-critical ISM simulations (purple). The network correlation function measures the fraction of particles remaining within the $n_c$ nearest neighbours after a time $t$ (see SI-III), hence it quantifies how quickly the interaction network reshuffles with time. In both natural swarms and ISM, the network decorrelates on the same time scale as the velocity, hence they are strongly active systems. Here $n_c=18$, which is the mean number of interacting neighbours in simulations; in the SI-III Fig.15 we show that in natural swarms the two timescales are the same over all spatial scales.
			{\bf c)} {\it Crossover of the flow}: A close up of the RG flow around the active inertial fixed point shows that when the flow starts at weak dissipation and strong activity, it first rapidly approaches the active inertial fixed point, staying in its neighbourhood for many RG iterations, and then it crosses-over to the overdamped regime  (red dots).
			{\bf d)} {\it Crossover of the critical exponent}: The RG evolution of the dynamical critical exponent $z$ and of the reduced dissipation, $\hat\eta=\eta/(1+\eta)$, along the crossover flow line depicted in panel (c). The RG crossover from underdamped to overdamped fixed points corresponds to an actual crossover in real space, such that  $z=1.35$ for $L\ll{\cal R}^{2/z}$ and  $z=1.73$ for $L\gg{\cal R}^{2/z}$.}
	}
	\label{fig:crossover}
\end{figure*}

\begin{figure*}[t!]
	\centering
	\includegraphics[width=0.9\textwidth]{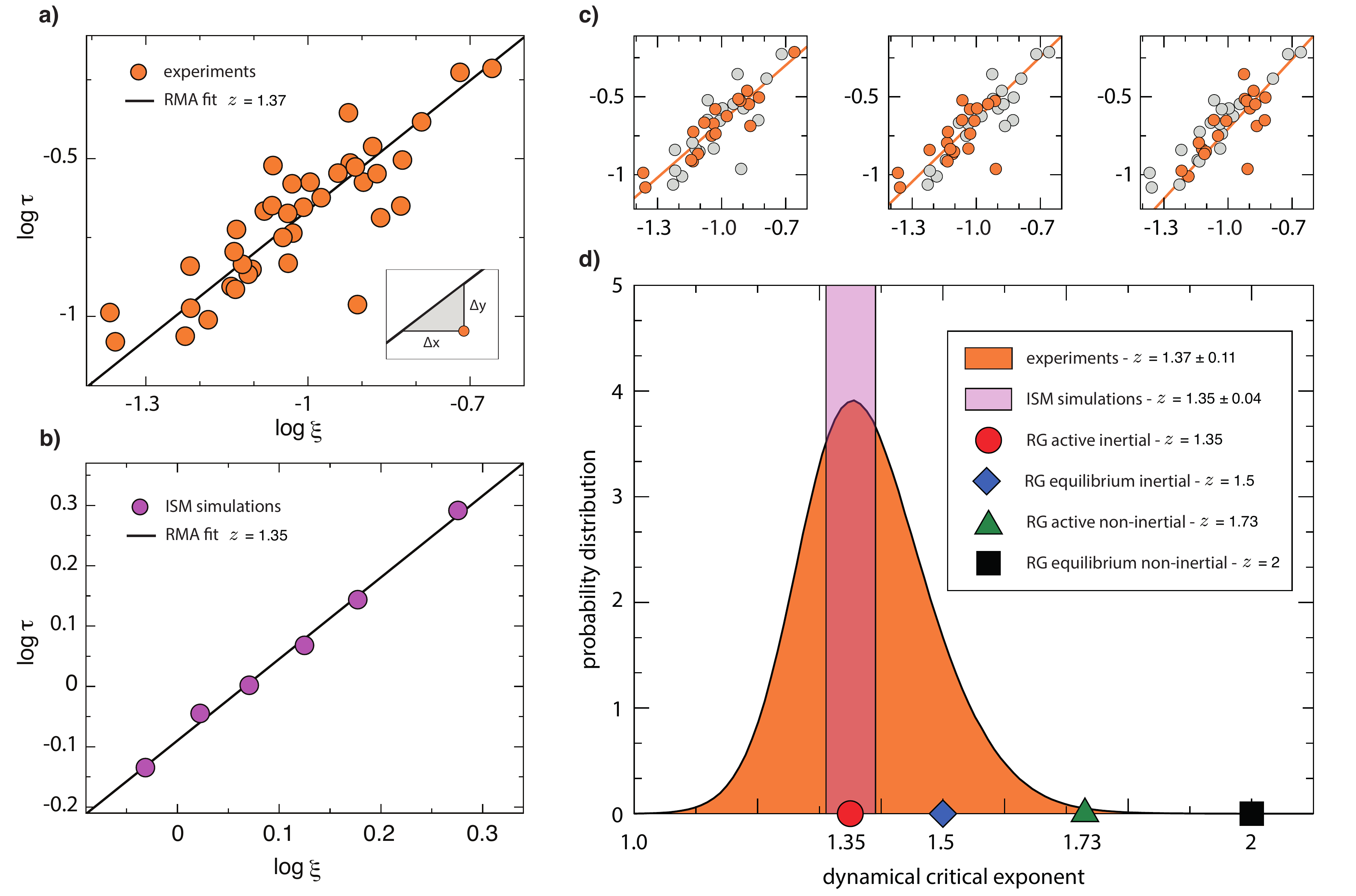}
	\caption{{\bf Experimental and numerical results.}  
		\textcolor{black}{
			\textbf{a)} {\it Experiments}: Logarithm of the relaxation time $\tau$ vs logarithm of the correlation length $\xi$ in natural swarms (logarithms are in base $10$); the critical exponent $z_\mathrm{exp}$ is the slope of the linear fit. Because experimental uncertainty affects both $\tau$ and $\xi$, standard Least Squares (LS) regression (which assumes no uncertainty on the abscissa) systematically underestimates the exponent. Reduced Major Axis (RMA) regression treats uncertainty on the two variables in a symmetric way by minimizing the sum of the areas of the triangles formed by each point and the fitted line (inset). RMA regression gives $z_\mathrm{exp}=1.37$.
			\textbf{b)}  {\it Numerical simulations}: $\log\tau$ vs $\log\xi$  in the Inertial Spin Model (ISM). Numerical errors are so small that LS and RMA give the same result, $z_\mathrm{sim}=1.35$; the LS error is $\delta z_\mathrm{sim} = 0.04$.
			\textbf{c)}	{\it Experimental resampling}: To estimate the error bar on the experimental exponent we use a resampling method: we randomly draw $10^7$ subsets with half the number of points and in each subset we determine $z$ using RMA; we report here three such random subsets (orange: selected point - grey: unselected point). Rare experimental fluctuations under resampling can produce an unphysical value of $z$ smaller than $1$; this, however, happens in only the $0.002\%$ of data subsets.
			\textbf{d)}  {\it Final comparison}: Probability distribution of the experimental critical exponent (orange) from the resampling method of (b); the standard deviation of this distribution gives the error on the experimental exponent, $\delta z_\mathrm{exp}= 0.11$. The vertical band (purple) indicates the position and error of the numerical critical exponent; coloured symbols indicate the various RG fixed points values of $z$.
		}
	}
	\label{fig:experiments}
\end{figure*}

\textcolor{black}{
The field theory we studied is the coarse-grained expression of the Inertial Spin Model (ISM - \cite{cavagna2015flocking}), in which the particles' velocities are rotated by the spins, while the spins are acted upon by the social alignment forces,
\begin{equation}
\label{microISM}
\begin{split}
\frac{d  \boldsymbol v_i }{dt} &= \frac 1 {\chi} \boldsymbol s_i \times \boldsymbol v_i \ , \\
\frac{d \boldsymbol s_i}{dt} &=  
\boldsymbol {v}_i \times 
\frac{J}{n_i} \sum_j n_{ij}(t)\; \boldsymbol {v}_j
 - \frac{\eta}{\chi} \boldsymbol {s_i} + \boldsymbol  {v}_i \times \boldsymbol{\zeta}_i  \ ,  \\
\frac{d \boldsymbol r_i}{dt} &= \boldsymbol v_i  \ ,
\end{split}
\end{equation}
with noise correlator, $\langle \boldsymbol{\zeta}_i (t) \cdot \boldsymbol{\zeta}_j(t')\rangle = 2dT\; \eta\; \delta_{ij}\delta(t-t')$; $\chi$ is the generalized  turning inertia, $J$ the alignment strength, $\eta$ the spin dissipation\footnote{
\textcolor{black}{With a small abuse of notation, we use the same symbol, $\eta$,  for both the microscopic dissipation and its mesoscopic counterpart.}}, and $T$ the noise amplitude (or  temperature); the adjacency matrix $n_{ij}(t)$ is defined by a metric interaction radius, $r_c$. 
We want to compare the numerical results with the incompressible RG calculation; hence, even though we do {\it not} impose incompressibility in the simulation, we employ a normalized alignment strength, $J/n_i=J/\sum_k n_{ik}$, 
a prescription known to make alignment-based models less prone to phase separation \cite{chepizhko2021revisiting}; moreover, we monitor each simulation to be sure that phase separation does not occur. 
We run three-dimensional simulations in the near-ordering scale-free regime, where $\xi \sim L$ (see SI-VB). 
In the overdamped limit, $\eta\to\infty$, the ISM converges to the non-inertial Vicsek model \cite{cavagna2015flocking}, exactly as our dynamical field theory converges to the non-inertial theory of \cite{chen2015critical}; but our aim is to check that in the {\it underdamped} regime the dynamics of a finite-size ISM simulation is ruled by the inertial fixed point; hence, the dissipation $\eta$ has been chosen small enough to yield inertial relaxation, as in natural swarms (see Fig.\ref{fig:crossover}a). On the other hand, the speed $v_0=|{\boldsymbol v}_i|$, has been chosen large enough to be in the active regime, namely to have a network relaxation time of the same order as the velocity relaxation time (see Fig.\ref{fig:crossover}b). 
Full details of the simulation are reported in the Methods and in the SI-V. Relaxation time vs correlation length is shown in Fig.\ref{fig:experiments}b; numerical errors are quite small, hence LS and RMA give the same value of $z$, and we can therefore calculate $\delta z$ simply through LS. The result is $z_\mathrm{sim}=1.35\pm 0.04$, in remarkable agreement with the RG theoretical prediction. This consistency also validates the idea that the incompressible theory can indeed be used to describe finite-size compressible systems, as long as density fluctuations are not strong.
}

\textcolor{black}{
A final comment is in order: of the two keystones of the Renormalization Group, rescaling and coarse-grainin, only the latter produces anomalous critical exponents, giving rise to non-trivial collective behaviours. The technical fingerprint of coarse-graining is the presence of Feynman diagrams: this is the case of the present calculation, which therefore probes the core element of RG. The consistency between theory, simulations and experiments attained here strongly supports the idea that the RG -- and its most fruitful consequence, universality -- may have an incisive impact also in biology.}

This work was supported by ERC Advanced Grant RG.BIO (n.785932) to AC. TSG was supported by grants from CONICET, ANPCyT and UNLP (Argentina). We thank E. Branchini, F. Cecconi, M. Cencini, M. Testa, G. Parisi, L. Peliti, J. Sethna, and V. Skultety for discussions.

\begin{center}
\textbf{Author Contributions}\\
\end{center}
AC, IG and TSG designed the study. AC, LDC, IG, TSG, GP and MS derived the structure of the dynamical field theory. LDC, GP and MS, coordinated by AC and TSG, carried out the RG calculation. LDC designed the code to evaluate the Feynman diagrams, which was further developed with the help of MS. \textcolor{black}{GP performed the numerical simulations.} SM and LP performed the $3D$ tracking of the experimental data and estimated the exponent; SM measured experimental swarm activity vs relaxation. MS, with the help of LDC and GP, wrote the Supplementary Information. AC wrote the paper.

\begin{center}
\textbf{Corresponding Authors}\\
\end{center}
Correspondence should be addressed to Luca Di Carlo (luca.dicarlo@uniroma1.it), \textcolor{black}{Giulia Pisegna (giulia.pisegna@uniroma1.it)} or Mattia Scandolo (mattia.scandolo@uniroma1.it).

\begin{center}
\textbf{Data Availability}\\
\end{center}
The data that support the plots within this paper and other findings of this study are available from the corresponding authors upon request.

\begin{center}
\textbf{Code Availability}\\
\end{center}
All codes used for the data processing and other findings of this study are available upon request.


\begin{center}
\textbf{METHODS}\\
\end{center}

\textcolor{black}{
{\bf Experiments.}
Data were collected in the field by acquiring video sequences using a multi-camera system of three synchronized cameras (IDT-M5) shooting at 170 fps. 
Two cameras (the stereometric pair) were at a distance between 3m and 6m depending on the swarm and on the environmental constraints. A third camera, placed at a distance of 25cm from the first camera was used to solve tracking ambiguities. 
We used Schneider Xenoplan 50mm f =2.0 lenses. Typical exposure parameters: aperture f =5.6, exposure time 3ms. Recorded events have a time duration between 0.88 and 15.8 seconds (see Table I of the SI-II). More details can be found in \cite{attanasi2014collective}. 
To reconstruct the $3d$ positions and velocities of individual midges we used the tracking method described in \cite{attanasi2015greta}. Our tracking method is accurate even on large moving groups and produces very low time fragmentation and very few identity switches, therefore allowing for accurate measurements of time-dependent correlations.}
\vskip 0.5 truecm

\textcolor{black}{
{\bf Fit of the dynamic critical exponent.}
Dynamic scaling states that the relaxation time $\tau_k$ at wavelength $k$ and correlation length $\xi$ are linked by the relation,
$\tau_k=\xi^z\; \Omega(k\xi)$, where $\Omega$ is a scaling function.
To infer the value of $z$ from experimental data, we measured the relaxation time $\tau_k$ of the mode at wavelength $k=\xi^{-1}$ in different swarming events. Experimental evaluation of $\tau_k$ and $\xi$ is discussed in SI-IV, and follows \cite{cavagna2017swarm}. 
Dynamic scaling in this case reduces to $\tau \sim \xi^z$ (where $\tau\equiv\tau_{k=\xi^{-1}}$); hence, $\log \tau = z \log \xi + c$.
In \cite{cavagna2017swarm} $z$ was fitted through a standard Least Squares (LS) regression, which gave $z_\mathrm{exp}=1.12\pm 0.16$ on the dataset of \cite{cavagna2017swarm}, and $z_\mathrm{exp}=1.16\pm 0.12$ on the current larger dataset; the problem with LS, though, is that it assumes that experimental uncertainty is only present in the dependent variable $y$, which is not true for our experimental data, as both $\tau$ and $\xi$ are subject to experimental uncertainty; when using it on a dataset where the error affects also $x$, LS systematically underestimate the slope \cite{sokal1995biometry}. Therefore, LS is not a good method in our case. Reduced Major Axis (RMA) regression, on the other hand, is 
a method that works under the hypothesis that both $x$ and $y$ are affected by uncertainties \cite{woollet1941method, samuelson1942note}.
RMA fits a set Gaussian variables $x_i$ and $y_i$, with homogeneous variance $\sigma_x^2$ and $\sigma_y^2$ to a regression line, $y=f(x)=\alpha x+\beta$,
and it  determines $\alpha$ and $\beta$ through the minimization of the sum of the areas of the triangles formed between each point and the regression line with sides parallel to the axis (see insert in Fig.\ref{fig:experiments} panel a). For each point, the area of this triangle is given by,
$\left|\Delta x_i \Delta y_i\right|/2$,
where,
\begin{align}
    \Delta x_i&=x_i - f^{-1}(y_i)=\frac{\alpha x_i+\beta-y_i}{\alpha}\\
    \Delta y_i&=y_i - f(x_i)=y_i-\alpha x_i-\beta  \ .
\end{align}
The function to be minimized is therefore,
\begin{equation}
     \Sigma\left(\alpha,\beta\right) = \frac{1}{N}\sum_{i=1}^N \frac{\left(y_i-\alpha x_i-\beta\right)^2}{\left|\alpha\right|} \ .
\end{equation}
The minimization equations, $\partial_\alpha\Sigma=\partial_\beta\Sigma=0$, give,
\begin{equation}
    \alpha^2=\frac{\E{y^2}-\E{y}^2}{\E{x^2}-\E{x}^2}\quad , \quad
    \beta=\E{y}-\alpha \E{x} \ ,
\end{equation}
where $\E{g(x,y)}=\frac{1}{N}\sum_{i=1}^N g(x_i,y_i)$.
The sign of $\alpha$ is the same as the sign of the correlation between $x$ and $y$.
A further benefit of RMA compared to other methods, such as Least Squares or Effective Variance (both discussed in SI-IVB), is that the fit is invariant under an interchange of variables, $y(x)$ vs $x(y)$. Moreover, RMA is also invariant under any scale change of the variables, hence it is not sensitive to the values of $\sigma_x$ and $\sigma_y$, at variance with other methods.
RMA is the only method, among those in which the fitted coefficient can be expressed in terms of elementary regression coefficients, that obeys both properties above \cite{samuelson1942note}.
In SI-IVB we also describe the Effective Variance (EV) regression method, which requires the experimental errors $\delta\tau$ and $\delta\xi$ as an input; we use EV with two different estimates of the (most problematic) experimental error $\delta\tau$, and obtain results compatible with RMA ($z_\mathrm{exp}=1.32 \pm 0.18$ and $z_\mathrm{exp}=1.34 \pm 0.18$). Given the significant difficulties in assigning a univocal experimental error on $\tau$ to each swarm (see SI-IVA3), we prefer to quote the RMA result $z_\mathrm{exp}=1.37\pm0.11$ -- which is error-neutral -- as our most confident determination of the exponent.
}

\textcolor{black}{ {\bf Numerical simulations.} Eqs \eqref{microISM} of the microscopic ISM are numerically implemented using the RATTLE algorithm to enforce the constraint $|\boldsymbol v_i(t)| = v_0$. Simulations are performed in $d=3$ in cubic boxes with periodic boundary conditions; average density is fixed at $\rho_0=1$ and the sizes explored are  $N=256, 384, 512, 729, 1024, 2048$, with $N$ the number of particles. The effective inertia is $\chi=1$ and the alignment strength is $J=18$; the metric interaction range is  $r_c=1.6$, corresponding (at this density) to an average number $n_c\sim 18$ of interacting neighbours; the microscopic spin dissipation is $\eta =1$ and the speed is $v_0=2$; this choice of $\eta$ and $v_0$ ensures that the dynamics is clearly underdamped and active (see main text). The temperature (or noise strength) $T$ serves as control parameter of the order-disorder transition; we explored the interval $T \in [1:8]$.  The time-step of integration is chosen as $dt=10^{-4}$. For each size $N$ we identify a finite-size `critical' point, $T_c(N)$: at every value of $T$  we run 5  independent samples, initialised with polarized configurations, of length $6-8 \times 10^5$ steps increasing with the system size; we compute the susceptibility $\chi$ and we take as $T_c(N)$ the point where this quantity reaches a maximum. We measure the correlation length with the inverse of the wave-number $k=1/\xi$ where the static correlation function at this temperature peaks, and we calculate the relaxation time $\tau$ of the velocity's  $C(k,t)$ in complete analogy with the analysis of experimental data. This procedure ensures that the systems are always in a near-critical scaling regime, with the correlation length scaling linearly with the linear system's size $L$ (see SI-VB). 
}
 
\putbib
\end{bibunit}

\onecolumngrid
\newpage

\begin{bibunit}
\large
\begin{center}
\textbf{Supplementary Information\\ Natural Swarms in $\bf 3.99$ Dimensions}\\
\end{center}
\normalsize

\tableofcontents

\section{The Renormalization Group calculation}

The calculation we are going to describe here is the final point of a rather long trajectory, so that it may be useful for the reader to keep in mind the sequence of the main works leading to the present result. The hydrodynamic theory originally developed by Toner and Tu for flocks \cite{tonertu1995}, can also be applied to swarms when studied in its critical phase; this was done in the incompressible case in \cite{chen2015critical}, which is our starting point for the active incompressible non-inertial branch of our theory. The crossover between the non-active (equilibrium) fixed point to the active (off-equilibrium) fixed point in this theory was studied in \cite{cavagna2020equilibrium}, where it was also established that the results of the incompressible RG calculation agree perfectly well with simulations of the finite-size compressible model, provided that density fluctuations are mild (no phase separation).  The experimental evidence of an inertial coupling between velocity and spin, and therefore of the need to go beyond Toner-Tu theory, was first found for flocks in \cite{attanasi2014information} and later for swarms in \cite{cavagna2017swarm}; the microscopic model of active self-propelled particles obeying such inertial mode-coupling dynamics is the Inertial Spin Model (ISM), which is defined in detail in \cite{cavagna2015flocking}; the Vicsek model \cite{vicsek1995novel} is the overdamped limit of the ISM \cite{cavagna2015flocking}: when spin dissipation is very large compared to inertia, the spin becomes irrelevant and one reduces to a theory for just one degree of freedom, the velocity (or polarization). A discussion of the connections between Vicsek model, Toner and Tu theory and the ISM can be found in \cite{cavagna2018correlations}.

The path from the microscopic inertial active model (the ISM) to the relative coarse-grained dynamical field theory that we study here has required some intermediate steps. First, the equilibrium version (i.e. fixed network, or non-active) of the ISM has a field-theoretical benchmark in Model E (planar case) and Model G ($3d$ case), according to the classic Halperin-Hohenberg classification of dynamical critical phenomena \cite{HH1977}; these equilibrium models are really a fundamental starting point for understanding the inertial mode-coupling dynamics of a primary field coupled to the generator of its rotations. These equilibrium non-active models have perfect conservation of the spin - namely zero dissipation - which is unlikely to be exactly true in biological systems, hence in \cite{cavagna2019short, cavagna2019long} we studied the effect of dissipation at equilibrium on this kind of theories, finding that for finite-size systems the inertial fixed point of the mode-coupling theories is still relevant. Finally, we assessed (again in the non-active case) how to impose the incompressibility constraint (which is a solenoidal constraint on a fixed network) on a theory where there is inertial coupling between primary field and spin in \cite{cavagna2021dynamical}. The present calculation finally analyses the active out-of-equilibrium incompressible case in presence of inertial coupling between velocity and spin.

\subsection{Derivation of the equations of motion}

\subsubsection{Describing activity: the Toner and Tu theory in the incompressible case}

The starting point of our field-theoretical description is given by the hydrodynamic theory developed by Toner and Tu \cite{tonertu1995}, representing the minimal description of an active system with rotational invariance.
This theory represents the hydrodynamic counterpart of the Vicsek model (VM) \cite{vicsek1995novel}, although it describes in general any model which shares the same symmetries and conservation laws.
The VM describes the dynamics of a collection of self-propelled particles with fixed speed $v_0$, whose velocities interact through a dynamic alignment as in equilibrium XY or Heisenberg models. The continuous theory of Toner and Tu represents a sort of moving ferromagnet, combining activity with Model A dynamics $\partial_t \vv=-\delta_\vv \Ham+\vecc{\theta}$ \cite{HH1977}, with $\Ham$ being the Landau-Ginzburg free-energy.
This makes the Toner and Tu theory fall into the class of overdamped Langevin equations, in which the {\it social} force $\delta_\vv \Ham$ acts directly on the time-evolution of the order parameter. In the incompressible limit - see Section \ref{sec:inc} - the equation of motion for the velocity field in the near-critical Toner and Tu theory is given by \cite{chen2015critical},
\begin{equation}
	\partial_t \vv + \gamma_v  \left(\vv \cdot\vnabla\right) \vv= \Gamma \nabla^2 \vv - \left(m + J v^2 \right)\vv  - \vnabla\mathcal{P} + \vecc{\theta}\ .
	\label{eq:ctl}
\end{equation}
where on the l.h.s. we can recognise the material derivative $D_t\vv=\partial_t\vv+\gamma_v \left(\vv \cdot\vnabla\right) \vv$.
Since the active force breaks Galilean invariance \cite{tonertu1998}, the coefficient $\gamma_v$ needs not to be equal to unity.
On the r.h.s. we have the alignment force $\nabla^2\vv$, the active force coming from a Landau potential, the pressure and a gaussian white noise of variance $2\tGamma$.
A RG analysis of Eq. \eqref{eq:ctl} near criticality predicts at one loop a dynamic critical exponent of $z=1.73$ in $d=3$ dimensions \cite{chen2015critical}.

When $\gamma_v=0$, namely in absence of advection, equilibrium Model A's critical behaviour is recovered, with the dynamic exponent being $z=2$ at one-loop.
A crossover thus occurs between the non-active equilibrium and the active off-equilibrium critical behaviour \cite{cavagna2020equilibrium}.
The microscopic parameter controlling this crossover is the microscopic speed $v_0$; to see this it is convenient to work with the coarse-grained orientation field $\vpsi$ (or local polarization) rather than with the velocity field $\vv$.
In models as the VM, in which each individual has the same speed $v_0$, the connection between these two fields is simply given by,
\begin{equation}
	\vv\left(\vx,t\right)=v_0 \vpsi\left(\vx,t\right) \ .
	\label{speed}
\end{equation}
Written in terms of the field $\vpsi$, equation \eqref{eq:ctl} becomes,
\begin{equation}
	\partial_t \vpsi + v_0 \gamma_v  \left(\vpsi \cdot\vnabla\right) \vpsi= \Gamma \nabla^2 \vpsi - \left(m + J \psi^2 \right)\vpsi  - \vnabla\mathcal{P} + \vecc{\theta}\ ,
	\label{eq:ctl_psi}
\end{equation}
in which the explicit dependence of the advection term on the microscopic parameter $v_0$ clarifies the mechanism underlying the crossover between equilibrium Model A and off-equilibrium active theory \cite{cavagna2020equilibrium}.

\subsubsection{Restoring inertia: equilibrium mode-coupling dynamics}\label{sec:MG}
Experimental evidence about swarms' temporal correlation function shows that an inertial structure - vanishing derivative for small times - absent in the incompressible Toner and Tu theory, is needed to describe natural swarms \cite{cavagna2017swarm}. This is confirmed by the discrepancies between the experimental value of the dynamic critical exponent $z$, \textcolor{black}{found to be $z\simeq1.37$ in natural swarms}, and the theoretical prediction of $z=1.73$ of the Toner and Tu incompressible theory in $d=3$ \cite{chen2015critical}.
As proposed in \cite{attanasi2014information} for the case of flocks, restoring inertia can be done by recognizing that, although Eq. \eqref{eq:ctl} is invariant under rotations, there is no trace of a conservation law associated with this symmetry.
According to Noether's theorem, when a theory is invariant under a given symmetry, the generator of this symmetry is conserved.
The presence of conservation laws heavily affects the critical properties of a system, leading to completely different dynamic behaviours.
Since Toner and Tu theory is built to be invariant under rotational symmetry, our aim is to couple \eqref{eq:ctl} with the conservation law associated with the rotational invariance, in order to restore inertial behaviour.

\textcolor{black}{A possible way to restore inertial behaviour has been proposed in \cite{cavagna2015flocking}, where a model named Inertial Spin Model (ISM) has been introduced to provide a theoretical explanation for information propagation in flocks \cite{attanasi2014information}. The ISM shares many common features with the Vicsek Model, but has one main (and crucial) difference: the aligning force does not act directly on $\partial_t \vv$, but is mediated by the generator of rotational symmetry. This new variable, in analogy with quantum mechanics, has been called {\it spin}, since it represents the generator of rotations in the {\it internal} space of the velocities. It must not be confused with angular momentum, which is the generator of rotations in positions' space. The spin is a measure of how much an individual is rotating around its own axis; more precisely, it is proportional to the curvature of the trajectory, namely to the inverse of its radius of curvature: all individuals sharing the same spin undergo equal radius turns rather than parallel-path turns \cite{attanasi2014information}.
Conservation (or slow dissipation) of the total spin has huge impacts on the dynamics of a swarm or a flock.
The presence of a spin-velocity coupling makes the spin responsible to carry information, giving rise to second-sound propagation. The presence of second-sound modes even close to the ordering transition (called `paramagnons' in condensed matter) was also found experimentally in natural swarms \cite{cavagna2017swarm}, supporting the idea that spin-velocity mode-coupling is an essential mechanism also of these system.}

When shifting our attention to hydrodynamics, inertia can be restored by dynamically coupling the velocity/polarization field with to spin field \cite{cavagna2019short}.
The global conservation of the spin allows it to fluctuate on space-time-scales comparable with those of critical fluctuations, thus making spin-velocity couplings relevant in the RG sense. To simplify the discussion, we will first discuss the effects of restoring inertia in absence of activity. At equilibrium, a mode-coupling interaction between an order parameter $\vpsi$ and its spin $\vs$ arises from their Poisson-bracket relation \cite{HH1977},
\begin{equation}
    \left\{ s_{\alpha\beta}\left(\vx\right) , \psi_\gamma\left(\vx'\right) \right\} = 2\, g\, \II_{\alpha\beta\gamma\nu} \psi_\nu\left(\vx\right) \delta\left(\vx-\vx'\right)\ ,
    \label{eq:poisson}
\end{equation}
which encodes the fact that $\vs$ is the generator of rotations of $\vpsi$ (repeated indices imply a summation over them). \textcolor{black}{The parameter $g$ is the reversible coupling regulating the symplectic structure, i.e. the inertial coupling between polarization and spin.}
In general, when the order parameter is a $n$-dimensional vector, the generator of its rotations $\vs$ is a $n\times n$ anti-symmetric tensor \cite{SSS1975}.
The tensor $\II$ represents the identity in the space of $s_{\alpha \beta}$, and it is given by,
\begin{equation}
	\II_{\alpha\beta\gamma\nu}=\frac{\delta_{\alpha\gamma} \delta_{\beta\nu} - \delta_{\alpha\nu} \delta_{\beta\gamma}}{2}\ ,
	\label{eq:II}
\end{equation}
with the factor $\frac{1}{2}$ ensuring that $\II_{\alpha\beta\gamma\nu}s_{\gamma\nu}=s_{\alpha\beta}$ and $\II_{\alpha\beta\sigma\tau}\II_{\sigma\tau\gamma\nu}=\II_{\alpha\beta\gamma\nu}$.

We work here with an order parameter of generic dimension $n$, although in the physical case we have $n=3$.
This choice might seem inconvenient at first glance.
When $n=3$, the spin $\vs$ can be written as a $3$-dimensional vector, lightening the tensorial structure and reducing the number of indices. This comes from the fact that when $n=3$, the plane on which the rotation occurs can be uniquely identified by the vector orthogonal to it, while this does not happen when $n\neq3$.
However, there is an important reason to work with a tensorial spin, rather than a vectorial one. In the following, we will impose incompressibility, which requires $\vv$ and therefore $\vpsi$, to have the same dimension $d$ as space.
Although the physical case is given by $n=d=3$, the RG perturbative expansion is performed by expanding $d$ near the upper critical dimension $d_c=4$, hence one is forced to work with an order parameter of dimension $n=d\sim d_{c}$ to correctly perform the RG perturbative expansion.
The spin associated with an $n$-dimensional order parameter, in generic dimension $n$, is represented by a $n\times n$ anti-symmetric tensor rather than a vector.
Therefore, we will need to work with this more generic form.

The equilibrium dynamics of a near-critical system in which $\vs$ is conserved, known as Model G in the physical case of a three dimensional order parameter ($n=3$) \cite{HH1977} and generalized by the Sasvari-Schwabl-Szepfalusy (SSS) model in $n$ dimensions \cite{SSS1975,SSS1977}, can be constructed following the classic Mori-Zwanzig formalism \cite{mori1974new,zwanzig1961memory} and it is given by,
\begin{equation}
	\partial_t \psi_\alpha = - \Gamma \dder{\Ham}{\psi_\alpha} + g\, \psi_\beta \dder{\Ham}{s_{\alpha\beta}} + \theta_\alpha\ ,
	\label{eq:psieq}
\end{equation}
\begin{equation}
	\partial_t s_{\alpha\beta} =- \Lambda_{\alpha\beta\gamma\nu} \dder{\Ham}{s_{\gamma\nu}} + 2 g \, \II_{\alpha\beta\gamma\nu} \psi_\gamma \dder{\Ham}{\psi_\nu}  + \zeta_{\alpha\beta}
	\label{eq:seq}
\end{equation}
Here the free-energy functional $\Ham$ is chosen to take the usual Landau-Ginzburg form for the critical field $\vpsi$ while it is gaussian for the spin field (we set to $1$ the inertia, which does not get any renormalization),
\begin{equation}
	\Ham = \int d^d x \left[ \frac{1}{2} \psi_\alpha \left(-\nabla^2+r\right) \psi_\alpha + \frac{u}{4} \left(\psi_\alpha \psi_\alpha\right)^2 + \frac{1}{4} s_{\alpha\beta}s_{\alpha\beta} \right] \ .
	\label{eq:ham}
\end{equation}
The square gradient enforces local alignment of the order parameter $\vpsi$, while $r\psi^2+u\psi^4$ is the modulus' confining potential.
At mean-field level, when $r<0$ the ground state exhibits symmetry breaking and an ordered phase is observed, while for $r>0$ the ground state is given by the disordered state with zero mean polarization.

Inertia is restored thanks to the presence of mode-coupling interactions that encode the conservative nature of the dynamics, arising as a consequence of the Poisson-bracket relation \eqref{eq:poisson}.
The term $\partial_t s\sim g \psi \delta_\psi\Ham$ represents the action of the {\it force} on the dynamics of the spin, rather than directly on the order parameter.
The indirect action of this force on the dynamics of $\vpsi$ is guaranteed by the term $\partial_t\psi\sim g \psi \delta_s \Ham$, which expresses the rotation of $\vpsi$ induced by the conservation of $\vs$.
This mode-coupling mechanism restores the inertial structure of the equations of motion, thus allowing to describe the behaviour observed experimentally in swarms in the field \cite{cavagna2017swarm}.
On the other hand, the terms $\partial_t\psi = -\Gamma \delta_\psi \Ham$ and $\partial_t s\sim -\Lambda \delta_s\Ham$ represent dynamic relaxations, giving rise to the diffusion and transport phenomenology typical of stochastic statistical systems. These relaxation terms are thus complemented by the white Gaussian noises $\vecc{\theta}$ and $\vecc{\zeta}$, whose variances are given by Einstein relations when the system is at equilibrium.
The dissipative constant (or kinetic coefficient) $\Gamma$ rules the relaxation of the order parameter and it is a crucial player in determining the dynamic exponent $z$ since it fixes the time-scale on which relaxation occurs.
Similarly, the kinetic tensor $\Lambda_{\alpha\beta\gamma\nu}$ rules the relaxation of the spin.
When the total spin is conserved, the tensor $\Lambda$ is proportional to $\nabla^2$  \cite{HH1977}, and in the isotropic theory it takes the form,
\begin{equation}
\Lambda_{\alpha\beta\gamma\nu}= - \lambda \,\nabla^2\,\II_{\alpha\beta\gamma\nu} \ .
\label{begabega}
\end{equation}
In this theory the total spin is conserved, 
\begin{equation}
\dot S_{\alpha\beta}(t) = \frac{d}{dt} \int d^dx\; s_{\alpha\beta}(x,t) = 0
\end{equation}
An RG analysis of this field theory shows that the critical dynamic exponent is given by $z=\frac{d}{2}$ \cite{HH1977,SSS1977,dedominicis1978field}, that is  $z=1.5$ in $d=3$.

\textcolor{black}{Is behavioural inertia the only way in which experimental observations of \cite{cavagna2017swarm} can be explained? The possibility that other mechanisms take part in explaning those observation cannot be excluded a priory. For example, another candidate which could account for the faster and more efficient information propagation, and hence a lower value of $z$, could be the presence of long-range interactions. However, midges in the field seem to interact mainly with sound-mediated interactions with an interaction range of only few centimeters \cite{fyodorova2003interactions,pennetier2010singing}, way smaller than the size of the swarms observed. The short-range nature of the interactions in swarms was also confirmed in \cite{attanasi2014collective}, where the radius of the effective aligning interaction was extimated to of $2-5\ \text{cm}$, in agreement with acoustic interactions. This seems to suggest that effective interactions in swarms are short range, and hence that the symmetry-based arguments of behavioural inertia are the most economic way to describe the large scale behaviour of natural swarms.}

\subsubsection{The role of dissipation}

\textcolor{black}{The spin-velocity coupling introduced in the previous section was mainly motivated by symmetry arguments, and associated conservation laws.
However, in real biological systems, information is not expected to be propagated forever with zero dissipation, as damping effects may become relevant over longer and longer distances. Thus, a (small) spin dissipation cannot be excluded in real biological systems. Note that the introduction of this friction does not violate the rotational symmetry of the problem, since all hydrodynamic equations are still invariant upon rotations. Moreover, we will show in sec. \ref{sec:diss} that our theory in the presence of large friction is equivalent to the incompressible Toner and Tu theory, which has been explicitly built up to obey rotational symmetry.}
 
\textcolor{black}{To understand why violation of spin conservation does not come from a weak violation of the symmetry, let us discuss an example in a case with which the reader might be more familiar. In a translational invariant system - say a collection of marbles -, Noether theorem states that the total linear momentum is conserved. If we had complete control of all degrees of freedom, namely position $q$ and momentum $p$ of each marble, we could describe the system through Hamilton equations $\dot{p_{i}}=-\pder{H}{q_{i}}\ , \ \dot{q_{i}}=\pder{H}{p_{i}}$. In this case, Noether theorem can be explicitly tested: since all the interactions in $H$ must obey translational invariance $q_{i}\to q_{i}+\delta q$, the time derivative of the total momentum $P=\sum_{i} p_{i}$ vanishes. In which cases can momentum conservation be violated? The first case is when the system is closed in a box. In this case, collisions with the walls of the box violate momentum conservation. This lack of conservation {\it is} due to the violation of the symmetry: the presence of a wall in a precise point of space manifestly violates translational invariance.}

\textcolor{black}{If instead our marbles were surrounded by some medium, say a fluid, exchange of momentum between the system and the medium is allowed. Hence, when observing the momentum of the collections of marbles only, it might be that violation of momentum conservation are detected. Does this mean that the symmetry is violated? Of course not: Noether theorem ensures conservation of the total momentum of all degrees of freedom, including those describing the medium. What happens to the total momentum of the marbles only depends on the interactions between fluid and marbles. When we want to describe only the degrees of freedom of the system of marbles, non-Hamiltonian effective interactions must be taken into account to describe the effect of those degrees of freedom we decided to coarse-graine (the medium). These interactions between the system and the medium can be effectively described as a dissipation of the momentum, apparently violating the conservation law associated with the symmetry. In fact, this effective dissipation arises as a consequence of not taking into account all the possible degrees of freedom; the symmetry and its associated conservation law are still in place.}

\textcolor{black}{Similarly, if the swarm were an isolated system, its total spin would be exactly conserved. Nevertheless, we expect midges in a swarm to interact not only among each other, but also with the surroundings. Due to the global rotational symmetry, in virtue of Noether’s theorem we expect the total spin of all degrees of freedom to be conserved. However, our field theory is an effective description of swarms, in which the interactions with the environment have been coarse-grained. Hence, some spin friction may arise as the effect of external forces on the swarms. Whether these interactions between midges and environment allow some spin exchange is out of our current knowledge, and hence we can not exclude them. What we do know is that if spin exchange was possible, it has to be weak, since the presence of inertial effects \cite{cavagna2017swarm} indicate that the spin dissipation is small.
}

\textcolor{black}{In order to introduce spin-dissipation, a $k$-independent term must be added to the kinetic coefficient of the spin,
$\Lambda_{\alpha\beta\gamma\nu}\to\Lambda_{\alpha\beta\gamma\nu}+\eta \II_{\alpha\beta\gamma\nu}$
where $\eta$ is the dissipative friction.
The new form of $\Lambda_{\alpha\beta\gamma\nu}$, in isotropic theories, is therefore given by
\begin{equation}
	\Lambda_{\alpha\beta\gamma\nu} =  \eta \II_{\alpha\beta\gamma\nu} - \lambda\nabla^{2}\II_{\alpha\beta\gamma\nu} \ .
\end{equation}
}
Although from a purely hydrodynamic perspective - i.e. at long wavelengths and long times - the existence of a spin friction $\eta$ would make the field $\vs$ a fast mode that can be dropped from the hydrodynamic description \cite{HH1977}, it was shown in \cite{cavagna2019short,cavagna2019long} that this is not the case for finite-size systems.
When the size of the system is finite, crossover phenomena that are usually ignored in the hydrodynamic limit may become relevant.
In the present case, if the dissipation is weak, $\eta\ll\lambda$, the system undergoes an RG crossover between the conservative regime of Model G ($\eta=0$) and the fully overdamped dissipative case of  Model A ($\eta\neq0$). Below a certain crossover length-scale ${\cal R}$ determined by the extent of dissipation, i.e. for modes with wavevector $k\gg {\cal R}^{-1}$, the critical dynamics has a inertial nature as if $\eta=0$, with $z=1.5$ \cite{cavagna2019short,cavagna2019long}.
On the other hand, on length-scales larger than ${\cal R}$, $k\ll {\cal R}^{-1}$, the dissipation overcomes and the dissipative result of Model A is recovered. This argument will discussed more thoroughly later on in Section \ref{sec:crossover}.

Natural swarms definitively have a finite size. Moreover, in natural swarms, the spin dissipation must be small enough to keep the system in its underdamped phase, as otherwise the temporal correlation functions of the theory would not reproduce the experimental ones \cite{cavagna2017swarm}.
This means having a crossover length-scale ${\cal R}$ larger than the system's size so that experimentally one observes the conservative inertial dynamics at all the accessible scales \cite{cavagna2019short}.
\textcolor{black}{For this reason, we will be particularly interested on the neighbourhood of the conservative $\eta=0$ plane, even though $\eta$ turns out to be a relevant perturbation in the RG sense.}

\subsubsection{Combining activity and inertial dynamics}
 \label{sec:cgeom}

Inspired by this equilibrium dynamic structure, we build the off-equilibrium theory describing inertial active matter by adding a term $g \vpsi \times \delta_{\vs} \Ham$ to the active field theory \eqref{eq:ctl_psi}, identifying the rotation of $\vpsi$ induced by the conservation of $\vs$.
The dynamics for $\vs$ is instead constructed from Eq. \eqref{eq:seq}, with advection added through the minimal substitution $\partial_t\to\mathcal{D}_t=\partial_t+\gamma_s \vv\cdot\vnabla$, encoding the fact that also the spin field is transported by the velocity. Here $\gamma_s$ is not necessary equal to $1$ since Galilean invariance is violated.
Thus, the resulting equations of motion take the following form:
\begin{equation}
\partial_t \psi_\alpha + v_0 \gamma_v  \left(\psi_\nu \partial_\nu \right) \psi_\alpha = - \Gamma \dder{\Ham}{\psi_\alpha} + g \psi_\beta \dder{\Ham}{s_{\alpha\beta}} - \partial_\alpha\mathcal{P} + \theta_\alpha\ ,
\label{eq:psi0}
\end{equation}
\begin{equation}
\partial_t s_{\alpha\beta} + v_0 \gamma_s   \left( \psi_\nu \partial_\nu \right) s_{\alpha\beta}  =-\Lambda_{\alpha\beta\gamma\nu} \dder{\Ham}{s_{\gamma\nu}} + 2 g \II_{\alpha\beta\gamma\nu} \psi_\gamma \dder{\Ham}{\psi_\nu}  + \zeta_{\alpha\beta}
\label{eq:s0}
\end{equation}
\textcolor{black}{Although in equilibrium systems $\Gamma$ needed to be a coefficient, rather than a matrix, because of the rotational symmetry, out of equilibrium the possibility of having non-symmetric interactions allows in principle for more complex tensorial structures, making $\Gamma_{\alpha\beta}$ a matrix rather than a real coefficient. However, non-symmetric linear couplings between the different components of $\psi_{\alpha}$ are typically known to lead to totally different phase transition phenomenology, very different from that in which we are interested \cite{vitelli2021non}. Hence, for the sake of simplicity, we shall assume that $\Gamma_{\alpha\beta}$ has no anti-symmetric component. Because of the rotational symmetry, the only symmetric form $\Gamma_{\alpha\beta}$ can take is that proportional to the identity: $\Gamma_{\alpha\beta}=\Gamma \delta_{\alpha\beta}$.}

The theory presented in Eq. \eqref{eq:psi0}-\eqref{eq:s0} will be studied in the incompressible case, therefore we omit all terms incompatible with this constraint, such as - for example - the two non-standard advective terms $\vv \left(\vnabla\cdot\vv\right)$ and $\vnabla v^2$ arising in the compressible Toner and Tu theory as a consequence of the absence of Galilean invariance \cite{tonertu1998}.
Incompressibility, however, does not forbid the presence of non-standard advection terms in the equation of motion for the spin and indeed we will demonstrate in Section \ref{sec:anomal} that the RG does generate two of these advection (adv) terms, namely,
\begin{align}
V^\mathrm{adv,1}_{\alpha\beta}= &      v_0 \, \mu_1\, {\gamma_s} \  \partial_\nu \left(s_{\alpha\nu}\psi_\beta - s_{\beta\nu}\psi_\alpha\right) \ ,
\\
V^\mathrm{adv,2}_{\alpha\beta}=  &      v_0\, \mu_2\, {\gamma_s}  \left[\partial_\alpha \left(\psi_\nu s_{\nu\beta}\right) - \partial_\beta \left(\psi_\nu s_{\nu\alpha}\right)\right] \ .
\label{sandapa}
\end{align}
Moreover, we will demonstrate that the RG also generates two of these anomalous mode-coupling (mc) terms in the equation of the spin (see Section \ref{sec:anomal}), 
\begin{align}
  V^\mathrm{mc,1}_{\alpha\beta}   = &  \phi_1\, g\,  \left[\partial_\alpha \left(\psi_\nu \partial_\nu \psi_\beta\right) - \partial_\beta \left(\psi_\nu \partial_\nu \psi_\alpha\right) \right]  \ ,
   \\
  V^\mathrm{mc,2}_{\alpha\beta}   = & \phi_2\, g\,  \partial_\nu \left[\psi_\nu \left(\partial_\alpha \psi_\beta-\partial_\beta \psi_\alpha\right)\right]  \ .
  \label{sandapo}
\end{align}
Crucially, each one of these anomalous terms is the divergence of a current, implying that the RG does not generate non-conserved ($k$-independent) spin dissipation: the conservation of the total spin, $\dot S_{\alpha\beta}(t)=0$,  hallmark of the mode-coupling theories \cite{HH1977}, is preserved even out of equilibrium. The novel vertices are accompanied by four new dimensionless couplings $\mu_1, \mu_2, \phi_1, \phi_2$.

\textcolor{black}{
Some other terms could be included in the calculation.}
It is the case of linear couplings between spin and velocity described in \cite{yang2015hydrodynamics}, which near criticality take the form $\partial_t \psi_\alpha\sim \partial_\beta s_{\alpha\beta}$ and $\partial_t s_{\alpha\beta}\sim \nabla^2 \left(\partial_\alpha\psi_\beta-\partial_\beta\psi_\alpha\right)$.
These terms modify the structure of linearized hydrodynamic equations, allowing the presence of propagators and correlation functions that mix the fields $\vs$ and $\vpsi$.
However, if such terms were included, the number of diagrams (which are not few even in the present calculation) would inevitably become enormous and impossible to manage.
Moreover, the presence of these new linear terms does not modify the dynamic critical exponent $z$ of the linear theory, while the presence of advection or inertial mode-coupling alone has a great impact on it.
Therefore, the presence of these additional linear terms is expected only to perturb the effect of non-linear interactions on the critical exponents.
Hence, we decide to focus on the study of spin-velocity couplings due to non-linear interactions only by working on the sub-manifold of the parameter space where such linear terms are not present.
\textcolor{black}{Because the RG calculation is in perfect agreement with numerical simulations even in the case in which these linear terms are ignored, we believe that including them from the beginning should not really affect the results we find here.}

The resulting equations of motion therefore become,
\begin{equation}
\partial_t \psi_\alpha + v_0 \gamma_v  \left(\psi_\nu \partial_\nu \right) \psi_\alpha = - \Gamma \dder{\Ham}{\psi_\alpha} + g \psi_\beta \dder{\Ham}{s_{\alpha\beta}} - \partial_\alpha\mathcal{P} + \theta_\alpha\ ,
\label{eq:psi}
\end{equation}
\begin{multline}
\partial_t s_{\alpha\beta} + v_0 \gamma_s   \left( \psi_\nu \partial_\nu \right) s_{\alpha\beta} + v_0 \, \mu_1\, {\gamma_s} \  \partial_\nu \left(s_{\alpha\nu}\psi_\beta - s_{\beta\nu}\psi_\alpha\right) + v_0\, \mu_2\, {\gamma_s}  \left[\partial_\alpha \left(\psi_\nu s_{\nu\beta}\right) - \partial_\beta \left(\psi_\nu s_{\nu\alpha}\right)\right] =\\
=-\Lambda_{\alpha\beta\gamma\nu} \dder{\Ham}{s_{\gamma\nu}} + 2 g \II_{\alpha\beta\gamma\nu} \psi_\gamma \dder{\Ham}{\psi_\nu} + \phi_1\, g\,  \left[\partial_\alpha \left(\psi_\nu \partial_\nu \psi_\beta\right) - \partial_\beta \left(\psi_\nu \partial_\nu \psi_\alpha\right) \right] + \phi_2\, g\,  \partial_\nu \left[\psi_\nu \left(\partial_\alpha \psi_\beta-\partial_\beta \psi_\alpha\right)\right] + \zeta_{\alpha\beta}
\label{eq:s}
\end{multline}
In principle, these equations should be coupled to an additional equation for the density field $\rho$.
However, as we will discuss in Section \ref{sec:inc}, we can get rid of the density field by studying incompressible systems.
Moreover, due to anisotropic effects caused by requiring the system to be incompressible, the kinetic tensor $\Lambda$ in \eqref{begabega} will have two different diffusive coefficients $\lambda^\perp\neq\lambda^\parallel$ for the longitudinal and transverse modes of $\vs$ \cite{cavagna2021dynamical}.

Although the system we are dealing with is out of equilibrium, it is possible to identify the truly non-equilibrium dynamic terms from those arising from a free-energy functional $\Ham$ that would survive also in the equilibrium limit $v_0\to0$.
In this limit, the theory resembles the dynamical structure of Model G \cite{HH1977}, therefore \eqref{eq:psieq}, \eqref{eq:seq} can be viewed as given by the merging of this equilibrium model \cite{HH1977} with Navier-Stokes equation \cite{FNS1977}. The latter takes into account the active motion of particles, as it happens for the Toner and Tu theory.

Because the system is out of equilibrium, Einstein relations between the kinetic coefficients and the corresponding noise variances are not expected to hold.
Therefore, $\vecc{\theta}$ and $\vecc{\zeta}$ of Eq. \eqref{eq:psi}, \eqref{eq:s} are white gaussian noises with zero mean $\langle\vecc{\theta}\rangle=\langle\vecc{\zeta}\rangle=0$ and variance given by
\begin{equation}
    \langle \theta_\alpha\left(\vx,t\right) \theta_\beta\left(\vx',t'\right) \rangle = 2 \tGamma \delta_{\alpha\beta} \delta\left(\vx-\vx'\right) \delta\left(t-t'\right)\ ,
\end{equation}
\begin{equation}
    \langle \zeta_{\alpha\beta}\left(\vx,t\right) \zeta_{\gamma\nu}\left(\vx',t'\right) \rangle = 4 \tLambda_{\alpha\beta\gamma\nu} \delta\left(\vx-\vx'\right) \delta\left(t-t'\right)\ ,
\end{equation}
where $\tGamma\neq\Gamma$ and the amplitude $\tLambda$ to take the same form of $\Lambda$ but with different coefficients ($\tlambda\neq\lambda$ and $\tilde\eta\neq\eta$).

All the other terms, which cannot be written as derivatives of a free energy functional, represent genuinely off-equilibrium interactions; these are advection and anomalous terms, which all occur as a consequence of the fact that individuals are not fixed on a network.
As we already said, the couplings of the advective terms $\gamma_v$ and $\gamma_s$ need not be equal to $1$ nor to each other, due to the absence of Galilean invariance \cite{tonertu1998}. Together with these active terms, we added also a pressure force $-\partial_\alpha\mathcal{P}$ to \eqref{eq:psi}, as it happens in Navier-Stokes, as well as in Toner-Tu equations.

\textcolor{black}{The equations of motion we just derived in the present section describe inertial active matter. By tuning the different parameters, we expect these equations to be able to describe many different phases of active matter. Since swarms have large, scale free correlations but no net global polarization \cite{attanasi2014collective,attanasi2014finite}, they are expected to be described by the near-critical regime of the present field-theory. The absence of any intrinsic length-scale in the correlations of swarms suggestes that the renormalized mass of the field theory has to vanish. This means that the bare mass $r$ is expected to be {\it small}. Does this arise as a consequence of some fine tuning of the parameters of the swarms in the field? Although no answer to this question has been given yet, we believe there are two main possibilities. One is of course that natural swarms do fine-tune their intrinsic parameters to achieve scale-invariance. This would however require midges to be able to change their interactions with neighbours, and tune them accordingly. A second possibility is that each swarm in the field has a given set of parameters, and tunes its size to maximize its susceptibility, namely the collective response. This mechanism, proposed in \cite{attanasi2014finite}, lies on a simple assumption: midges gather in swarms only when it is convenient, namely when they do maximize their ability to behave collectively. This could happen due to interactions we are not aware of, that make the swarms unstable whenever its size is too large, naturally breaking it into smaller swarms until the collective response is maximal. Note that, whatever is the correct scenario, based on the results in \cite{attanasi2014collective,attanasi2014finite}, swarms can be effectively described at a field-theoretical level as a system near a critical point, namely with a small mass $r$.}


\subsubsection{Enforcing incompressibility}\label{sec:inc}
In an active system, individuals are not fixed on a lattice but are free to move, allowing fluctuations in the local number density to arise.
When the total number of individuals is conserved, these fluctuations occur on large space and time-scales, making the local density $\rho\left(\vx,t\right)$ one of the slow-modes characterizing the hydrodynamic behaviour of the system \cite{HH1977}.
The time-evolution of the coarse-grained density field is thus given by the continuity equation,
\begin{equation}
	\partial_t \rho + v_0\vnabla\cdot\left(\rho\vpsi\right)=0\ ,
	\label{eq:rho}
\end{equation}
where $v_0\vpsi = \vv$ is the velocity field.
When considering systems with effective alignment interactions, 
the presence of density fluctuations plays a crucial role in determining the phenomenology of the ordering transition.
While in equilibrium ferromagnetic systems the transition is known to be continuous, things radically change when activity is added \cite{chate2008collective,gregoire2004,bertin2006boltzmann}.
The presence of density fluctuations generates instabilities when the transition is approached from the ordered state, thus leading to a discontinuous transition from finite to zero polarization.
On the contrary, a continuous transition has been shown to arise if density fluctuations are suppressed \cite{chen2015critical}.
\textcolor{black}{The first-order nature of the transition in compressible active matter is hence induced by the presence of density fluctuations. A recent renormalization group study has highlighted that a crossover between second and first order phenomenology is present in a modification of the Toner and Tu theory \cite{dicarlo2022evidence}, with the former belonging to the incompressible universality class. Moreover, also recent numerical simulation of the compressible Vicsek Model show that a continuous, second-order phenomenology is observed when density fluctuations are mild, namely when no phase-separation arises \cite{cavagna2020equilibrium}, with scaling laws ruled by the incompressible exponents found in \cite{chen2015critical}.}
In {\it finite-size} systems the transition may thus acquire a continuous phenomenology as a consequence of finite-size effects \cite{gregoire2004}, with density fluctuations not being strong enough to destabilize the second-order transition typical of equilibrium models \cite{vicsek1995novel}. Natural swarms have been shown to exhibit static and dynamic scaling laws typical of systems near to a continuous transition \cite{attanasi2014collective,cavagna2017swarm}, thus suggesting density fluctuations are not strong in determining the collective state.
\textcolor{black}{Following \cite{dicarlo2022evidence}, we believe that the incompressible universality class might describe the exponents of natural swarms. Hence, incompressibility will be enforced from the very beginning of our calculation.}
This is achieved by requiring a homogenous and constant density in eq \eqref{eq:rho}, namely  $\rho\left(\vx,t\right)=\rho_0$, thus dropping it from the hydrodynamic description. Incompressibility radically decreases the technical intricacy of the RG calculation, to a level that can be managed. Therefore, in the present work, we shall get rid of density fluctuations and study the theory at fixed density, namely assuming the system to be incompressible.

In an incompressible system Eq. \eqref{eq:rho} becomes a constraint on the field $\vv$ and, consequently, on the polarization $\vpsi$:
\begin{equation}
	\vnabla\cdot\vpsi=0
\end{equation}
In Fourier-space this constraint translates into the following two equivalent statements,
\begin{equation}
	k_\alpha \psi_\alpha\left(\vk\right) = 0 \quad \ , \quad P_{\alpha\beta}^\perp \left(\vk\right) \psi_\beta\left(\vk\right) = \psi_\alpha\left(\vk\right) \ ,
	\label{eq:inc}
\end{equation}
where we have defined an object which is rather central to this calculation, namely the projector onto the subspace orthogonal to $\vk$, 
\begin{equation}
	P_{\alpha\beta}^\perp\left(\vk\right)=\delta_{\alpha\beta}-\frac{k_\alpha k_\beta}{k^2}
	\label{pandino}
\end{equation}
and where,
\begin{equation}
	\psi_\alpha\left(\vx\right)=\int_{\vk} \eu^{\iu \vx\cdot\vk} \psi_\alpha\left(\vk\right)\ .
\end{equation}
Here and in the following, we will use the notation,
\begin{equation}
 \int_{\vk} = \int\limits_{\left|\vk\right|<\Lambda} \frac{\di^dk}{\left(2\pi\right)^d}
 \end{equation}
where $\Lambda$ is the ultraviolet cutoff of the theory, of the order of the inverse of the microscopic inter-particle distance.
Summation over repeated indices is always understood.

In order to enforce incompressibility in equations \eqref{eq:psi} and \eqref{eq:s}, two steps are necessary. The first one is rather intuitive, and it consists in projecting the 
 equation of motion for $\vpsi$ \eqref{eq:psi} onto the subspace transverse to $\vk$, which is a standard procedure \cite{FNS1977,chen2015critical}; this  is equivalent to say that the pressure term $\partial_\alpha \mathcal{P}$ enforces the constraint. The second step is less intuitive and it has been discovered in the equilibrium case $v_0=0$ \cite{cavagna2021dynamical}: the presence of a solenoidal constraint on the order parameter $\vpsi$ requires to project also the {\it force} $\delta_\vpsi \Ham$ that appears in the mode-coupling term of equation \eqref{eq:s} \cite{cavagna2021dynamical}. This second projection leads to the presence of an additional non-linear interaction in the equation of motion for $\vs$, the so-called DYnamic-Static (DYS) vertex, first found in \cite{cavagna2021dynamical},
\begin{equation}
\partial_t s_{\alpha\beta}(\vk) \sim 2 \kappa \, \II_{\alpha\beta\gamma\nu} \int_{\vq,\vp,\vh} \psi_\gamma (\vq) P_{\nu\rho}(\vq) \psi_\sigma(\vp)\psi_\sigma(\vh)\psi_\rho(\vk-\vq-\vp-\vh) \ .
\end{equation}
This vertex mixes the effects of the dynamic mode-coupling term and the static ferromagnetic $\psi^4$ interaction. At equilibrium the coupling constant $\kappa$ must obey the relation $\kappa=g\,u$, a crucial result that allows the equilibrium theory to be closed under renormalization \cite{cavagna2021dynamical}.
However, off-equilibrium effects may lead to a violation of the relation between $\kappa$, $g$ and $u$ meaning that, in general, one can have, 
\begin{equation}
    \kappa\neq g\,u  \ .
    \label{eq:2u}
\end{equation}
We shall demonstrate that at the new off-equilibrium inertial fixed point, $u$ and $g$ remain finite, while $\kappa$ vanishes. For this reason, in the main text, we omitted altogether the DYS interaction in the equations of motion to facilitate reading. However, in the actual calculation described here, this interaction will be kept for two reasons. First, it allows maintaining a connection with the equilibrium theory of \cite{cavagna2021dynamical}, in particular, recovering the same result as in equilibrium when $v_0\to 0$ is an important consistency check in such a complicated calculation. Secondly, the presence of this vertex is crucial for an additional reason: the high dimensionality of the parameter space (16 dimensions) and the intricate form of the $\beta$-functions will not allow us to find analytically the RG fixed point. To perform a successful numerical integration of the RG flow equations, the initial condition will be chosen in a region of the parameters space close to the equilibrium theory with solenoidal constraint. For the RG flow to go smoothly from the equilibrium to the off-equilibrium novel fixed point, it is technically crucial to keep this DYS interaction in the calculation, even though it eventually flows to zero at the new RG fixed point. In other words, although the DYS vertex is not relevant at the novel fixed point so that it does not contribute to the new value of the dynamical critical exponent, the DYS vertex is technically relevant to {\it find} the new fixed point in the large parameter space.

\subsubsection{The equations of motion of incompressible inertial active swarms}

Incompressibility implies that the field $\vpsi$ is only allowed to fluctuate in the direction perpendicular to $\vk$, thus generating an anisotropic behaviour of the field $\vs$ that acquires two different relaxation rates for its longitudinal and transverse components with respect to $\vk$ \cite{cavagna2021dynamical}.
Therefore the tensor $\Lambda_{\alpha\beta\gamma\nu}$, and consequently also $\tLambda$, takes the form,
\begin{equation}
\Lambda_{\alpha\beta\gamma\nu} = {\color{black}\eta\,\II_{\alpha\beta\gamma\nu} +} \left( \lambda^\perp \PP_{\alpha\beta\gamma\nu}^\perp +
\lambda^\parallel (\II-\PP^\perp)_{\alpha\beta\gamma\nu} \right)   k^2 \ ,
\label{bazooka}
\end{equation}
where $\PP_{\alpha\beta\gamma\nu}^\perp$ is the projection operator in the anti-symmetric space of $\vs$ \cite{cavagna2021dynamical}, which in $k$-space is given by
\begin{equation}
    \PP_{\alpha\beta\gamma\nu}^\perp\left(\vk\right)=\II_{\alpha\beta\gamma\nu}-\II_{\alpha\beta\sigma\tau}P_{\sigma\gamma}^\perp\left(\vk\right)P_{\tau\nu}^\perp\left(\vk\right)\ ,
    \label{begagigante}
\end{equation}
and we recall that,
\begin{equation}
    \II_{\alpha\beta\gamma\nu}=\frac{\delta_{\alpha\gamma} \delta_{\beta\nu} - \delta_{\alpha\nu} \delta_{\beta\gamma}}{2}\ .
    \label{lazonamorta}
\end{equation}
After symmetrizing terms containing powers of the same field, the incompressible equations of motion in $k$-space, finally become,
\begin{equation}
\begin{split}
\partial_t \psi_\alpha\left(\vk,t\right) 
+  \left(k^2 \Gamma+m\right) \psi_\alpha\left(\vk,t\right) =& \,\theta_\alpha
- v_0 \frac{\iu\gamma_v}{2} P_{\alpha\beta\gamma}\left(\vk\right) \int_{\vq} \psi_\beta\left(\vq,t\right) \psi_\gamma\left(\vk-\vq,t\right)\\
&- \frac{J}{3} Q_{\alpha\beta\gamma\nu}\left(\vk\right) \int_{\vq,\vh} \psi_\beta\left(\vq,t\right) \psi_\gamma\left(\vh,t\right) \psi_\nu\left(\vk-\vq-\vh,t\right)\\
&+ g P_{\alpha\rho}^\perp\left(\vk\right) \II_{\rho\beta\gamma\nu} \int_{\vq} \psi_\beta\left(\vk-\vq,t\right) s_{\gamma\nu}\left(\vq,t\right)\ ,
\end{split}
\label{eq:Ppsi}
\end{equation}

\begin{equation}
\begin{split}
    \partial_t  s_{\alpha\beta}\left(\vk,t\right) + \Lambda_{\alpha\beta\gamma\nu} s_{\gamma\nu} \left(\vk,t\right)=&
    \,\zeta_{\alpha\beta}
    - v_0\, \iu\, \gamma_s\, \II_{\alpha\beta\gamma\nu}k_\rho \int_{\vq} s_{\gamma\nu}\left(\vq,t\right)\psi_\rho\left(\vk-\vq,t\right)\\
    & - 2 \,v_0\, \iu\, \mu_1\, \gamma_s\, \II_{\alpha\beta\rho\eta}\II_{\rho\tau\gamma\nu} k_\tau \int_{\vq} s_{\gamma\nu}\left(\vq,t\right)\psi_\eta\left(\vk-\vq,t\right)\\
    & - 2 \,v_0\, \iu\, \mu_2\, \gamma_s\, \II_{\alpha\beta\rho\sigma}\II_{\rho\eta\gamma\nu} k_\sigma \int_{\vq} s_{\gamma\nu}\left(\vq,t\right)\psi_\eta\left(\vk-\vq,t\right)\\
    & + 2 \,g \,\II_{\alpha\beta\gamma\nu} \int_{\vq} \vk\cdot\vq\, \psi_\gamma\left(-\vq+\vk/2,t\right) \psi_\nu\left(\vq+\vk/2,t\right)\\
    & + 2 \,\Phi_1 \, g\, \II_{\alpha\beta\rho\sigma} \II_{\rho\tau\gamma\nu} \int_{\vq} k_\sigma q_\tau \psi_\gamma\left(-\vq+\vk/2,t\right) \psi_\nu\left(\vq+\vk/2,t\right)\\
    & + 2 \, \Phi_2\, g\,  \II_{\alpha\beta\rho\sigma} \II_{\rho\tau\gamma\nu} \int_{\vq} k_\tau q_\sigma \psi_\gamma\left(-\vq+\vk/2,t\right) \psi_\nu\left(\vq+\vk/2,t\right)\\
    & +\frac{\kappa}{6} \int_{\vq,\vh,\vp} K_{\alpha\beta\gamma\nu\sigma\tau}\left(\vk,\vq,\vh,\vp,\vk-\vq-\vh-\vp\right) \times\\
    &\qquad\qquad\times\psi_\gamma\left(\vq,t\right)\psi_\nu\left(\vh,t\right)\psi_\sigma\left(\vp,t\right)\psi_\tau\left(\vk-\vq-\vh-\vp,t\right)\ ,
\end{split}
\label{eq:Ps}
\end{equation}
where the following tensors were introduced,
\begin{equation}
    P_{\alpha\beta\gamma}\left(\vk\right) = k_\beta P_{\alpha\gamma}^\perp\left(\vk\right) + k_\gamma P_{\alpha\beta}^\perp\left(\vk\right)\ ,
\end{equation}
\begin{equation}
    Q_{\alpha\beta\gamma\nu}\left(\vk\right) = P_{\alpha\beta}^\perp\left(\vk\right) \delta_{\gamma\nu} + P_{\alpha\gamma}^\perp\left(\vk\right) \delta_{\beta\nu} +
    P_{\alpha\nu}^\perp\left(\vk\right) \delta_{\beta\gamma}\ ,
\end{equation}
\begin{equation}
\begin{split}
K_{\alpha\beta\gamma\nu\sigma\tau}\left({\vk},{\vp}_1,{\vp}_2,{\vp}_3,{\vp}_4\right)=&
\II_{\alpha\beta\gamma\rho}Q_{\rho\nu\sigma\tau}\left(\vk-\vp_1\right)+
\II_{\alpha\beta\nu\rho}Q_{\rho\gamma\sigma\tau}\left(\vk-\vp_2\right)+\\
+&\II_{\alpha\beta\sigma\rho}Q_{\rho\gamma\nu\tau}\left(\vk-\vp_3\right)+
\II_{\alpha\beta\tau\rho}Q_{\rho\gamma\nu\sigma}\left(\vk-\vp_4\right)\ ,
\end{split}
\end{equation}
and the noises have correlations,
\begin{equation}
    \langle \theta_\alpha\left(\vk,t\right) \theta_\beta\left(\vk',t'\right) \rangle = 2 \left(2\pi\right)^d \tGamma P_{\alpha\beta}^\perp\left(\vk\right) \delta^{(d)}\left(\vk+\vk'\right) \delta\left(t-t'\right)\ ,
\end{equation}
\begin{equation}
    \langle \zeta_{\alpha\beta}\left(\vk,t\right) \zeta_{\gamma\nu}\left(\vk',t'\right) \rangle = 4 \left(2\pi\right)^d \tLambda_{\alpha\beta\gamma\nu}\left(\vk\right) \delta^{(d)}\left(\vk+\vk'\right) \delta\left(t-t'\right)\ ,
\end{equation}
Finally, to simplify the notation, in \eqref{eq:Ppsi}, \eqref{eq:Ps} the following reduced parameters have been defined,
\begin{align}
    J&=\Gamma u & \Phi_1&=-2\left(\phi_1+\phi_2\right)\ ,
    \label{sumo}
    \\
    m&=\Gamma r & \Phi_2&=-2 \phi_2\ .
    \label{suma}
\end{align}

\subsection{Setting up the stage for the diagrammatic expansion}

\subsubsection{The Martin-Siggia-Rose-Janssen-De Dominicis action}\label{sec:MSR}

In order to employ RG to study the critical dynamics of our model we follow the method proposed by Martin, Siggia, Rose \cite{martin1973statistical}, Janssen \cite{janssen1976on} and De Dominicis \cite{de1976techniques}. The Martin-Siggia-Rose-Janssen-De Dominicis (MSRJD) formalism allows to describe the behaviour of fields evolving according to stochastic differential equations in terms of a field theory formulated using path integrals. The dynamic behaviour of the field $\vphi$, defined by the following Ito stochastic differential equation,
\begin{equation}
	\vecc{\mathcal{F}}\left[\vphi\right]-\vecc{\theta}=0
	\label{eq:stochastic_equation}
\end{equation}
is reproduced by the field-theoretical action $\mathcal{S}$ given by,
\begin{equation}
	\mathcal{S}[\vphi,\hvphi]=\int\di\vx\; \di t\, \hphi_\alpha \mathcal{F}_\alpha\left[\vphi\right] - \hphi_\alpha L_{\alpha\beta}\hphi_\beta
	\label{eq:eff_action}
\end{equation}
Here $\vecc{\mathcal{F}}$ is the deterministic evolution operator, $\vecc{\theta}$ a white gaussian noise with variance $2 L_{\alpha\beta}$.
The introduction of the hatted field $\hvphi$ in the action is the price that has to be paid to exploit the path integral formulation, using the standard rules of static renormalization and writing the perturbative series in terms of Feynman diagrams.
The field theoretical description reproduces the stochastic dynamics in the sense that, for a given observable $\mathcal O\left[\vphi\right]$,
\begin{equation}
	\langle \mathcal O \rangle=\langle \mathcal O \rangle_{\mathcal S}
\end{equation}
where $\langle \mathcal O \rangle$ is the average value of $\mathcal O$ over all possible realizations of the noise $\vecc{\theta}$, while 
\begin{equation}
	\langle \mathcal O \rangle_{\mathcal S}=\frac{1}{\mathcal Z} \int\mathcal{D}\vphi\int\mathcal{D}\hvphi\, O\left[\vphi\right]
	\eu^{-\mathcal{S}[\vphi,\hvphi]}
\end{equation}
Thanks to this equivalence, the critical dynamics can be investigated by studying the action $\mathcal S$ through RG techniques.
The gaussian part of the action $\mathcal{S}$ derives from the linear dynamics, namely the linear part of the operator $\mathcal{F}$, while the interactions derive from non-linear terms. Within this formalism, an external source $\vecc{h}$ introduced in the dynamical equation of $\vphi$ is coupled to $\hvpsi$ in the effective action. Therefore, the response function, known also as Green function or propagator, can be written as,
\begin{equation}
	\dder{\langle\phi_\alpha\left(\vx,t\right)\rangle}{h_\beta\left(\vx',t'\right)}=\langle\phi_\alpha\bigl(\vx,t\bigr)\hat{\phi}_\beta\bigl(\vx',t'\bigr)\rangle
	\label{eq:response}
\end{equation}
For this reason, $\hvphi$ takes the name `response field'.

The MSRDJ action $\mathcal{S}$ for the stochastic equations \eqref{eq:Ppsi} and \eqref{eq:Ps} depends upon four fields: $\vpsi$, $\hvpsi$, $\vs$ and $\hvs$.
The action can be split in the following terms,
\begin{equation}
    \mathcal{S}[\vpsi,\hvpsi,\vs,\hvs]=\mathcal{S}_{0,\psi}[\vpsi,\hvpsi]+\mathcal S_{0,s}[\vs,\hvs] +  \mathcal S_I [\vpsi,\hvpsi,\vs,\hvs]
\end{equation}
where $\mathcal{S}_{0,\psi}$ and $\mathcal{S}_{0,s}$ are the gaussian parts of the action, respectively coming from the linear dynamic terms of the equations of motion of $\vpsi$ and $\vs$, while $\mathcal{S}_I$ is the interacting part.
From Eq. \eqref{eq:eff_action} we have,
\begin{equation}
    \mathcal{S}_{0,\psi}[\vpsi,\hvpsi]=
    \int_{\tvk}
    \hpsi_\alpha(-\tvk) \left[ -\iu\omega + \Gamma k^2+ m\right] \psi_\alpha(\tvk) -
    \tGamma \hpsi_\alpha(-\tvk) P_{\alpha\beta}^\perp\left(\vk\right) \hpsi_\beta(\tvk)
    \label{eq:S0psi}
\end{equation}
\begin{equation}
    \mathcal{S}_{0,s}[\vs,\hvs]=
    \frac{1}{2}\int_{\tvk}
    \hs_{\alpha\beta}(-\tvk) \left[-\iu\omega \II_{\alpha\beta\gamma\nu} +
    \Lambda_{\alpha\beta\gamma\nu}\right] s_{\gamma\nu}(\tvk)
    -\hs_{\alpha\beta}(-\tvk)
    \tLambda_{\alpha\beta\gamma\nu}
    \hs_{\gamma\nu}(\tvk)
    \label{eq:S0s}
\end{equation}
\begin{equation}
    \begin{split}
        \mathcal S_I [\vpsi,\hvpsi,\vs,\hvs]=&
        -g
        \int_{\tvk,\tvq}
        \hpsi_\alpha(-\tvk) P_{\alpha\rho}^\perp\left(\vk\right)\II_{\rho\beta\gamma\nu}
        \psi_\beta(\tvk-\tvq) s_{\gamma\nu}(\tvq)+\\
        - & g\, \II_{\alpha\beta\rho\sigma} \II_{\rho\tau\gamma\nu}
        \int_{\tvk,\tvq}
        \hs_{\alpha\beta}(-\tvk)
        \left[ \vk\cdot\vq\, \delta_{\sigma\tau} +
        \Phi_1 k_\sigma q_\tau+
        \Phi_2 q_\sigma k_\tau \right]
        \psi_\gamma(-\tvq+\tvk/2) \psi_\nu(\tvq+\tvk/2)-\\
        + & v_0 \frac{\iu \gamma_v}{2}
        \int_{\tvk,\tvq}
        \hpsi_\alpha(-\tvk) P_{\alpha\beta\gamma}\left(\vk\right)
        \psi_\beta(\tvq)\psi_\gamma(\tvk-\tvq)-\\
        + & v_0 \frac{\iu \gamma_s}{2} \II_{\alpha\beta\rho\sigma} \II_{\rho\tau\gamma\nu}
        \int_{\tvk,\tvq}
        \hs_{\alpha\beta}(-\tvk) 
        \left[k_\eta \delta_{\sigma\tau} +
        2 \mu_1   k_\tau \delta_{\sigma\eta}+
        2 \mu_2 k_\sigma \delta_{\tau\eta}\right]
        s_{\gamma\nu}(\tvq) \psi_\eta(\tvk-\tvq)-\\
        + & \frac{J}{3}
        \int_{\tvk,\tvq,\tvh}
        \hpsi_\alpha(-\tvk) Q_{\alpha\beta\gamma\nu}\left(\vk\right)
        \psi_\beta(\tvq)\psi_\gamma(\tvh)\psi_\nu(\tvk-\tvq-\tvh)+\\
        - & \frac{\kappa}{12}
        \int_{\tvk,\tvq,\tvh,\tvp}
        \hs_{\alpha\beta}(-\tvk) K_{\alpha\beta\gamma\nu\sigma\tau}\left(\vk,\vq,\vh,\vp,\vk-\vq-\vh-\vp\right)
        \psi_\gamma(\tvq)\psi_\nu(\tvh)\psi_\sigma(\tvp)\psi_\tau(\tvk-\tvq-\tvh-\tvp)
        \end{split}
        \label{eq:SI}
\end{equation}
We wrote the effective action in $k$ and $\omega$ space, where the generic field $\phi$ is given by
\begin{equation}
    \phi\left(\vx,t\right)=\int_{\tvk} \eu^{\iu \left(\vx\cdot\vk-t\omega\right)} \phi(\tvk)\ ,
\end{equation}
with $\tvk=\left(\vk,\omega\right)$ and,
\begin{equation}
\int_{\tvk} = \int\limits_{\left|\vk\right|<\Lambda} \frac{\di^dk}{\left(2\pi\right)^d} \int\limits_{-\infty}^{\infty}\frac{\di \omega}{2\pi} \ .
\end{equation}
Notice that there is no cutoff in the frequency $\omega$.

\subsubsection{Free theory: propagators and correlation functions}\label{sec:lin}

The starting point to build the perturbative expansion of the equations of motion is the free theory, obtained by setting to zero all the dynamic non-linear couplings, namely $g$, $\gamma_v$, $\gamma_s$, $J$ and $\kappa$.
From the gaussian part of the action, given by Eqs.~\eqref{eq:S0psi} and \eqref{eq:S0s}, we can derive the expressions for the bare propagators and correlation functions for the effective field theory, which are the same as Model G with solenoidal constraint \cite{cavagna2021dynamical}, and are given by,
\begin{align}
\langle \psi_\alpha(\tvk)\hpsi_\beta(\tvq) \rangle_0&=\mathbb{G}_{\alpha\beta}^{0,\psi}(\tvk) \tdelta(\tvk+\tvq)\ , &
\langle s_{\alpha\beta}(\tvk) \hs_{\gamma\nu}(\tvq)\rangle_0&=\mathbb{G}_{\alpha\beta\gamma\nu}^{0,s}(\tvk)\tdelta(\tvk+\tvq)\ ,\\
\langle \psi_\alpha(\tvk)\psi_\beta(\tvq)\rangle_0&=\mathbb{C}_{\alpha\beta}^{0,\psi}(\tvk)\tdelta(\tvk+\tvq)\ , &
\langle s_{\alpha\beta}(\tvk) s_{\gamma\nu}(\tvq)\rangle_0&=\mathbb{C}_{\alpha\beta\gamma\nu}^{0,s}(\tvk)\tdelta(\tvk+\tvq)\ ,
\end{align}
where $\tdelta(\tvh)=(2\pi)^{d+1}\delta^{(d)}(\vh)\delta(\omega_h)$.
The subscripted zeros on thermal averages indicate that they are computed within the non-interacting theory.
The tensors $\mathbb{G}$ and $\mathbb{C}$ are given by,
\begin{gather}
\mathbb{G}_{\alpha\beta}^{0,\psi}(\tvk)=G_{0,\psi}(\tvk)\delta_{\alpha\beta}\label{eq:greenpsi}\\
\mathbb{C}_{\alpha\beta}^{0,\psi}(\tvk)=C_{0,\psi}(\tvk) P_{\alpha\beta}^\perp (\vk)\label{eq:corrpsi}\\
\mathbb{G}_{\alpha\beta\gamma\nu}^{0,s}(\tvk) = G_{0,s}^\perp(\tvk) \PP_{\alpha\beta\gamma\nu}^\perp(\vk) + G_{0,s}^\parallel(\tvk) (\II-\PP^\perp)_{\alpha\beta\gamma\nu}(\vk)\label{eq:greens}\\
\mathbb{C}_{\alpha\beta\gamma\nu}^{0,s}(\tvk) = C_{0,s}^\perp(\tvk) \PP_{\alpha\beta\gamma\nu}^\perp(\vk) + C_{0,s}^\parallel(\tvk) (\II-\PP^\perp)_{\alpha\beta\gamma\nu}(\vk)\label{eq:corrs}
\end{gather}
\textcolor{black}{
In Eq.~\eqref{eq:greenpsi}, \eqref{eq:greens}, \eqref{eq:corrpsi} and \eqref{eq:corrs} we have,
\begin{align}
G_{0,\psi}\left(\vk,\omega\right)&=\frac{1}{-\iu\omega+\Gamma k^2+m} &
C_{0,\psi}\left(\vk,\omega\right)&=\frac{2\tGamma}{\omega^2+\left( m + \Gamma k^2 \right)^2}\\
G_{0,s}^\perp\left(\vk,\omega\right)&=\frac{2}{-\iu\omega+\eta+\lambda^\perp k^2}&
C_{0,s}^\perp\left(\vk,\omega\right)&=\frac{4\tlambda^\perp k^2}{\omega^2+(\eta+\lambda^\perp k^2)^2}\\
G_{0,s}^\parallel\left(\vk,\omega\right)&=\frac{2}{-\iu\omega+\eta+\lambda^\parallel k^2}&
C_{0,s}^\parallel\left(\vk,\omega\right)&=\frac{4\tlambda^\parallel k^2}{\omega^2+(\eta+\lambda^\parallel k^2)^2}
\end{align}
}
In the diagrammatic framework, the fields $\vpsi$ and $\hvpsi$ are represented with a solid line, while the fields $\vs$ and $\hvs$ are represented with wavy lines.
Bare propagators and correlation functions thus take the following graphical representation
\begin{align}
	\langle \psi_\alpha\hpsi_\beta\rangle_0\quad=&\quad
	\feynmandiagram [small,horizontal=c to b] {
        		c -- [fermion] b,
        	};
    	&
	\langle s_{\alpha\beta}\hs_{\gamma\nu}\rangle_0\quad=&\quad
	\feynmandiagram [small, horizontal=c to b] {
        		c -- [charged boson] b,
        	};
	\\
	\langle \psi_\alpha\psi_\beta\rangle_0\quad=&\quad
	\feynmandiagram [small, horizontal=c to b] {
        		c -- b,
        	};
    	&
    	\langle s_{\alpha\beta}s_{\gamma\nu}\rangle_0\quad=&\quad
	\feynmandiagram [small, horizontal=c to b] {
        		c -- [boson] b,
        	};
\end{align}
where the arrows in the propagators always point in the direction of the response field.

\subsubsection{Non-linear terms: the vertices}\label{sec:nonlin}
The six terms that compose $\mathcal{S}_I$ represent the non-linear interactions in the equations of motion.
Each interaction involves one response field, identifying the equation of motion in which the corresponding non-linearity appears: $\hvpsi$, if the vertex comes from a non-linearity in the equation of $\vpsi$; $\hvs$, if it comes from a non-linearity in the equation of $\vs$.
In the diagrammatic framework, these interactions are graphically represented by vertices, in which different lines merge, each representing one of the fields involved in the interaction.
We remind that full lines represent $\hvpsi$ and $\vpsi$ fields, while wavy lines represent $\hvs$ and $\vs$ fields. Moreover, an \textit{entering} arrow is used to identify the leg representing the response field.
We shall choose vertices to have opposed signs with respect to the interactions; the convenience of this choice is that vertices play a crucial role in building Feynman diagrams, which come from the expansion of $\exp(-\mathcal{S})$.

The first vertex involving $\hvpsi$ represents the mode coupling non-linearity proportional to the reversible dynamic coupling $g$, 
\begin{equation}
    \begin{tikzpicture}[baseline=(a.base)]
		\begin{feynman}[small]
			\vertex (a) at (0,0) {\(\hpsi_\alpha(-\tvk)\)};
			\vertex (b) [dot] at (1.5,0) {};
			\vertex (c) at (2.25,1.3) {\(\psi_\beta(\tvq)\)};
			\vertex (d) at (2.25,-1.3) {\(s_{\gamma\nu}(\tvp)\)};
			\diagram* {
				(a) -- [fermion] (b) -- (c),
				(b) -- [boson] (d),
			};
		\end{feynman}
	\end{tikzpicture}
    {\color{black}\quad : \quad} g P_{\alpha\rho}^\perp\left(\vk\right)\II_{\rho\beta\gamma\nu}\tdelta(\tvk-\tvq-\tvp)\ , 
    \label{fd:mcv_vertex}
\end{equation}
The second interaction involving $\hvpsi$ is the self-propulsion (or advection) interaction coming from the convective derivative in the equation of motion.
This vertex is proportional to $v_0 \gamma_v$, and it vanishes when the microscopic speed does.
Graphically, this interaction is represented by
\begin{equation}
    \begin{tikzpicture}[baseline=(a.base)]
		\begin{feynman}[small]
			\vertex (a) at (0,0) {\(\hpsi_\alpha(-\tvk)\)};
			\vertex (b) [dot,red] at (1.5,0) {};
			\vertex (c) at (2.25,1.3) {\(\psi_\beta(\tvq)\)};
			\vertex (d) at (2.25,-1.3) {\(\psi_\gamma(\tvp)\)};
			\diagram* {
				(a) -- [fermion] (b) -- (c),
				(b) -- (d),
			};
		\end{feynman}
	\end{tikzpicture}
    {\color{black}\quad : \quad} - v_0 \frac{\iu\gamma_v}{2} P_{\alpha\beta\nu}\left(\vk\right) \tdelta(\tvk-\tvq-\tvp)\ ,
    \label{fd:sppv_vertex}
\end{equation}
The third vertex involving $\hvpsi$ derives from the ferromagnetic $\psi^3$ Landau-Ginzburg interaction, proportional to $J$.
It is represented by the term,
\begin{equation}
	\begin{tikzpicture}[baseline=(a.base)]
		\begin{feynman}[small]
			\vertex (a) at (0,0) {\(\hpsi_\alpha(-\tvk)\)};
		    \vertex (b)[square dot] at (1.5,0) {};
			\vertex (c) at (2.25,1.3) {\(\psi_\beta(\tvq)\)};
			\vertex (d) at (2.75,0) {\(\psi_{\gamma}(\tvp)\)};
			\vertex (e) at (2.25,-1.3) {\(\psi_{\gamma}(\tvh)\)};
			\diagram* {
				(a) -- [fermion] (b) -- (c),
				(b) -- (d),
				(b) -- (e),
			};
		\end{feynman}
	\end{tikzpicture}
    	{\color{black}\quad : \quad} -\frac{J}{3}Q_{\alpha\beta\gamma\nu}\left(\vk\right) \tdelta(\tvk-\tvq-\tvp-\tvh)\ ,
    \label{fd:ferro_vertex}
\end{equation}
The other three vertices involve one field $\hvs$ and derive from the equation for the spin.
The first one represents the dynamic mode-coupling interaction proportional to $g$ with the addition of the two mode-coupling anomalous terms, with different tensorial structure,
\begin{equation}
	\begin{tikzpicture}[baseline=(a.base)]
		\begin{feynman}[small]
			\vertex (a) at (0,0) {\(\hs_{\alpha\beta}(-\tvk)\)};
			\vertex (b)[empty dot] at (1.7,0) {};
			\vertex (c) at (2.45,1.3) {\(\psi_\gamma(\tvh)\)};
			\vertex (d) at (2.45,-1.3) {\(\psi_\nu(\tvp)\)};
			\diagram* {
				(a) -- [charged boson] (b) -- (c),
				(b) -- (d),
			};
		\end{feynman}
	\end{tikzpicture}
    	{\color{black}\quad : \quad} \frac{g}{2}\II_{\alpha\beta\rho\sigma}\II_{\rho\tau\gamma\nu} \left[\left(p^2-h^2\right)\,\delta_{\sigma\tau} +
        \Phi_1  p_\sigma^{(+)} p_\tau^{(-)} +
        \Phi_2 p_\sigma^{(-)} p_\tau^{(+)} \right]\tdelta(\tvk-\tvh-\tvp)\ ,
    \label{fd:mcs_vertex}
\end{equation}
where $\vp^{(+)}=\vp+\vh$ while $\vp^{(-)}=\vp-\vh$.
This vertex vanishes when $\vecc{h}=\vecc{p}$, guaranteeing that this interaction does not contribute to the dynamics of the total spin $S\left(t\right)=s\left(\vk=0,t\right)$.
The anomalous mode coupling terms are those proportional to $\Phi_{1,2}$.
From a technical point of view, it is essential to note that this vertex can be rewritten in an alternative form; by using the delta function, together with a symmetric distribution of the momenta, one has, 
\begin{equation}
	\begin{tikzpicture}[baseline=(a.base)]
		\begin{feynman}[small]
			\vertex (a) at (0,0) {\(\hs_{\alpha\beta}(-\tvk)\)};
			\vertex (b)[empty dot] at (1.7,0) {};
			\vertex (c) at (2.45,1.3) {\(\psi_\gamma(\tvk/2-\tvq)\)};
			\vertex (d) at (2.45,-1.3) {\(\psi_\nu(\tvk/2+\tvq)\)};
			\diagram* {
				(a) -- [charged boson] (b) -- (c),
				(b) -- (d),
			};
		\end{feynman}
	\end{tikzpicture}
    	{\color{black}\quad : \quad} {g}\ \II_{\alpha\beta\rho\sigma}\II_{\rho\tau\gamma\nu} \left[
	\vecc{k}\cdot \vecc{q}\,\delta_{\sigma\tau} +
        \Phi_1 \, k_\sigma q_\tau +
        \Phi_2 \, q_\sigma k_\tau
        \right]
        \ .
    \label{fd:mcs_vertex_alt}
\end{equation}
This second form is convenient for two reasons: first, it has a simpler structure, which makes it easier to recognize corrections to the coupling constant $g$; secondly, in this form, it is more transparent to demonstrate the diagrammatic origin of the anomalous terms, a derivation that we will see later on in Section \ref{sec:anomal}. On the other hand, the first form of this same vertex, equation \eqref{fd:mcs_vertex}, is handier when calculating diagrams in which $\hs_{\alpha\beta}$ appears as an internal leg.


The second vertex involving $\hvs$ is the self-propulsion interaction, coming from the fact that $\vs$ is advected by the velocity $v_0\vpsi$, and it is proportional to $v_0 \gamma_s$. Graphically, this interaction is represented by,
\begin{equation}
    \begin{tikzpicture}[baseline=(a.base)]
		\begin{feynman}[small]
			\vertex (a) at (0,0) {\(\hs_{\alpha\beta}(-\tvk)\)};
			\vertex (b) [dot,green] at (1.5,0) {};
			\vertex (c) at (2.25,1.3) {\(s_{\gamma\nu}(\tvq)\)};
			\vertex (d) at (2.25,-1.3) {\(\psi_\eta(\tvp)\)};
			\diagram* {
				(a) -- [charged boson] (b) -- [boson] (c),
				(b) -- (d),
			};
		\end{feynman}
	\end{tikzpicture}
    {\color{black}\quad : \quad} - v_0 \frac{\iu\gamma_s}{2} \II_{\alpha\beta\rho\sigma} \II_{\rho\tau\gamma\nu} \left[ k_\eta \delta_{\sigma\tau}+
        2 \mu_1  k_\tau \delta_{\sigma\eta}+
        2 \mu_2  k_\sigma \delta_{\tau\eta}\right] \tdelta(\tvk-\tvq-\tvp)\ ,
    \label{fd:spps_vertex}
\end{equation}
Also this interaction vanishes when $\vecc{q}=\vecc{p}$, and therefore does not contribute to the dynamics of the total spin either. This vertex takes into account also the anomalous advection of $\vs$ through the terms proportional to $\mu_{1,2}$.

The last interaction term is the DYnamic-Static (DYS) vertex \cite{cavagna2021dynamical}.
This interaction mixes the ferromagnetic-like interaction and the mode-coupling dynamic term, as a consequence of the presence of incompressibility.
It represents the effects of the Landau confining potential on the dynamics of $\vs$, mediated by the mode-coupling dynamic interaction.
It takes the following form,
\begin{equation}
	\begin{tikzpicture}[baseline=(a.base)]
		\begin{feynman}[small]
			\vertex (a) at (0,0) {\(\hs_{\alpha\beta}(-\tvk)\)};
			\vertex (b) [crossed dot] at (1.7,0) {};
			\vertex (c) at (1.3,1.425) {\(\psi_\gamma(\tvp_1)\)};
			\vertex (d) at (2.91,0.88) {\(\psi_\nu(\tvp_2)\)};
			\vertex (e) at (2.91,-0.88) {\(\psi_\sigma(\tvp_3)\)};
			\vertex (f) at (1.3,-1.425) {\(\psi_\tau(\tvp_4)\)};
			\diagram* {
				(a) -- [charged boson] (b) -- (c),
				(b) -- (d),
				(b) -- (e),
				(b) -- (f),
			};
		\end{feynman}
	\end{tikzpicture}
    		{\color{black}\quad : \quad} \frac{\kappa}{12} K_{\alpha\beta\gamma\nu\sigma\tau}\left({\vk},\vp_1,\vp_2,\vp_3,\vp_4\right)\tdelta(\tvk-\tvp_1-\tvp_2-\tvp_3-\tvp_4)\ .
    	\label{fd:mixed_vertex}
\end{equation}
This interaction does not vanish when $\vk=0$, and thus it contributes to the dynamics of the total spin.
However, we shall be careful not to confuse the absence of conservation with the presence of dissipation.
If the spin was largely dissipated, it would become a non-hydrodynamic variable whose behaviour does not affect that of the order parameter.
As shown in \cite{cavagna2021dynamical}, where the fixed network approximation of the incompressible theory developed here is analyzed, even though the spin may not be globally conserved due to the DYS interaction, it represents still an hydrodynamic slow mode.
This is because the DYS vertex does not involve the spin, and thus represents the effects of the slow-mode $\vpsi$. Even if this effect remains finite as $\vk\to0$, it may only be responsible for slow variations of the total spin $\vs$.
Hence, although the total spin is not conserved, it is not even dissipated and thus it undergoes a generalized precession caused by the DYS vertex \cite{cavagna2021dynamical}.


\subsection{The diagrammatic expansion}

\subsubsection{Dynamical RG in momentum shell: general procedure}\label{sec:RGt}

The momentum shell RG scheme \cite{wilson1971renormalization1,goldenfeld_lectures_1992,parisi_book,wilson1974renormalizarion,cardy1996scaling} provides an explicit method to calculate critical exponents. By integrating out the short-wavelength modes iteratively, the RG generates a flow in the parameter space that eventually converges towards a fixed point. The study of the linearized flow equations near an attractive fixed point allows computing explicitly the critical exponents of the theory. Universality, in this RG context, means that one single fixed point rules the long-wavelength behaviour of a large class of theories, each one identified by a different initial point in the parameter space. The RG flow is defined by a set of recursive relations, obtained by iterating a RG transformation of the effective action of the field theory. The RG transformation consists of two steps: \textit{i)} the probability distribution of the fields is marginalized by integrating short-wavelength modes on the shell $b^{-1}\Lambda<k< \Lambda$, with $b>1$, hence effectively decreasing the cutoff in momentum space; \textit{ii)} space and time are rescaled, so to formally restore the same cutoff as the original theory.
The action obtained after one RG transformation has different parameters and it describes the system when it is observed on a larger scale.
However, since the partition function remains the same up to a multiplicative constant, the physical observables are left unchanged.
The flow equations are obtained under the form of recursive relations, describing how the parameters at the iteration $(l+1)$ can be obtained starting from those at step $l$.

In the gaussian theory, namely when all interaction terms vanish, the shell integration is harmless since modes at different wavelengths are independent \cite{goldenfeld_lectures_1992}, and thus the RG flow is trivial: essentially each parameter rescales according to naive dimensional analysis.
However, when non-gaussian interactions are present, the shell integration couples long and short wavelength modes, generating nontrivial corrections to the bare action.
These corrections can be computed within a perturbative expansion in powers of $\epsilon=d_c-d$, where $d_c$ is the upper critical dimension, namely the dimension above which mean-field theory is exact, while $d$ is the spatial dimension.
This expansion method to compute the critical exponents, known as $\epsilon$-expansion, was proposed by K.G. Wilson, and first explored in the seminal paper {\it ``Critical exponents in 3.99 dimensions''} \cite{wilson1972critical}, in honour of which we chose the title of the present work.

In order to perform the shell integration, it is convenient to split the fields in their Infra-Red (IR) and Ultra-Violet (UV) modes \cite{goldenfeld_lectures_1992}, and write the partition function as,
\begin{equation}
	\mathcal Z =\int \mathcal D \vphi\, e^{-\mathcal{S}[\vphi]}=\int  \mathcal D \vphi^< \mathcal D \vphi^> \, e^{-\mathcal{S}[\vphi^<+\vphi^>]}
\end{equation}
where $\vphi$ stands for all the fields of the theory, while superscripts $<$ and $>$ indicate whether the field has momenta lower (IR modes) or higher (UV modes) than $\Lambda/b$ respectively.
The action $\mathcal{S}$ can then be written in the following form,
\begin{equation}
	\mathcal{S}[\vphi^<+\vphi^>]= \mathcal{S}[\vphi^<]+\mathcal{S}_0[\vphi^>]-\mathcal{V}[\vphi^<,\vphi^>]
\end{equation}
where $\mathcal{S}_0$ is the gaussian part of the action while $\mathcal{V}$ represents all the interactions between UV and IR modes.
Integrating out short wavelength details, namely performing the integration over the UV modes $\vphi^>$ with wavevector on the shell, leads to,
\begin{equation}
	\mathcal Z = \mathcal{Z}_0^>\int \mathcal D \vphi^< e^{-\mathcal{S}[\vphi^<]-\Delta \mathcal{S}[\vphi^<]}
\end{equation}
where,
\begin{equation}
	e^{-\Delta \mathcal{S}[\vphi^<]}=\frac{1}{\mathcal{Z}_0^>}\int \mathcal D \vphi^> e^{-\mathcal{S}_0[\vphi^>]} e^{\mathcal{V}[\vphi^<,\vphi^>]}
\end{equation}
is the gaussian average of $e^{\mathcal{V}[\vphi^<,\vphi^>]}$ over the UV fields with on-shell momentum, and $\Delta \mathcal{S}[\vphi^<]$ represents the corrections to the bare action $\mathcal{S}$ due to the shell integration.
By assuming the interaction couplings \textit{small} -- assumption that will be verified {\it a posteriori} at the stable fixed point-- it is possible to expand $e^{\mathcal{V}}$ in powers of the couplings.
The expansion of $\Delta \mathcal{S}$ can be graphically represented as an expansion in Feynman diagrams, composed only by connected diagrams \cite{goldenfeld_lectures_1992}.
This shell integration is usually performed under a thin-shell approximation, namely taking $b\sim 1$: in this limit $\Delta \mathcal{S}$ is proportional to the shell thickness $1-b^{-1}\simeq \ln b$.
The final result  after the integration over the UV modes $\vphi^>$ is an effective action with a new cutoff $\Lambda/b$ and modified parameters, namely,
\begin{equation}
	\mathcal{P}_0\to\mathcal{P}_0+\mathcal{P}_0\delta\mathcal{P}\ln b=\mathcal{P}_0\left(1+\delta\mathcal{P}\ln b\right)\simeq \mathcal{P}_0 b^{\delta\mathcal{P}}
	\label{corro}
\end{equation}
where we used the relation $b^a\simeq1+a\ln b$ when $b$ is close to $1$. The aim of the perturbative theory is to calculate the corrections to the parameters, namely the values of $\delta\mathcal{P}$.

In order to recover a theory with the same cutoff $\Lambda$ as the bare theory momenta must be rescaled, which in turns requires to rescale also frequency and fields \cite{widom1965equation,kadanoff1966scaling,Ferrell1967,HH1967scaling},
\begin{align}
	\vk =& b^{-1}\vk_b & \omega =& b^{-z} \omega_b & {\color{black}\vphi(\vk,\omega)=}& {\color{black} b^{-\chi_\phi} \vphi(\vk_b,\omega_b)}
\end{align}
where $z$ is the dynamical critical exponent and ${\chi_\phi}$ is the scaling dimension of the field. Once this rescaling is done, 
the action takes the same form as the bare one, but with new renormalized parameters and couplings, which will be denoted with a subscript $b$. These new values of the parameters, defined in order to absorb all the powers of $b$ in front of them, can be expressed as functions of the bare parameters through the following relation,
\textcolor{black}{
\begin{equation}
    \mathcal{P}_{b}=b^{\chi_{\mathcal{P}}}  \mathcal{P}_0
    \label{recuppo}
\end{equation}
}
where $\chi_{\mathcal{P}}$ defines the total scaling dimension of $\mathcal{P}$, which takes into account both the naive physical dimension, coming from dimensional analysis, and the anomalous scaling dimension due to the RG coupling of IR and UV modes.
We may therefore write $\chi_{\mathcal{P}}$ as,
\textcolor{black}{
\begin{equation} 
\chi_{\mathcal{P}}=d_{\mathcal{P}}+\delta\mathcal{P} \ ,
\end{equation}
}
where $d_{\mathcal{P}}$ is the \textcolor{black}{naive physical dimension of $\mathcal{P}$ in units of momentum $k$}, while $\delta\mathcal{P}$ is the anomalous scaling dimension defined in Eq \eqref{corro}.
Thanks to this transformation, the partition function can be written as,
\begin{equation}
	\mathcal Z \propto\int \mathcal D \vphi^< e^{-(\mathcal{S}[\vphi^<]+\Delta \mathcal{S}[\vphi^<])}\propto\int \mathcal D \vphi e^{-\mathcal{S}_b[\vphi]}
\end{equation}
where $\mathcal{S}_b[\vphi]$ is the {\it renormalized action}, obtained after integration over a shell of thickness $\ln b$ and after rescaling.


\subsubsection{Self-energies and vertex corrections}\label{sec:pertRG}

Following the RG scheme introduced in the previous Section we now derive the recursive relations for the parameters and coupling constants. As a consequence of the shell integration, the bare gaussian action $\mathcal{S}_0$ of \eqref{eq:S0psi} and \eqref{eq:S0s} acquires some corrections, that it is customary to write in the following way \cite{HH1977},
\begin{equation}
\begin{split}
    \Delta \mathcal S_0 & = \int \hpsi_\alpha(-\tvk) \Sigma_{\alpha\beta}(\tvk) \psi_\beta(\tvk) - \hpsi_\alpha(-\tvk) \tSigma_{\alpha\beta}(\tvk) \hpsi_\beta(\tvk)+\\
    &  + \int \hs_{\alpha\beta}(-\tvk) \Pi_{\alpha\beta\gamma\nu}(\tvk) s_{\gamma\nu}(\tvk) - \hs_{\alpha\beta}(-\tvk) \tPi_{\alpha\beta\gamma\nu}(\tvk) \hs_{\gamma\nu}(\tvk)\ ,
\end{split}
\label{eq:dS0}
\end{equation}
where all momenta are integrated off-shell, $k<\Lambda/b$, while frequency integrals still run form $-\infty$ to $\infty$. The new quantities
$\Sigma$, $\tSigma$, $\Pi$ and $\tPi$ are the {\it self-energies}, which contribute to the perturbative corrections of the gaussian parameters of the original action.
On the other hand, the interacting part of the action \eqref{eq:SI} acquires the corrections, 
\begin{equation}
\begin{split}
    \Delta \mathcal S_I & = \int \hpsi_\alpha(-\tvk) V_{\alpha\beta\gamma\nu}^{\hpsi \psi s}(\tvk,\tvq) \psi_\beta(\tvk-\tvq)s_{\gamma\nu}(\tvq)+\\
    & + \int \hs_{\alpha\beta}(-\tvk) V_{\alpha\beta\gamma\nu}^{\hs \psi\psi}(\tvk,\tvq) \psi_\gamma(-\tvq+\tvk/2) \psi_\nu(\tvq+\tvk/2)+\\
    & +\int \hpsi_\alpha(-\tvk) V_{\alpha\beta\gamma}^{\hpsi \psi \psi}(\tvk,\tvq) \psi_\beta(\tvq)\psi_\gamma(\tvk-\tvq)+\\
    & + \int \hs_{\alpha\beta}(-\tvk) V_{\alpha\beta\gamma\nu\eta}^{\hs s \psi}(\tvk,\tvq) s_{\gamma\nu}(\tvq) \psi_\eta(\tvk-\tvq)+\\
    & +\int \hpsi_\alpha(-\tvk)
    V_{\alpha\beta\gamma\nu}^{\hpsi \psi \psi \psi}(\tvk,\tvq,\tvh) \psi_\beta(\tvq)\psi_\gamma(\tvh) \psi_\nu(\tvk-\tvq-\tvh)+\\
    & + \int \hs_{\alpha\beta}(-\tvk) V_{\alpha\beta\gamma\nu\sigma\tau}^{\hs \psi\psi\psi \psi}(\tvk,\tvq,\tvh,\tvp)  \psi_\gamma(\tvq)\psi_\nu(\tvh)\psi_\sigma(\tvp)\psi_\tau(\tvk-\tvq-\tvh-\tvp)\ ,
\end{split}
\end{equation}
where the various {\it vertex-functions} $V$s give perturbative corrections to the coupling constants of the non-linear interactions.
We notice that all the corrections are proportional to the volume of the momentum shell, which is given by $1-b^{-1}\simeq\ln b$.
In what follows, we will compute the self-energies and the vertex functions by using perturbation theory.
Evaluation of perturbative corrections is done using a Feynman diagram expansion and considering only the first order in $\epsilon$.
Since we are interested in the dynamic behaviour near criticality, the mass $m$ is set to $0$ in all diagrams contributing to self-energies and vertex functions \cite{goldenfeld_lectures_1992}\textcolor{black}{, except for those giving corrections to $m$ itself.}


As mentioned earlier, the self-energies represent the perturbative corrections to the gaussian part of the effective action. We distinguish four different self energies, one for each combination of fields appearing in $\mathcal{S}_0$, namely $\hpsi\psi$, $\hpsi\hpsi$, $\hs s$ and $\hs \hs$.
\textcolor{black}{From a diagrammatic point of view, each self-energy is given by the sum of all amputated 1-particle irreducible diagrams with external fields $\hpsi\psi$, $\hpsi\hpsi$, $\hs s$ and $\hs \hs$ respectively. Graphically, they are represeded by the blobs in the following diagrammatic scheme}
\begin{align}
    \Sigma_{\alpha\beta}(\tvk) {\color{black}\quad : \quad}&
	\begin{tikzpicture}[baseline=(a.base)]
		\begin{feynman}
			\vertex at (0,0) (a) {\(\hpsi_\alpha(-\tvk)\)};
			\vertex at (2.2,0) [blob] (b){};
			\vertex at (4,0) (c) {\(\psi_\beta(\tvk)\)};
			\diagram* {
				(a) -- [fermion] (b) -- (c),
			};
		\end{feynman}
	\end{tikzpicture}
	\label{sissi}
	\\
	\tSigma_{\alpha\beta}(\tvk){\color{black}\quad : \quad}&
	\begin{tikzpicture}[baseline=(a.base)]
		\begin{feynman}
			\vertex at (0,0) (a) {\(\hpsi_\alpha(-\tvk)\)};
			\vertex at (2.2,0) [blob] (b){};
			\vertex at (4,0) (c) {\(\hpsi_\beta(\tvk)\)};
			\diagram* {
				(a) -- [fermion] (b) -- [anti fermion] (c),
			};
		\end{feynman}
	\end{tikzpicture}
	\\
	\Pi_{\alpha\beta\gamma\nu}(\tvk){\color{black}\quad : \quad}&
	\begin{tikzpicture}[baseline=(a.base)]
		\begin{feynman}
			\vertex at (0,0) (a) {\(\hs_{\alpha\beta}(-\tvk)\)};
			\vertex at (2.2,0) [blob] (b){};
			\vertex at (4,0) (c) {\(s_{\gamma\nu}(\tvk)\)};
			\diagram* {
				(a) -- [charged boson] (b) --  [boson] (c),
			};
		\end{feynman}
	\end{tikzpicture}
	\\
	\tPi_{\alpha\beta\gamma\nu}(\tvk){\color{black}\quad : \quad}&
	\begin{tikzpicture}[baseline=(a.base)]
		\begin{feynman}
			\vertex at (0,0) (a) {\(\hs_{\alpha\beta}(-\tvk)\)};
			\vertex at (2.2,0) [blob] (b) {};
			\vertex at (4,0) (c) {\(\hs_{\gamma\nu}(\tvk)\)};
			\diagram* {
				(a) -- [charged boson] (b) --  [anti charged boson] (c),
			};
		\end{feynman}
	\end{tikzpicture}
\end{align}

To identify perturbative corrections of the gaussian parameters, we need to expand the self energies in $\omega$ and $\vk$ and keep only the leading terms.
By comparing the corrections $\Delta\mathcal{S}_0$ in Eq. \eqref{eq:dS0}. with the form of the bare action $\mathcal{S}_0$ in Eq. \eqref{eq:S0psi}-\eqref{eq:S0s}, we can define the perturbative corrections to the gaussian parameters through the relations,
\begin{align}
    \Sigma_{\alpha\beta}(\tvk)&=\left[-\iu \omega\, {\color{black}\delta\Omega} + m_0 \,\delta m  + k^2 \Gamma_0 \,\delta \Gamma \right] \ln b \ P_{\alpha\beta}^\perp(\vk) + \dots
    \label{signora}
    \\
    \tSigma_{\alpha\beta}(\tvk)&= \tGamma_0 \,\delta \tGamma \ln b \ P_{\alpha\beta}^\perp(\vk) + \dots\\
    \Pi_{\alpha\beta\gamma\nu}(\tvk)&= \frac{1}{2} k^2 \lambda_0^\perp\, \delta\lambda^\perp \ln b \ \PP_{\alpha\beta\gamma\nu}^\perp (\vk) +
    \frac{1}{2} k^2 \lambda_0^\parallel\,\delta\lambda^\parallel \ln b \  (\II-\PP^\perp)_{\alpha\beta\gamma\nu} (\vk) +\dots\label{pi}\\
    \tPi_{\alpha\beta\gamma\nu}(\tvk)&= \frac{1}{2} k^2 \tlambda_0^\perp\, \delta\tlambda^\perp  \ln b \ \PP_{\alpha\beta\gamma\nu}^\perp (\vk) +
    \frac{1}{2} k^2 \tlambda_0^\parallel\,\delta\tlambda^\parallel \ln b  \ (\II-\PP^\perp)_{\alpha\beta\gamma\nu} (\vk) +\dots\label{tpi}
\end{align}
where we denoted the bare parameters with the subscript $0$ and  where the ellipses stand for higher orders terms in $\omega$ and $\vk$, which are irrelevant in determining the critical behaviour at first order in $\epsilon$. The $\ln b$ factors present in all terms reflect the fact that perturbative corrections are proportional to the volume of the momentum shell.
After the shell integration, the gaussian action takes the following form,
\begin{equation}
\begin{split}
    \mathcal S_{\Lambda/b} & = \int\limits_{\left|\vk\right|<\Lambda/b} \frac{\di^dk}{\left(2\pi\right)^d} \int\limits_{-\infty}^{\infty}\frac{\di \omega}{2\pi}
    \left[ -\iu \omega \left(1+{\color{black}\delta\Omega}\ln b\right)\hat \psi \psi  + \Gamma_0 ( 1+\delta \Gamma \ln b ) k^2 \hat \psi \psi + m_0 (1 +\delta m \ln b) \hat \psi \psi \right] -\\
    - & \int\limits_{\left|\vk\right|<\Lambda/b} \frac{\di^dk}{\left(2\pi\right)^d} \int\limits_{-\infty}^{\infty}\frac{\di \omega}{2\pi} \, \tGamma_0 \left(1+\delta\tGamma \ln b\right) \hpsi \hpsi+
    \\
    + & \frac{1}{2} \int\limits_{\left|\vk\right|<\Lambda/b} \frac{\di^dk}{\left(2\pi\right)^d} \int\limits_{-\infty}^{\infty}\frac{\di \omega}{2\pi} \left[ -\iu \omega \hat s s+ \eta_0 \hat s s+  \lambda^\perp _0( 1 + \delta \lambda^\perp \ln b)k^2 \hat s s + \lambda^\parallel_0 ( 1+\delta \lambda^\parallel \ln b) k^2 \hat s s \right] - \\
    - &  \frac{1}{2}\int\limits_{\left|\vk\right|<\Lambda/b} \frac{\di^dk}{\left(2\pi\right)^d} \int\limits_{-\infty}^{\infty}\frac{\di \omega}{2\pi}  \left[\tilde\eta_0 \hat s\hat s+ \tlambda^\perp _0( 1 + \delta \tlambda^\perp \ln b)k^2 \hat s\hat s+ \tlambda^\parallel_0 ( 1+\delta \tlambda^\parallel \ln b) k^2\hat s\hat s \right]
\end{split}
\label{corrado}
\end{equation}
where we omitted the tensorial structure of the action to facilitate the reading.
\textcolor{black}{The self-energies thus give perturbative corrections to the gaussian parameters $\Gamma$, $m$, $\tGamma$, $\lambda^{\perp/\parallel}$ and $\tlambda^{\perp/\parallel}$. Moreover, the self-energy $\Sigma$ has also a non-vanishing term linear in $\omega$, namely $\delta \Omega$, which gives a perturbative correction to $-\iu\omega \hpsi\psi$; because this term is not multiplied by any parameter in the original bare action, the only way to reabsorb this correction will be through a modification of the scaling dimension of the fields $\hpsi$ and $\psi$. 
On the other hand, analogous perturbative terms do not arise in the self-energy $\Pi$, because all one-loop diagrams are at least of order $\vk$ as a consequence of the particular properties of the vertices \eqref{fd:mcs_vertex} and \eqref{fd:spps_vertex}, which vanish at $\vk=0$, so that the term $-\iu\omega \hs s$ acquires only $\vk$-dependent perturbative corrections that vanish when $\vk= 0$. 
Another -- rather crucial -- consequence of the fact that all corrections to $\Pi$ and $\tilde\Pi$ are proportional to $\vk$, is that there are no perturbative corrections to neither $\eta_0\hs s$ nor $\tilde\eta_0\hs\hs$, and hence no $\delta\eta$ and $\delta\tilde\eta$ corrections are present in Eq. \eqref{pi} and \eqref{tpi}. This is no coincidence, but it is a result deeply related to the presence of the underlaying (rotational) symmetry of the problem. Although terms explicitly violating the conservation of the spin $s$ are present, namely the dissipative terms $\eta$ and $\tilde\eta$, the symmetry is still at work and the non-linear interactions are not able to generate any perturbative corrections to the dissipative coefficients $\eta$ and $\tilde\eta$.}

The standard way to explicitly perform this shell integration, and to compute the corrections to the bare parameters of the model, is using perturbation theory. The corrections $\delta \mathcal P$ are thus computed using a Feynman diagram expansion.
At order $\epsilon$, the non-vanishing diagrams contributing the self-energies are listed in Section \ref{sec:selfen}.

Once the shell integration is performed, we are left with an effective action with a cutoff of $\Lambda/b$, a coefficient different from $1$ in front of the $-\iu\omega\hpsi\psi$ term and modified parameters, namely
\begin{equation}
    \mathcal{P}_0\to\mathcal{P}_0 (1+\delta\mathcal{P}\ln b)\simeq \mathcal{P}_0 b^{\delta\mathcal{P}}
\end{equation}
Following Section \ref{sec:RGt}, we perform the following rescalings
\begin{align}
    {\color{black}\vk =}&{\color{black} b^{-1}\,\vk_b} & {\color{black}\omega =}& {\color{black}b^{-z} \omega_b}\label{pino}\\
    \vpsi(\vk,\omega) =& b^{-\chi_\psi} \vpsi(\vk_b,\omega_b) &
    \hvpsi(\vk,\omega) =& b^{-\chi_{\hpsi}} \hvpsi(\vk_b,\omega_b)\label{scala}\\
    \vs(\vk,\omega) =& b^{-\chi_s} \vs(\vk_b,\omega_b) &
    \hvs(\vk,\omega) =& b^{-\chi_{\hs}} \hvs(\vk_b,\omega_b)\label{abete}
\end{align}
After this rescaling, we end up with an action with the same cutoff $\Lambda$ but with new renormalized parameters and couplings, which will be denoted with a subscript $b$.

\textcolor{black}{Let us at first focus on what happens to the $-\iu\omega$ terms under rescaling. Once the shell integration is performed, the rescaling procedure defined by Eq. \eqref{pino}-\eqref{abete} transforms these terms as follows
\begin{multline}
	\int\limits_{\left|\vk\right|<\Lambda/b} d^{d}k \int\limits_{-\infty}^{+\infty} d \omega \left[-\iu\omega\left(1+{\color{black}\delta\Omega} \ln b\right)\hpsi(-\vk,-\omega)\psi(\vk,\omega)\right] \quad\Rightarrow\\
	 \Rightarrow\quad b^{-\chi_{\hpsi}-\chi_{\psi}-d-2z+{\color{black}\delta\Omega}} \int\limits_{\left|\vk_{b}\right|<\Lambda} d^{d}k_{b} \int\limits_{-\infty}^{+\infty} d \omega_{b} \left[-\iu\omega_{b}\hpsi(-\vk_{b},-\omega_{b}) \psi(\vk_{b},\omega_{b})\right]
	 \label{reno}
\end{multline}
\begin{equation}
	\int\limits_{\left|\vk\right|<\Lambda/b} d^{d}k \int\limits_{-\infty}^{+\infty} d \omega \left[-\iu\omega\hs(-\vk,-\omega)s(\vk,\omega)\right] \quad\Rightarrow\quad b^{-\chi_{\hs}-\chi_{s}-d-2z} \int\limits_{\left|\vk_{b}\right|<\Lambda} d^{d}k_{b} \int\limits_{-\infty}^{+\infty} d \omega_{b} \left[-\iu\omega_{b}\hs(-\vk_{b},-\omega_{b}) s(\vk_{b},\omega_{b})\right]\label{senna}
\end{equation}
where we remind that the $-\iu\omega \hs s$ term has no perturbative corrections because of the rotational symmetry.
As previously mentioned, these terms lack of a coupling constant to redefine. Hence, the scaling dimensions of the fields must be properly chosen to restore the same structure of the bare action, namely
\begin{align}
    \chi_{\hpsi}+\chi_\psi = -d-2z+{\color{black}\delta\Omega}& & \chi_{\hs}+\chi_s=-d-2z& \label{eq:b_omega}
\end{align}
With this choice, the renormalized theory will still have a coefficient equal to one in front of the time derivative of the equation of motion. This requirement allows to fix the scaling dimension of the response fields $\chi_{\hpsi}$ and $\chi_{\hs}$, 
\begin{align}
	\chi_{\hpsi}&=-\chi_\psi-d-2z+{\color{black}\delta\Omega} & \chi_{\hs}&=-\chi_s-d-2z \,\label{eq:chi_field}\ .
\end{align}}
\textcolor{black}{Rescaling can be performed on all the other terms of the action in the same way as it was done in Eq.~\eqref{reno}-\eqref{senna}. In these cases however, the powers of $b$ appearing after rescaling can be reabsorbed by defining new {\it renormalized} parameters of the theory.} These new values of the parameters can be expressed as functions of the bare parameters, \textcolor{black}{through the perturbative corrections}, the dynamic exponent $z$ and the \textcolor{black}{scaling dimensions} $\chi$ of the fields.
\textcolor{black}{Scaling dimensions} for time and fields are not known {\it a priori}, but can be determined by imposing additional conditions.
For what concernes the gaussian action, the renormalized parameters are given by
\textcolor{black}{
\begin{align}
    \Gamma_b&=\Gamma_0 b^{\chi_\Gamma} & 
    \chi_\Gamma&=-\chi_{\hpsi}-\chi_\psi-d-2-z+\delta\Gamma\\
    \tilde{\Gamma}_b&= \tilde{\Gamma}_0 b^{\chi_{\tilde{\Gamma}}} & 
    \chi_{\tilde{\Gamma}}&=-2\chi_{\hpsi}-d-z+\delta\tGamma\\
    \lambda_b^{\perp/\parallel}&=\lambda_0^{\perp/\parallel} b^{\chi_{\lambda^{\perp/\parallel}}} &
    \chi_{\lambda^{\perp/\parallel}}&=-\chi_{\hs}-\chi_s-d-2-z+\delta\lambda^{\perp/\parallel}\\
    \tilde{\lambda}_b^{\perp/\parallel}&=\tilde{\lambda}_0^{\perp/\parallel} b^{\chi_{\tilde{\lambda}^{\perp/\parallel}}} &
    \chi_{\tilde{\lambda}^{\perp/\parallel}}&=-2\chi_{\hs}-d-2-z+\delta\tlambda^{\perp/\parallel}\\
    \eta_b&=\eta_0 b^{\chi_{\eta}} &
    \chi_{\eta}&=-\chi_{\hs}-\chi_s-d-z\\
    \tilde\eta_b&=\tilde\eta_0 b^{\chi_{\tilde\eta}} &
    \chi_{\tilde\eta}&=-2\chi_{\hs}-d-z\\
    m_b&=m_0 b^{\chi_m} & 
    \chi_m&=-\chi_{\hpsi}-\chi_\psi-d-z+\delta m
\end{align}}
\textcolor{black}{By taking advantage of Eq. \eqref{eq:chi_field} it is furthermore possible to write all the scaling dimensions of the parameters in terms of the scaling dimensions of the frequency and physical fields only, namely of $z$, $\chi_\psi$ and $\chi_s$:
\begin{align}
    	\chi_\Gamma&=z-2+\delta\Gamma-{\color{black}\delta\Omega} &
    	\chi_{\tilde{\Gamma}}&=2\chi_{\psi}+d+3z+\delta\tGamma-2{\color{black}\delta\Omega}
	\label{eq:scaling_gamma}\\
    	\chi_{\lambda^{\perp/\parallel}}&=z-2+\delta\lambda^{\perp/\parallel} &
    	\chi_{\tlambda^{\perp/\parallel}}&=2\chi_{s}+d+3z-2+\delta\tlambda^{\perp/\parallel}
	\label{eq:scaling_lambda}\\
    	\chi_{\eta}&=z&
    	\chi_{\tilde\eta}&=2\chi_{s}+d+3z
	\label{eq:scaling_eta}\\
	\chi_{m}&=z+\delta m-{\color{black}\delta\Omega} &
\end{align}
Let us note here a crucial fact: since the dynamic critical exponent $z$ is always positive, so is the scaling dimension $\chi_{\eta}$ of the dissipation $\eta$. Due to this, $\eta$ grows exponentially when the RG transformation is iterated over and over, eventually diverging. This makes $\eta$ a relevant parameter, with a role similar to that of the mass $m$: while $m$ is the relevant parameter that drives the system away from the critical manifold, similarly $\eta$ drives the system away from the conservative plane (while remaining on the critical manifold). Hence, independentely of all the other parameters, we expect $\eta$ to diverge under the RG flow whenever its bare value is non-zero. This will have important consequences that we will discuss later on.}

Now that we have defined the perturbative corrections to the gaussian parameters, we shall switch our attention to the coupling constants of the non-linear interactions.
Perturbative corrections to the interactions are known as {\it vertex functions}. \textcolor{black}{These vertex functions can be diagrammatically expressed through the sum of all the amputated connected diagrams having as external fields the same fields of the vertex they are correcting.
The six vertex functions of our theory, one for each bare vertex, can be graphycally represented through the blobs in the following expressions}
\begin{align}
    V_{\alpha\beta\gamma\nu}^{\hpsi \psi s}(\tvk,\tvq){\color{black} \quad : \quad}&
	\begin{tikzpicture}[baseline=(a.base)]
		\begin{feynman}[small]
			\vertex (a) at (0,0) {\(\hpsi_\alpha(-\tvk)\)};
			\vertex (b) [blob] at (1.6,0) {};
			\vertex (c) at (2.35,1.3) {\(\psi_\beta(\tvk-\tvq)\)};
			\vertex (d) at (2.35,-1.3) {\(s_{\gamma\nu}(\tvq)\)};
			\diagram* {
				(a) -- [fermion] (b) -- (c),
				(b) -- [boson] (d),
			};
		\end{feynman}
	\end{tikzpicture}
    &
    V_{\alpha\beta\gamma\nu}^{\hs \psi\psi}(\tvk,\tvq){\color{black} \quad : \quad}&
    \begin{tikzpicture}[baseline=(a.base)]
		\begin{feynman}[small]
			\vertex (a) at (0,0) {\(\hs_{\alpha\beta}(-\tvk)\)};
			\vertex (b)[blob] at (1.7,0) {};
			\vertex (c) at (2.45,1.3) {\(\psi_\gamma(\tvk/2-\tvq)\)};
			\vertex (d) at (2.45,-1.3) {\(\psi_\nu(\tvk/2+\tvq)\)};
			\diagram* {
				(a) -- [charged boson] (b) -- (c),
				(b) -- (d),
			};
		\end{feynman}
	\end{tikzpicture}\\
    V_{\alpha\beta\gamma}^{\hpsi \psi \psi}(\tvk,\tvq){\color{black} \quad : \quad}&
    \begin{tikzpicture}[baseline=(a.base)]
		\begin{feynman}[small]
			\vertex (a) at (0,0) {\(\hpsi_\alpha(-\tvk)\)};
			\vertex (b) [blob] at (1.6,0) {};
			\vertex (c) at (2.35,1.3) {\(\psi_\beta(\tvq)\)};
			\vertex (d) at (2.35,-1.3) {\(\psi_\gamma(\tvk-\tvq)\)};
			\diagram* {
				(a) -- [fermion] (b) -- (c),
				(b) -- (d),
			};
		\end{feynman}
	\end{tikzpicture}
	&
	V_{\alpha\beta\gamma\nu\eta}^{\hs s \psi}(\tvk,\tvq){\color{black} \quad : \quad}&
	\begin{tikzpicture}[baseline=(a.base)]
		\begin{feynman}[small]
			\vertex (a) at (0,0) {\(\hs_{\alpha\beta}(-\tvk)\)};
			\vertex (b) [blob] at (1.7,0) {};
			\vertex (c) at (2.45,1.3) {\(s_{\gamma\nu}(\tvk/2-\tvq)\)};
			\vertex (d) at (2.45,-1.3) {\(\psi_\eta(\tvk/2+\tvq)\)};
			\diagram* {
				(a) -- [charged boson] (b) -- [boson] (c),
				(b) -- (d),
			};
		\end{feynman}
	\end{tikzpicture}\\
	V_{\alpha\beta\gamma\nu}^{\hpsi \psi \psi \psi}(\tvk,\tvq,\tvh){\color{black} \quad : \quad}&
	\begin{tikzpicture}[baseline=(a.base)]
		\begin{feynman}[small]
			\vertex (a) at (0,0) {\(\hpsi_\alpha(-\tvk)\)};
		    \vertex (b)[blob] at (1.6,0) {};
			\vertex (c) at (2.35,1.3) {\(\psi_\beta(\tvq)\)};
			\vertex (d) at (2.85,0) {\(\psi_{\gamma}(\tvh)\)};
			\vertex (e) at (2.35,-1.3) {\(\psi_{\gamma}(\tvk-\tvq-\tvh)\)};
			\diagram* {
				(a) -- [fermion] (b) -- (c),
				(b) -- (d),
				(b) -- (e),
			};
		\end{feynman}
	\end{tikzpicture}
	&
	V_{\alpha\beta\gamma\nu\sigma\tau}^{\hs \psi\psi\psi\psi} (\tvk,\tvq,\tvh,\tvp){\color{black} \quad : \quad}&
	\begin{tikzpicture}[baseline=(a.base)]
		\begin{feynman}[small]
			\vertex (a) at (0,0) {\(\hs_{\alpha\beta}(-\tvk)\)};
			\vertex (b) [blob] at (1.7,0) {};
			\vertex (c) at (1.3,1.425) {\(\psi_\gamma(\tvq)\)};
			\vertex (d) at (2.91,0.88) {\(\psi_\nu(\tvh)\)};
			\vertex (e) at (2.91,-0.88) {\(\psi_\sigma(\tvp)\)};
			\vertex (f) at (1.3,-1.425) {\(\psi_\tau(\tvk-\tvq-\tvh-\tvp)\)};
			\diagram* {
				(a) -- [charged boson] (b) -- (c),
				(b) -- (d),
				(b) -- (e),
				(b) -- (f),
			};
		\end{feynman}
	\end{tikzpicture}
\end{align}
At one-loop, the non-vanishing diagrams contributing to the vertices are listed in Section \ref{sec:vertexfun}. Each vertex function contributes to the corrections of couplings and parameters of Eqs. \eqref{fd:mcv_vertex}-\eqref{fd:mixed_vertex} in the following way,
\begin{gather}
     V_{\alpha\beta\gamma\nu}^{\hpsi \psi s}(\tvk,\tvq) = g\, \delta g_\psi\, P_{\alpha\rho}^\perp\left(\vk\right)\II_{\rho\beta\gamma\nu} \label{eq:corr_gv}\\
     V_{\alpha\beta\gamma}^{\hpsi \psi \psi}(\tvk,\tvq) = - v_0 \frac{\iu\gamma_v}{2} \delta\gamma_v P_{\alpha\beta\nu}\left(\vk\right)\\
     V_{\alpha\beta\gamma\nu}^{\hpsi \psi \psi \psi}(\tvk,\tvq,\tvh) = -\frac{J}{3} \delta J \,Q_{\alpha\beta\gamma\nu}\left(\vk\right)\\
    V_{\alpha\beta\gamma\nu}^{\hs \psi\psi}(\tvk,\tvq) = g \, \II_{\alpha\beta\rho\sigma}\II_{\rho\tau\gamma\nu} \left[
    \delta g_s \, \vk\cdot\vq \, \delta_{\sigma\tau} +
    \Phi_1 \, \delta g_{s1} \,  k_\sigma q_\tau +
    \Phi_2 \, \delta g_{s2} \, q_\sigma k_\tau
    \right] \label{eq:corr_gs}\\
    V_{\alpha\beta\gamma\nu\eta}^{\hs s \psi}(\tvk,\tvq) = - v_0 \frac{\iu \gamma_s}{2} \II_{\alpha\beta\rho\sigma} \II_{\rho\tau\gamma\nu} \left[
    \delta \gamma_s \,k_\eta \delta_{\sigma\tau}+
    2 \mu_1\, \delta \gamma_{s1} \,  k_\tau \delta_{\sigma\eta} +
    2 \mu_2\, \delta \gamma_{s2} \, k_\sigma \delta_{\tau\eta} \right]\label{eq:corr_gammas}\\
    V_{\alpha\beta\gamma\nu\sigma\tau}^{\hs \psi\psi\psi\psi} (\tvk,\tvq,\tvh,\tvp) = \frac{\kappa}{12} \delta\kappa \, K_{\alpha\beta\gamma\nu\sigma\tau}\left({\vk},\vq,\vh,\vp,\vk-\vq-\vh-\vp\right)
\end{gather}
Higher orders in $\omega$ and $\vk$ turn out to be irrelevant in determining the critical behaviour at first order in $\epsilon$. \textcolor{black}{The perturbative corrections to the anomalous terms have been defined as, 
\begin{gather}
\delta g_{s1}\equiv\delta\left(g\, \Phi_{1}\right) \quad , \quad \delta g_{s2}\equiv\delta\left(g\, \Phi_{2}\right) \\
\delta \gamma_{s1}\equiv\delta\left(\gamma_{s}\, \mu_{1}\right) \quad , \quad \delta \gamma_{s2}\equiv\delta\left(\gamma_{s}\, \mu_{2}\right)
\end{gather} 
(because they represent corrections to the {\it products} $g\,\Phi_{1,2}$ and $\gamma_{s}\,\mu_{1,2}$, they were not  simply called
$\delta\Phi_{1,2}$ and $\delta\mu_{1,2}$).}
After rescaling momenta, frequency and fields, we obtain the following RG transformations,
\textcolor{black}{
\begin{align}
    {\gamma_\psi}_b&={\gamma_\psi}_0 b^{\chi_{\gamma_\psi}} & 
    \chi_{\gamma_\psi}&=-\chi_{\hpsi}-2\chi_\psi-2d-1-2z+\delta\gamma_\psi\\
    {\gamma_s}_b&={\gamma_s}_0 b^{\chi_{\gamma_s}}  & 
    \chi_{\gamma_s}&=-\chi_{\hs}-\chi_s-\chi_\psi-2d-1-2z+\delta\gamma_s\\
    {\mu_1}_b&= {\mu_1}_0 b^{\chi_{\mu_1}} &
    \chi_{\mu_1}&=\delta\gamma_{s1}-\delta\gamma_s\\
    {\mu_2}_b&= {\mu_2}_0 b^{\chi_{\mu_2}} &
    \chi_{\mu_2}&=\delta\gamma_{s2}-\delta\gamma_s\\
    g_b^{\left(\psi\right)}&=g_0 b^{\chi_{g^{\left(\psi\right)}}} & 
    \chi_{g^{\left(\psi\right)}}&=-\chi_{\hpsi}-\chi_\psi-\chi_s-2d-2z+\delta g_\psi\label{eq:b_gpsi}\\
    g_b^{\left(s\right)}&=g_0 b^{\chi_{g^{\left(s\right)}}} & 
    \chi_{g^{\left(s\right)}}&=-\chi_{\hs}-2\chi_\psi-2d-2-2z+\delta g_s\label{eq:b_gs}\\
    {\Phi_1}_b&= {\Phi_1}_0 b^{\chi_{\Phi_1}} &
    \chi_{\Phi_1}&=\delta g_{s1}-\delta g_s \\
    {\Phi_2}_b&= {\Phi_2}_0 b^{\chi_{\Phi_2}} &
    \chi_{\Phi_2}&= \delta g_{s2}-\delta g_s \\
    J_b&=J_0 b^{\chi_J} &
    \chi_J&=-\chi_{\hpsi}-3\chi_{\psi}-3d-3z+\delta J\\
    \kappa_b&=\kappa_0 b^{\chi_\kappa} &
    \chi_\kappa&=-\chi_{\hs}-4\chi_\psi - 4d - 4z + \delta\kappa
\end{align}
}
By using Eq. \eqref{eq:chi_field} we can write all the scaling dimensions of the interaction couplings in terms of the scaling dimensions of the frequency and physical fields, namely of $z$, $\chi_\psi$ and $\chi_s$:
\textcolor{black}{
\begin{align}
    	\chi_{\gamma_\psi}&=-\chi_\psi-d-1+\delta \gamma_\psi-{\color{black}\delta\Omega} &
	\chi_{\gamma_s}&=-\chi_\psi-d-1+\delta\gamma_s\label{eq:scaling_gamma2}\\
    	\chi_{\mu_1}&= \delta\gamma_{s1} - \delta\gamma_s &
	\chi_{\mu_2}&= \delta\gamma_{s2} - \delta\gamma_s\\
    	\chi_{g^{\left(\psi\right)}}&=-\chi_s-d+\delta g_\psi-{\color{black}\delta\Omega} &
	\chi_{g^{\left(s\right)}}&=\chi_{s}-2\chi_\psi-d-2+\delta g_s\\
    	 \chi_{\Phi_1}&=\delta g_{s1} - \delta g_s &
	 \chi_{\Phi_2}&=\delta g_{s2} - \delta g_s\\
    	\chi_J&=-2\chi_{\psi}-2d-z+\delta J - {\color{black}\delta\Omega} &
	\chi_\kappa&=\chi_s-4\chi_\psi -3d - 2z + \delta\kappa \label{eq:scaling_ferro}
\end{align}
}
By looking at Eqs. \eqref{eq:b_gpsi} and \eqref{eq:b_gs} it may seem that we have {\it two} different mode-coupling constants; however, this is not the case.
We defined two different renormalized coupling constants $g^{\left(\psi\right)}$ and $g^{\left(s\right)}$ only because the perturbative corrections $\delta g_\psi$ and $\delta g_s$ arising in Eqs. \eqref{eq:corr_gv} and \eqref{eq:corr_gs} are completely independent from each other, but this does not mean that there are two different {\it physical} constants. In fact, there must be only one mode-coupling constant, because $g$ arises in the derivation of the equations of motion as the consequence of a symmetry, encoded by the Poisson-bracket relation $\{ s, \psi\}\propto g \psi$, stating that $\vs$ is the generator of the rotations of $\vpsi$.
The mode-coupling terms in both the equations of motion derive from this one Poisson-bracket relation; therefore, the existence of two different couplings would mean losing the connection with the underlying symmetry and Poisson structure. As we mentioned before, the scaling behaviour of the physical fields $\vpsi$ and $\vs$ is arbitrary, which allows a certain freedom in determining the scaling dimensions $\chi_\psi$ and $\chi_s$, which enter Eqs. \eqref{eq:b_gpsi} and \eqref{eq:b_gs}.
Hence, it is possible to use this freedom to restore the identity of the two mode-coupling constants, 
\begin{equation}
    {g_b^{\left(\psi\right)}}={g_b^{\left(s\right)}}=g_b
\end{equation}
by simply asking that $\chi_s$ is chosen in such a way that,
\begin{equation}
    \chi_{g^{\left(\psi\right)}}=\chi_{g^{\left(s\right)}}
\end{equation}
Essentially, we are requiring the field $\vs$ to scale in such a way that the coupling regulating the symmetry has one unique scaling behaviour.


\subsubsection{A primer of the diagrammatic expansion}
\label{sec:primer}

The one-loop Feynman diagrams generated by the action's expansion, and necessary to calculate all the corrections $\delta\mathcal{P}$ to the parameters and coupling constants, are listed in Section \ref{sec:diagrams}; not only they are many, but they are also quite complicated due to the tensorial structure of the theory. Hence, we will perform in this Section the explicit calculation of one diagram, hoping that this may help the interested reader in picking up the general technique. The diagram we calculate is a contribution to the self-energy $\Sigma$ defined in \eqref{sissi},
\begin{equation} D_{\alpha \beta}  ( \vk,\omega_k) \quad {\color{black} :}\quad
\begin{tikzpicture}[baseline=(a.base)]
\begin{feynman}[layered layout,small]
\vertex (a) at (0,0) {$-\vk,-\omega_k$};
\vertex (v1) [dot] at (1.5,0) {};
\vertex (v2) [dot] at (3.5,0) {};
\vertex (b) at (5,0) {$\vk,\omega_k$};
\vertex (ind1) at (0.1,0.5) {$\alpha$};
\vertex (ind2) at (5,0.5) {$\beta$};
\vertex (mom1) at (2.8,1.3) {$\vk /2 - \vp,  \omega_k/2-\omega_p $};
\vertex (mom1) at (2.8,-1.3) {$\vk /2 + \vp, \omega_k/2+\omega_p $};
\diagram* {
(a) -- [fermion] (v1) -- [half left, fermion] (v2) -- (b),
(v2) -- [half left, boson ] (v1),
};
\end{feynman}
\end{tikzpicture}
\label{exampleDiagram}
\end{equation}
\textcolor{black}{To facilitate the reader to follow the calculation and understand it better, we shall set $\eta=\tilde{\eta}=0$ in the following paragraphs. Generalization of the following calculation in the case with $\eta$ and $\tilde{\eta}$ is straightforward.}
The integral expression corresponding to this diagram can written following the definitions of lines and vertices given in Sections \ref{sec:lin} and \ref{sec:nonlin},
 \begin{equation}
D_{\alpha \beta} ( \vk,\omega_k) =  g^2  \int  \frac {d^d p}{( 2 \pi)^d }  \frac{ d\omega_p}{ 2 \pi} \
P_{\alpha \theta}(\vk) \II_{\theta\gamma\rho\sigma}  \  \mathbb{G}_{\gamma \nu}^{0,\psi}\left(\frac{\vk}{2}-\vp ,
\frac{\omega_k}{2}- \omega_p \right)\mathbb{C}_{\rho \sigma  \mu \lambda}^{0,s}\left(\frac{\vk}{2}+\vp, \frac{\omega_k}{2}+ \omega_p \right) P_{\nu \eta}\left(\frac{\vk}{2}-\vp\right) \II_{\eta\beta\mu\lambda}
\label{exampleDiagramStep0}
\end{equation}
While the momentum integral is restricted to the momentum shell, $\Lambda/b < p < \Lambda$, the frequency integral is extended over the whole spectrum $(-\infty, +\infty)$. Furthermore, we notice that the integrand \eqref{exampleDiagramStep0} depends on the frequency $\omega_p$ only through the correlation functions $\mathbb G^{0,\psi}$ and $\mathbb C^{0,s}$ defined in Section \ref{sec:lin}; this is in general true for any Feynman diagram present in this work. All the correlation functions \eqref{eq:greenpsi}-\eqref{eq:corrs} have at most two poles in the complex plane, and it is always possible to compute the frequency integral using the residue theorem.
In the case of this diagram the integration over $\omega_p$ yields, 
\begin{equation}
\begin{split}
D_{\alpha \beta}(\vk,\omega_k) = g^2  \int  \frac {d^d p}{( 2 \pi)^d } P_{\alpha \theta}(\vk) \II_{\theta\gamma\rho\sigma} P_{\nu \eta}(\vk/2-\vp) \II_{\eta\beta\mu\lambda}  \left[ \frac{2\tilde \lambda^\perp  \ \delta_{\gamma \nu}\mathbb P ^\perp_{\rho \sigma \mu \lambda}(\vk/2-\vp) }{\lambda^\perp[\Gamma (\vk/2-\vp)^2+\lambda^\perp (\vk/2+\vp)^2-\iu \omega_k+m]} \right. \\
\left. + \frac{2\tilde \lambda^\parallel  \ \delta_{\gamma \nu}(\mathbb P ^\perp_{\rho \sigma \mu \lambda}(\vk/2-\vp) - \II_{\rho\sigma\mu\lambda} )}{\lambda^\parallel [\Gamma (\vk/2-\vp)^2+\lambda^\parallel (\vk/2+\vp)^2-\iu \omega_k+m]} \right]
\end{split}
\label{explicitDiagramstep2}
\end{equation}
Since we are going to work at first order in $\epsilon = 4-d$ (one loop), we can drop the $m$ dependence in the Feynman diagrams, because this would lead to higher order corrections. Hence we set $m=0$ in the following. The integral over the modulus of the momentum becomes trivial if we assume that the RG transformation is infinitesimal, namely if $b\simeq 1$; this means that the thickness of the momentum shell is infinitesimal and it is possible to approximate the integral of a generic function $f(\vp)$ as follows,
\begin{equation}
\int_{\Lambda/b}^{\Lambda} d^d p \ f(\vp)  =  \int  f(p) d \Omega_d p^{d-1}dp  = \ln b \ \Lambda^d \int f(p) \biggl |_{ |p| = \Lambda} d \Omega_d  
\end{equation}
which corresponds to approximating the integral as the value of the function at $|p| = \Lambda$ times the volume of the momentum shell $\Lambda^d ( 1-1/b) \simeq \ln b$.  After this step, the integrand may still depend on the direction of the momentum $\hat \vp=\vp/|p|$, which is a unit versor. Applying this procedure to equation \eqref{explicitDiagramstep2} leads to,
\begin{equation}
\begin{split}
D_{\alpha \beta}(\vk,\omega_k) = g^2  \int  \frac {d \Omega_d}{( 2 \pi)^d } \Lambda^d \ln b\  P_{\alpha \theta}(\vk) \II_{\theta\gamma\rho\sigma}& P_{\nu \eta}(\vk/2-\Lambda \hat \vp) \II_{\eta\beta\mu\lambda}  \left[ \frac{2\tilde \lambda^\perp  \ \delta_{\gamma \nu}\mathbb P ^\perp_{\rho \sigma \mu \lambda}(\vk/2-\Lambda \hat \vp) }{\lambda^\perp [\Gamma (\vk/2-\Lambda \hat \vp)^2+\lambda^\perp (\vk/2+\Lambda \hat \vp)^2-\iu \omega_k]} \right. \\&
\left. + \frac{2\tilde \lambda^\parallel  \ \delta_{\gamma \nu}(\mathbb P ^\perp_{\rho \sigma \mu \lambda}(\vk/2-\Lambda \hat \vp) - \II_{\rho\sigma\mu\lambda} )}{\lambda^\parallel [\Gamma (\vk/2-\Lambda \hat \vp)^2+\lambda^\parallel (\vk/2+\Lambda \hat \vp)^2-\iu \omega_k]} \right]
\end{split}
\label{explicitDiagramstep3}
\end{equation}
where the versor $\hat \vp$ is integrated over the $d$-dimensional sphere $\Omega_d$. The main advantage of having an expression of this kind is that all the dependence on the cut-off $\Lambda$ is made explicit.

In order to recognize within this diagram the corrections to the various parameters of the original action, one has to expand in small external momenta and frequency, as we did in \eqref{signora}. The order at which it is convenient to perform this expansion depends on the particular Feynman diagram we are considering and the specific parameters or coupling constants that it corrects. For example, the diagram we are studying here gives rise to perturbative corrections to the $\hat \psi \psi$ term in the Gaussian action \eqref{eq:S0psi}; hence we must expand it up to the second order in the external momentum $\vk$ and up to the first order in the external frequency $\omega_k$. In this example we will calculate explicitly only the contribution of order $\omega_k$ of this diagram, even though the diagram gives also contributions of $\mathcal O(1)$ and of $\mathcal O(k^2)$. The diagram at order $\omega_k$ is,
\begin{equation}
D_{\alpha \beta} (\vk,\omega_k) = \iu \omega_k
g^2 \Lambda^{d-4} \ln b \ \int  \frac {d \Omega_d }{( 2 \pi)^d }   P_{\alpha \theta}(\vk) \II_{\theta\gamma\rho\sigma} P_{\nu \eta}(\hat \vp) \II_{\eta\beta\mu\lambda} \delta_{\gamma \nu} \left[2  \tilde \lambda^\perp \frac{\mathbb P ^\perp_{\rho \sigma \mu \lambda}(\hat  \vp)}{\lambda^\perp(\Gamma +\lambda^\perp)^2}
+  2\tilde \lambda ^\parallel \frac{ \mathbb P ^\perp_{\rho \sigma \mu \lambda}(\hat \vp)- \II_{\rho\sigma \mu\lambda}}{\lambda^\parallel(\Gamma +\lambda^\parallel)^2}
\right]
\label{explicitDiagramstep4}
\end{equation}
Expanding all the tensors in equation \eqref{explicitDiagramstep4}, using the definitions \eqref{pandino}, \eqref{begagigante} and \eqref{lazonamorta}, we obtain,
\begin{equation}
D_{\alpha\beta} (\vk,\omega_k) = \iu \omega_k g^2  \Lambda^{d-4}\ln b\  P_{\alpha\theta}(\vk)\int \frac{d \Omega_d}{(2\pi)^d} \left [  \frac{  \tilde \lambda^\perp}{ \lambda^\perp ( \Gamma + \lambda^\perp) ^2 }     (1-d) \hat p_\beta \hat p_\theta   +    \frac{  \tilde \lambda^\parallel}{ \lambda^\parallel ( \Gamma + \lambda^\parallel) ^2 }    (d-2) (\hat  p_\beta \hat  p_\theta -\delta_{\alpha\beta})\right]
\end{equation}
We can now perform the integral over the $d$-dimensional sphere $d\Omega_d$, which can be done using the following two relations,
\begin{equation}
\left \langle \hat p_\alpha \hat p_\beta\right \rangle_{\hat p}=\frac{1}{d} \delta_{\alpha\beta}\ ,\qquad \quad
\left \langle \hat p_\alpha \hat p_\beta \hat p_\gamma \hat  p_\nu \right \rangle_{\hat p}=\frac{1}{d(d+2)}\left(\delta_{\alpha\beta} \delta_{\gamma\nu} + \delta_{\alpha\gamma} \delta_{\beta\nu}+\delta_{\alpha\nu} \delta_{\beta\gamma}\right)\ ,
\end{equation}
where the brackets indicate the average over the $d$-dimensional sphere,
\begin{equation}
\langle \cdot \rangle = \frac 1 {\Omega_d}  \int \ \cdot \ d\Omega_d
\end{equation}
While the average of an even number of versors $\hat \vp$ is nonzero, the angular average of an odd number of momenta vanishes by symmetry. This leads to the following expression for the diagram \eqref{exampleDiagram},
\begin{equation}
D_{\alpha\beta}(\vk,\omega_k) = - \iu \omega_k g^{2}\ \Lambda^{d-4} \frac{d-1}{d} \left[
\frac{\tilde \lambda ^\perp}{\lambda^\perp ( \Gamma+\lambda^\perp)^2}   +  \frac{\tilde \lambda^\parallel}{\lambda^\parallel ( \Gamma+\lambda^\parallel)^2}  
(d-2) \right]\ \ln b \ P_{\alpha\beta}^\perp(\vk)
\end{equation}
Since we are performing an $\epsilon$-expansion around the upper critical dimension $d_c=4$, we can set $d=4$ in all the Feynman diagrams when results at first order in $\epsilon$ are concerned. In the end, by comparing this expression to the expansion of the self-energy in \eqref{signora}, we can finally read the correction to the scaling dimension of the fields coming from this one diagram, namely,
\begin{equation}
{\color{black}\delta\Omega} =  g^{2} \left[
\frac{3}{4}\frac{\tilde \lambda ^\perp}{\lambda^\perp (  \Gamma+\lambda^\perp)^2}  + \frac{3}{2} \frac{\tilde \lambda^\parallel}{\lambda^\parallel ( \Gamma+\lambda^\parallel)^2}  
 \right] + \dots
\end{equation}
where the dots indicates the corrections from {\it all} other diagrams contributing to the self-energy $\Sigma$, which are listed in Section \ref{sec:diagrams}.


\subsubsection{Generation of the anomalous terms}
\label{sec:anomal}
The equations of motion proposed initially in \eqref{eq:psi0}, \eqref{eq:s0}, in which the anomalous terms proportional to $\mu_1$, $\mu_2$, $\Phi_1$ or $\Phi_2$ do not appear, have a (relatively) clear physical interpretation. However, as previously pointed out, in the absence of some additional terms, which we call {\it anomalous}, the equations of motion are not RG-invariant; basically, what happens is that the RG generates some terms that were not present in the original action. To see how this happens, we perform here a shell integration starting from a theory with bare coefficients ${\Phi_1}_0={\Phi_2}_0={\mu_1}_0={\mu_2}_0=0$, and we show that these anomalous terms are spontaneously generated by the renormalization group transformation. We also hope that this further direct analysis of some highly nontrivial diagrams may provide the reader with a more robust technique to perform the entire calculation. \textcolor{black}{Since the generation of these non-linear terms does not rely on the presence of $\eta$ and $\tilde\eta$, we shall set them both to $0$. Once again, we hope this makes the calculation easier to follow.}

\vskip 0.5 truecm
{\it *** Anomalous Mode-Coupling terms}
\vskip 0.2 truecm

Without anomalies, the non Gaussian action \eqref{eq:SI}  has  only one term proportional to $\hat s \psi \psi$, corresponding to the following vertex,
\begin{equation}
\begin{tikzpicture}[baseline=(a.base)]
\begin{feynman}[small]
\vertex (a) at (0,0) {\(\hs_{\alpha\beta}(-\tvk)\)};
\vertex (b)[empty dot] at (1.7,0) {};
\vertex (c) at (2.45,1.3) {\(\psi_\gamma(\tvk/2-\tvq)\)};
\vertex (d) at (2.45,-1.3) {\(\psi_\nu(\tvk/2+\tvq)\)};
\diagram* {
(a) -- [charged boson] (b) -- (c),
(b) -- (d),
};
\end{feynman}
\end{tikzpicture}
\quad {\color{black} :}\quad g \II_{\alpha\beta\gamma\nu} \vk\cdot\vq
\label{fd:vertex_mcs_simply}
\end{equation}
which is the same mode coupling vertex as \eqref{fd:mcs_vertex_alt}, but without the anomalous terms. To compute the perturbative corrections to this vertex we must consider all the diagrams with an incoming $\hat s$ line and two outcoming $\psi$ lines; these are $12$ diagrams, listed in Fig.(\ref{fig:Vsvv}) of Section \ref{sec:diagrams}, most of which produce anomalous corrections. Here, as an example, we limit ourselves to consider only the first {\it two} of all these diagrams namely,
\begin{equation}
D^{\rm{mc},1}  _{\alpha\beta\gamma\nu} \quad {\color{black} :}  \quad - \tvk,\alpha\beta \ \
\begin{tikzpicture}[baseline=-\the\dimexpr\fontdimen22\textfont2\relax]
\begin{feynman}
\vertex (aa) at (-0.2,0); 
\vertex (mom1)  at ( 1,1) {$\frac{\tvk}{2} +\frac{ \tvq}{2} + \tvp$};
\vertex (mom2)  at ( 1,-1) {$\frac{\tvk}{2} -\frac{ \tvq}{2} - \tvp$};
\vertex (mom3)  at ( 3.3,0) {$   \tvp-\frac{ \tvq}{2} $};
\vertex (a) at (0,0) ;
\vertex (b) at (3,1.73) {$\qquad \tvk/2-\tvq, \gamma$};
\vertex (c) at(3,-1.73){$\qquad \tvk/2+\tvq, \nu$};
\vertex (v1)[empty dot] at (1,0){};
\vertex (v2)[black, dot] at (2.5,0.866){};
\vertex (v3) [black, empty dot] at (2.5,-0.866){};
\diagram*{
(a) --[charged boson](v1)--[fermion](v2)--[charged boson](v3)--[black](v1);
(v2)--[black](b);
(v3)--[black](c)
};
\end{feynman}
\end{tikzpicture}
\quad\ , \quad
D^{\rm{mc},2}  _{\alpha\beta\gamma\nu} \quad {\color{black} :}  \quad - \tvk,\alpha \beta\ \
\begin{tikzpicture}[baseline=-\the\dimexpr\fontdimen22\textfont2\relax]
\begin{feynman}
\vertex (mom1)  at ( 1,1) {$\frac{\tvk}{2} +\frac{ \tvq}{2} + \tvp$};
\vertex (mom2)  at ( 1,-1) {$\frac{\tvk}{2} -\frac{ \tvq}{2} - \tvp$};
\vertex (mom3)  at ( 3.3,0) {$  \tvp-\frac{ \tvq}{2} $};
\vertex (aa) at (-0.2,0);
\vertex (a) at (0,0) ;
\vertex (b) at (3,1.73) {$\qquad \tvk/2-\tvq, \gamma$};
\vertex (c) at(3,-1.73){$\qquad \tvk/2+\tvq, \nu$};
\vertex (v1)[empty dot] at (1,0){};
\vertex (v2)[black, dot] at (2.5,0.866){};
\vertex (v3) [black, dot] at (2.5,-0.866){};
\diagram*{
(a) --[charged boson](v1)--[fermion](v2)--[boson](v3)--[anti fermion](v1);
(v2)--[black](b);
(v3)--[black](c)
};
\end{feynman}
\end{tikzpicture}\quad\ ,
\label{eq:andiag}
\end{equation}
where we explicitly write also the momenta of  internal field lines. These diagrams can be converted into integral expressions following the usual  Feynman rules. It is important to note, though, that this is exactly one of the cases in which it is more convenient to write the spin mode-coupling vertex as in \eqref{fd:mcs_vertex} (but of course with $\Phi_1 = \Phi_2 = 0$), rather than as in \eqref{fd:mcs_vertex_alt}. The result is, 
\begin{equation}
\begin{split}
D^{\rm{mc},1} _{\alpha\beta\gamma\nu}
=& g^3
\int\limits_{\Lambda/b<|\vp|<\Lambda}\frac{\di^dp}{(2\pi)^d}
\left[\left(\vp+\frac{\vk}{2}+\frac{\vq}{2}\right)^2-\left(\vp-\frac{\vk}{2}+\frac{\vq}{2}\right)^2\right]
\left[\left(\vq+\frac{\vk}{2}\right)^2-\left(\vp-\frac{\vk}{2}+\frac{\vq}{2}\right)^2\right]  \times\\
&\times  \II_{\alpha\beta\sigma\rho} P_{\mu,\eta}^{\perp}\left(\vp+\frac{\vk}{2}+\frac{\vq}{2}\right)
\II_{\eta\gamma\xi\zeta} \II_{\phi\chi\tau\nu}
\int\limits_{-\infty}^{\infty}\frac{\di \omega_p}{2\pi}
\mathbb{G}_{\rho\mu}^{0,\psi}\left(\tvp+\frac{\tvk}{2}+\frac{\tvq}{2}\right)
\mathbb{G}_{\xi\zeta\phi\chi}^{0,s}\left(\tvp-\frac{\tvq}{2}\right)
\mathbb{C}_{\sigma\tau}^{0,\psi}\left(\tvp-\frac{\tvk}{2}+\frac{\tvq}{2}\right)\ ,
\end{split}
\end{equation}
\begin{equation}
\begin{split}
D^{\rm{mc},2} _{\alpha\beta\gamma\nu}
=& g^3
\int\limits_{\Lambda/b<|\vp|<\Lambda}\frac{\di^dp}{(2\pi)^d}
\left[\left(\vp+\frac{\vk}{2}+\frac{\vq}{2}\right)^2-\left(\vp-\frac{\vk}{2}+\frac{\vq}{2}\right)^2\right]  \II_{\alpha\beta\sigma\rho}
P_{\mu,\eta}^{\perp}\left(\vp+\frac{\vk}{2}+\frac{\vq}{2}\right) \II_{\eta\gamma\xi\zeta}
P_{\sigma,\delta}^{\perp}\left(\vp-\frac{\vk}{2}+\frac{\vq}{2}\right) \II_{\phi\chi\tau\nu}\times\\
&\times  
\int\limits_{-\infty}^{\infty}\frac{\di \omega_p}{2\pi}
\mathbb{G}_{\rho\mu}^{0,\psi}\left(\tvp+\frac{\tvk}{2}+\frac{\tvq}{2}\right)
\mathbb{C}_{\xi\zeta\phi\chi}^{0,s}\left(\tvp-\frac{\tvq}{2}\right)
\mathbb{G}_{\delta\tau}^{0,\psi}\left(-\tvp+\frac{\tvk}{2}-\frac{\tvq}{2}\right)\ .
\end{split}
\label{anomalousMCgeneration}
\end{equation}
Their sum is given by,
\textcolor{black}{
\begin{equation} D^{\rm{mc},1}_{\alpha\beta\gamma\nu}+D^{\rm{mc},2}_{\alpha\beta\gamma\nu} = g \II_{\alpha\beta\rho\sigma} \II_{\rho\tau\gamma\nu} [\delta g_s \  \vk\cdot\vq \delta_{\sigma\tau} \ + A  \     k_\sigma q_\tau   + B \   q_\sigma k_\tau  ] \ln b\quad \mbox{,}
\end{equation}}
where the constants $\delta g_s$, $A$ and $B$ are,
\begin{equation}
\begin{split}
&  \delta g_s= g^2 w\frac{{\tlambda^\perp} w (w+1)^2+\tGamma x (w (x-5)-3 x-1)-{\tlambda^\parallel} (w-3) w x (w+x)}{12 \Gamma ^3 (w+1)^2 x (w+x)}
\\
& A =-g^2 w\frac{-3 {\tlambda^\perp} w (w+1)^2+\tGamma x (w (6 w+3 x+11)+5 x+3)-{\tlambda^\parallel} w (3 w+5) x (w+x)}{12 \Gamma ^3 (w+1)^2 x (w+x)}
\\
& B = g^2 w\frac{-5 {\tlambda^\perp} w (w+1)^2+\tGamma x (w (7-5 x)-3 x+5)+{\tlambda^\parallel} w (5 w+3) x (w+x)}{12 \Gamma ^3 (w+1)^2 x (w+x)}
\end{split}
\end{equation}
with $w = \Gamma/\lambda^\parallel $ and $x = \lambda^\perp  / \lambda^\parallel$. These two Feynman diagrams correct the interacting part of the action relative to the mode-coupling of the spin as follows,
\textcolor{black}{
\begin{equation}
\Delta \mathcal S_{\rm{mc}} = \int_{0}^{\Lambda/b} d \tvk  \ d \tvq  \ g \ \II_{\alpha\beta\rho\sigma} \II_{\rho\tau\gamma\nu}\ \hat s_{\alpha \beta} (-\tvk) [
\underbrace{
\delta g_s \ \vk\cdot\vq \ \delta_{\sigma\tau}}_{\rm{bare \ structure}}+  \underbrace{A \ k_\sigma q_\tau  + B \  q_\sigma k_\tau}_{\rm{absent \ in \ the \ bare\  theory }}] \psi _\gamma(\tvk /2 -\tvq)\psi _\nu(\tvk /2 +\tvq)   \ln b \quad  \mbox{,}
\label{zonzo}
\end{equation}}
\textcolor{black}{Note that, in terms of the corrections defined in Eq. \eqref{eq:corr_gs}, the terms $A$ and $B$ contribute to $\Phi_{1}\delta g_{s1}=A+\dots$ and $\Phi_{2}\delta g_{s2}=B+\dots$, where the ellipses stand for contributions coming from diagrams other than $D^{\rm{mc},1}$ and $D^{\rm{mc},2}$.}

The core idea of the renormalization group is that the perturbative contributions generated by  Feynman diagrams can be reabsorbed into a redefinition of the model's parameters. Comparing equations \eqref{fd:vertex_mcs_simply} and \eqref{zonzo} it is evident  that the first term,  proportional to $\II_{\alpha\beta\gamma\nu} \vk\cdot \vq$, has the same form as the bare vertex, hence it can be reabsorbed in the coupling, $g \to g(1+\delta g_s\ln b)$. However, we cannot reabsorb the second and third terms of equation \eqref{zonzo} as correction of any pre-existing parameters, because of the different tensorial structure.
For this reason two novel terms, proportional to $ \II_{\alpha\beta\rho\sigma}  \II_{\rho\tau\gamma\nu}\ k_\sigma q_\tau  $ and $ \II_{\alpha\beta\rho\sigma} \II_{\rho\tau\gamma\nu}\ q_\sigma k_\tau $ respectively, must be included in the action.  
These two terms coincide with the two \textit{anomalous} terms $\Phi_1$ and $\Phi_2$ in equations \eqref{eq:SI}.
We remind that most of the diagrams listed in Fig.(\ref{fig:Vsvv}) of Section \ref{sec:diagrams} {\it generate} the same anomalous terms with different coefficients $A$ and $B$.
It is crucial to note that, if the model is at equilibrium, $v_0=0$, $\tilde \Gamma = \Gamma$ and $\tilde \lambda ^{\perp, \parallel} = \lambda^{\perp,\parallel}$, all these perturbative contributions  vanish: $A = B = \delta g_s  = 0$, so that no anomalous terms are generated in the non-active case.

\vskip 0.5 truecm
{\it *** Anomalous Advection terms}
\vskip 0.2 truecm

The same procedure can be used to prove that the two anomalous advection terms, proportional to $\mu_1$ and $\mu_2$, are fundamental to guarantee the closure of the theory under the RG transformation. We use the same strategy: we assume that the anomalous terms are zero in the bare theory, $\mu_1 = \mu_2 = 0$, then they are spontaneously generated by the renormalization group transformation. If we assume that $ \mu_1 = \mu_2 = 0$ , there is only one term proportional to $\hat s s \psi$ in the non Gaussian action \eqref{eq:SI}, corresponding to the advective derivative. In the diagrammatic expansion, this corresponds to modifying the vertex,
\begin{equation}
\begin{tikzpicture}[baseline=(a.base)]
\begin{feynman}[small]
\vertex (a) at (0,0) {\(\hs_{\alpha\beta}(-\tvk)\)};
\vertex (b) [dot,green] at (1.7,0) {};
\vertex (c) at (2.45,1.3) {\(s_{\gamma\nu}(\tvk/2-\tvq)\)};
\vertex (d) at (2.45,-1.3) {\(\psi_\eta(\tvk/2+\tvq)\)};
\diagram* {
(a) -- [charged boson] (b) -- [boson] (c),
(b) -- (d),
};
\end{feynman}
\end{tikzpicture}
\quad {\color{black} :}\quad - v_0 \frac{\iu\gamma_s}{2} \II_{\alpha\beta\gamma\nu} k_\eta
\label{green2}
\end{equation}
The term proportional to $\hat s s \psi$ in the action is corrected by the Feynman diagrams with an incoming $\hat s $, one outcoming $s$ and $\psi$ line. 
These are only $3$ diagrams, listed in Fig.(\ref{fig:Vssv}) of Section \ref{sec:diagrams}, so we calculate here all the three of them, namely,
\begin{equation}
\begin{split}
&D^{\rm{adv,1}}\quad {\color{black} :}\quad - \tvk,\alpha\beta
\begin{tikzpicture}[baseline=-\the\dimexpr\fontdimen22\textfont2\relax]
\begin{feynman}
\vertex (aa) at (-0.2,0); 
\vertex (a) at (0,0) ;
\vertex (mom1)  at ( 1,1) {$\frac{\tvk}{2} +\frac{ \tvq}{2} + \tvp$};
\vertex (mom2)  at ( 1,-1) {$\frac{\tvk}{2} -\frac{ \tvq}{2} - \tvp$};
\vertex (mom3)  at ( 3.3,0) {$  \tvp-\frac{ \tvq}{2} $};
\vertex (b) at (3,1.73) {$\qquad \tvk/2-\tvq,\gamma\nu$};
\vertex (c) at(3,-1.73) {$\qquad \tvk/2+\tvq,\eta $};
\vertex (v1)[green, dot] at (1,0){};
\vertex (v2)[black, dot] at (2.5,0.866) {};
\vertex (v3) [black, dot] at (2.5,-0.866) {};
\diagram*{
(a) --[charged boson](v1)--[fermion](v2)--[fermion](v3)--[boson](v1);
(v2)--[boson](b);
(v3)--[black](c)
};
\end{feynman}
\end{tikzpicture}
\quad\ , \quad
D^{\rm{adv},2}\quad {\color{black} :}\quad -\tvk,\alpha\beta
\begin{tikzpicture}[baseline=-\the\dimexpr\fontdimen22\textfont2\relax]
\begin{feynman}
\vertex (aa) at (-0.2,0); 
\vertex (a) at (0,0) ;
\vertex (mom1)  at ( 1,1) {$\frac{\tvk}{2} +\frac{ \tvq}{2} + \tvp$};
\vertex (mom2)  at ( 1,-1) {$\frac{\tvk}{2} -\frac{ \tvq}{2} - \tvp$};
\vertex (mom3)  at ( 3.3,0) {$  \tvp-\frac{ \tvq}{2} $};
\vertex (b) at (3,1.73) {$\qquad \tvk/2-\tvq,\gamma\nu$};
\vertex (c) at(3,-1.73) {$\qquad \tvk/2+\tvq,\eta $};
\vertex (v1)[green, dot] at (1,0){};
\vertex (v2)[black, dot] at (2.5,0.866){};
\vertex (v3) [black, empty dot] at (2.5,-0.866){};
\diagram*{
(a) --[charged boson](v1)--[fermion](v2)--[black](v3)--[anti charged boson](v1);
(v2)--[boson](b);
(v3)--[black](c)
};
\end{feynman}
\end{tikzpicture} \quad\ , \quad
\\
& \hspace{4.5 truecm}
D^{\rm{adv},3}\quad {\color{black} :}\quad -\tvk,\alpha\beta
\begin{tikzpicture}[baseline=-\the\dimexpr\fontdimen22\textfont2\relax]
\begin{feynman}
\vertex (aa) at (-0.2,0); 
\vertex (a) at (0,0) ;
\vertex (mom1)  at ( 1,1) {$\frac{\tvk}{2} +\frac{ \tvq}{2} + \tvp$};
\vertex (mom2)  at ( 1,-1) {$\frac{\tvk}{2} -\frac{ \tvq}{2} - \tvp$};
\vertex (mom3)  at ( 3.3,0) {$  \tvp-\frac{ \tvq}{2} $};
\vertex (b) at (3,1.73) {$\qquad \tvk/2-\tvq,\gamma\nu$};
\vertex (c) at(3,-1.73) {$\qquad \tvk/2+\tvq,\eta $};
\vertex (v1)[green, dot] at (1,0){};
\vertex (v2)[black, dot] at (2.5,0.866){};
\vertex (v3) [black, empty dot] at (2.5,-0.866){};
\diagram*{
(a) --[charged boson](v1)--[black](v2)--[anti fermion](v3)--[anti charged boson](v1);
(v2)--[boson](b);
(v3)--[black](c)
};
\end{feynman}
\end{tikzpicture}\quad\ .
\label{eq:andiag-adv}
\end{split}
\end{equation}
The integral expressions of these Feynman diagrams are:
\begin{equation}
\begin{split}
D^{\rm{adv},1}
=& -\iu v_0 \frac{\gamma_s g^2}{2}
\int\limits_{\Lambda/b<|\vp|<\Lambda}\frac{\di^dp}{(2\pi)^d}
\II_{\alpha\beta\xi\zeta} k_\rho P_{\mu\iota}^\perp\left(\vp+\frac{\vk}{2}+\frac{\vq}{2}\right)\II_{\iota\sigma\gamma\nu}P_{\tau\delta}^\perp\left(\vp-\frac{\vq}{2}\right)\II_{\delta\eta\phi\chi}
\times \qquad \qquad \\
&\times
\int\limits_{-\infty}^{\infty}\frac{\di \omega_p}{2\pi}
\mathbb{G}_{\rho\mu}^{0,\psi}\left(\tvp+\frac{\tvk}{2}+\frac{\tvq}{2}\right)
\mathbb{G}_{\sigma\tau}^{0,\psi}\left(\tvp-\frac{\tvq}{2}\right)
\mathbb{C}_{\xi\zeta\phi\chi}^{0,s}\left(\tvp-\frac{\tvk}{2}+\frac{\tvq}{2}\right)\ ,
\end{split}
\end{equation}
\begin{equation}
\begin{split}
D^{\rm{adv},2}
=& -\iu v_0 \frac{\gamma_s g^2}{2}
\int\limits_{\Lambda/b<|\vp|<\Lambda}\frac{\di^dp}{(2\pi)^d}
\II_{\alpha\beta\xi\zeta} k_\rho P_{\mu\iota}^\perp\left(\vp+\frac{\vk}{2}+\frac{\vq}{2}\right)\II_{\iota\sigma\gamma\nu}\left[\left(\vq+\frac{\vk}{2}\right)^2-\left(\vp-\frac{\vq}{2}\right)^2\right]
\II_{\phi\chi\tau\eta}
\times\\
&\times
\int\limits_{-\infty}^{\infty}\frac{\di \omega_p}{2\pi}
\mathbb{G}_{\rho\mu}^{0,\psi}\left(\tvp+\frac{\tvk}{2}+\frac{\tvq}{2}\right)
\mathbb{C}_{\sigma\tau}^{0,\psi}\left(\tvp-\frac{\tvq}{2}\right)
\mathbb{G}_{\xi\zeta\phi\chi}^{0,s}\left(-\tvp+\frac{\tvk}{2}-\frac{\tvq}{2}\right)\ ,
\end{split}
\end{equation}
\begin{equation}
\begin{split}
D^{\rm{adv},3}
=& -\iu v_0 \frac{\gamma_s g^2}{2}
\int\limits_{\Lambda/b<|\vp|<\Lambda}\frac{\di^dp}{(2\pi)^d}
\II_{\alpha\beta\xi\zeta} k_\rho P_{\sigma\iota}^\perp\left(-\vp+\frac{\vq}{2}\right)\II_{\iota\mu\gamma\nu}\left[\left(\vq+\frac{\vk}{2}\right)^2-\left(\vp-\frac{\vq}{2}\right)^2\right]
\II_{\phi\chi\tau\eta}
\times\\
&\times
\int\limits_{-\infty}^{\infty}\frac{\di \omega_p}{2\pi}
\mathbb{C}_{\rho\mu}^{0,\psi}\left(\tvp+\frac{\tvk}{2}+\frac{\tvq}{2}\right)
\mathbb{G}_{\sigma\tau}^{0,\psi}\left(-\tvp+\frac{\tvq}{2}\right)
\mathbb{G}_{\xi\zeta\phi\chi}^{0,s}\left(-\tvp+\frac{\tvk}{2}-\frac{\tvq}{2}\right)\ ,
\end{split}
\end{equation}
their sum is,
\textcolor{black}{
\begin{equation}
D^{\rm{adv},1}+D^{\rm{adv},2}+D^{\rm{adv},3} = - v_{0} \frac{\iu\gamma_{s}}{2}
\II_{\alpha\beta\rho\sigma} \II_{\rho\tau\gamma\nu} \left[
    \delta \gamma_s \,k_\eta \delta_{\sigma\tau}+
    C \,  k_\tau \delta_{\sigma\eta} +
    D \, k_\sigma \delta_{\tau\eta} \right]\ln b
\label{sppAnomalousS}
\end{equation}}
with the constants $\delta \gamma_s$, $C$ and $D$, given by,
\begin{equation}
\begin{split}
\delta \gamma_s =&  -  g^2\frac{-x(x-1)(1+2w+x)\tGamma + w x (w+x)^2{\tlambda^\parallel}-w(1+w)^2{\tlambda^\perp}}{12\Gamma^3(1+w)^2 x (w+x)^2}\\
C =  & - g^2 \frac{-x(5+10w+12w^2+14wx+7x^2)\tGamma+7wx(w+x)^2{\tlambda^\parallel}+5w(1+w)^2{\tlambda^\perp}}{12\Gamma^3(1+w)^2 x (w+x)^2}\\
D = &  - g^2\frac{-x(x-1)(1+2w+x)\tGamma + w x (w+x)^2{\tlambda^\parallel}-w(1+w)^2{\tlambda^\perp}}{12\Gamma^3(1+w)^2 x (w+x)^2}\ ,
\end{split}
\end{equation}
These three Feynman diagrams correct the interacting part of the action relative to the advection of the spin as follows,
\textcolor{black}{
\begin{equation}
\Delta \mathcal S_{\rm{adv}} = \int_{0}^{\Lambda/b} d \tvk  \ d \tvq  \ v_0 \frac{-\iu\gamma_s}{2} \ \II_{\alpha\beta\rho\sigma} \II_{\rho\tau\gamma\nu} \ \hat s_{\alpha \beta}(-\tvk)   [\underbrace{\delta \gamma_s  k_\eta \delta_{\sigma\tau}
}_{\rm{bare\ structure}}
+  \underbrace{C  \ k_\tau\ \delta_{\sigma\eta} + D \ k_\sigma \delta_{\tau\eta}}
_{\rm{absent \ in\  the\ bare\ theory }}  ] s_{\gamma \nu}(\tvk/2-\tvq) \psi_\eta(\tvk/2+\tvq)  \quad  \mbox{,}
\label{segasega}
\end{equation}}
Comparing  equations \eqref{segasega} and \eqref{green2} it is evident that the term proportional to $ \II_{\alpha\beta\rho\sigma} \II_{\rho\tau\gamma\nu} k_\eta \delta_{\sigma\tau}= \II_{ \alpha\beta\gamma\nu} k_\eta$ is the same as in the bare case and it can be reabsorbed as a correction of the advection coupling, $\gamma_s$. On the other hand, the two terms proportional to $\II_{\alpha\beta\rho\sigma} \II_{\rho\tau\gamma\nu} \ k_\tau\ \delta_{\sigma\eta}$ and $ \II_{\alpha\beta\rho\sigma} \II_{\rho\tau\gamma\nu} \ k_\sigma \delta_{\tau\eta}$ have tensorial structure different from the bare action, so that they require the addition of two new anomalous advection terms, which are exactly the ones proportional to $\mu_1$ and $\mu_2$ in equation \eqref{eq:SI}. \textcolor{black}{In terms of the corrections of Eq. \eqref{eq:corr_gammas}, the terms $C$ and $D$ thus contribute to $2\mu_{1}\delta\gamma_{s1}=C$ and $2\mu_{2}\delta\gamma_{s2}=D$.}


\vskip 0.5 truecm
{\it *** Summary of the anomalous terms}
\vskip 0.2 truecm

We have shown explicitly that some vertex correction diagrams generate the following mode-coupling and advective anomalous terms in the equation of motion for the spin,  
\begin{align}
\rm{MC} \ \   &
\qquad  \quad
\begin{cases}
&\psi _\gamma(\tvk /2 -\tvq)\psi _\nu(\tvk /2 +\tvq)
\II_{\alpha\beta\rho\sigma}  k_\sigma q_\tau \II_{\rho\tau\gamma\nu} \\
&   \psi _\gamma(\tvk /2 -\tvq)\psi _\nu(\tvk /2 +\tvq)
\II_{\alpha\beta\rho\sigma}  q_\sigma k_\tau \II_{\rho\tau\gamma\nu}
\end{cases}
\quad \longrightarrow \quad
\begin{cases}
&\left[\partial_\alpha \left(\psi_\nu \partial_\nu \psi_\beta\right) - \partial_\beta \left(\psi_\nu \partial_\nu \psi_\alpha\right) \right]   \\
& \partial_\nu \left[\psi_\nu \left(\partial_\alpha \psi_\beta-\partial_\beta \psi_\alpha\right)\right]  
\end{cases}
 \ ,\\
\rm{ADV} & \qquad \quad
\begin{cases}
&     s_{\gamma \nu}(\tvk /2 -\tvq) \psi_\eta(\tvk /2 +\tvq)
\II_{\alpha\beta\rho\sigma} \II_{\rho\tau\gamma\nu} \ k_\tau \delta_{\sigma\eta}    \hspace{0.2 truecm}
\longrightarrow \hspace{0.9 truecm}
\partial_\nu \left(s_{\alpha\nu}\psi_\beta - s_{\beta\nu}\psi_\alpha\right)
\\
& s_{\gamma \nu}( \tvk /2 -\tvq) \psi_\eta(\tvk /2 +\tvq)
\II_{\alpha\beta\rho\sigma} \II_{\rho\tau\gamma\nu} \ k_\sigma \delta_{\tau\eta}   \hspace{0.2 truecm}
\longrightarrow \hspace{0.9 truecm}
\left[\partial_\alpha \left(\psi_\nu s_{\nu\beta}\right) - \partial_\beta \left(\psi_\nu s_{\nu\alpha}\right)\right]
\ .
\end{cases}
\end{align}
which we have summarized here together with the real-space form that we anticipated in \eqref{sandapa} and \eqref{sandapo}. Notice a complication: while each one of the two advection terms in $k$ space gives rise to one term in $x$ space, both mode-coupling anomalous terms in $k$ space give contribute to both terms in $x$ space. For this same reason the couplings of the MC anomalous terms in momentum space, $\Phi_1, \Phi_2$ are a linear combination of those in real space, $\phi_1, \phi_2$ (see equations \eqref{sumo} and \eqref{suma}).

The crucial result of the diagrammatic calculation is that {\it all} diagrams generating anomalous terms (including all diagrams that we have not explicitly analyzed in this Section), generate {\it only} terms with one of these four tensorial structures. In this sense, the RG calculation is closed once we include these four vertices in the action. Notice also that the total perturbative amplitudes of these terms are not simply given by the coefficients $A,B,C, D$ calculated here, indeed because anomalous corrections with the same tensorial structure arise in many more diagrams than the five computed in this Section.


\subsubsection{Effective couplings}
As it is always the case in RG calculations, it is now possible to define a set of effective parameters whose scaling behaviour does not depend on $z$, $\chi_\psi$ and $\chi_s$, so that all physical quantities turn out to depend on the parameters of the theory only via these effective couplings. The effective couplings can be found by looking at the scaling dimensions of the parameters of the theory, given in Eqs. \eqref{eq:scaling_gamma}-\eqref{eq:scaling_eta}, \eqref{eq:scaling_gamma2}-\eqref{eq:scaling_ferro}.
By recalling that the scaling dimension of the product of two couplings is given by the sum of the respective scaling dimensions, namely,
\begin{equation}
    {\color{black}\mathcal{C}_b = \mathcal{A}_b \mathcal{B}_b = b^{\chi_\mathcal{A}+\chi_\mathcal{B}} \mathcal{A}_0 \mathcal{B}_0 = b^{\chi_\mathcal{C}}\mathcal{C}_0}\qquad \rightarrow \qquad \chi_\mathcal{C}=\chi_\mathcal{A}+\chi_\mathcal{B}
\end{equation}
it is always possible to find a set of combinations of coupling constants and parameters of the theory that have a scaling behaviour which is independent from $z$, $\chi_\psi$ and $\chi_s$ \cite{dedominicis1978field}. 
For example, let us consider the coupling constant $J$: its scaling, given by Eq. \eqref{eq:scaling_ferro}, depends on $\chi_\psi$ and $z$.
To compensate the dependence on $\chi_\psi$ we can multiply it by the noise strength $\tGamma$, which also has a $\chi_\psi$ scaling dependence; thus, the product $\tGamma J$ does not depend anymore on $\chi_\psi$.
Similarly, the dependence on $z$ can be compensated by dividing by $\Gamma^2$, making the scaling behaviour of the combination $\tGamma J/\Gamma^2$ depending only on $d$ and perturbative corrections, which are computed through Feynman diagrams.
A similar procedure can be applied at any other constant; multiplying it by appropriate powers of $\tlambda^{\perp/\parallel}$ or $\tGamma$ to compensate the dependence on $\chi_s$ or $\chi_\psi$ respectively, and by appropriate powers of $\Gamma$, $\lambda^{\perp/\parallel}$ to compensate the dependence on $z$.
There is some arbitrariness on the choice of whether to use $\tlambda^{\perp}$ or $\tlambda^{\parallel}$ to compensate $\chi_s$ and whether to use $\Gamma$, $\lambda^{\perp}$ or $\lambda^{\parallel}$ to compensate $z$. However, as long as these parameters are not singular (and they will not be at the fixed point), this choice is irrelevant.

The effective parameters for \eqref{eq:psi}-\eqref{eq:s} are,
\begin{equation}
w=\frac{\Gamma}{\lambda^\parallel}\ , \quad  
x=\frac{\lambda^\perp}{\lambda^\parallel} \ ,\quad 
\theta^{\perp}=\frac{\tGamma \lambda^{\perp}}{\Gamma\tlambda^{\perp}} \frac{g^{\left(s\right)}}{g^{\left(\psi\right)}}
\ , \quad  
\theta^{\parallel}=\frac{\tGamma \lambda^{\parallel}}{\Gamma\tlambda^{\parallel}} \frac{g^{\left(s\right)}}{g^{\left(\psi\right)}} \ ,
\label{sanremo}
\end{equation}
where $\theta^{\perp,\parallel} \neq 1$ if the system is out-of-equilibrium.
Although the technical definition of $\theta^{\perp/\parallel}$ contains the ratio $g^{\left(s\right)}/g^{\left(\psi\right)}$, in the physical case we expect $g^{\left(s\right)}=g^{\left(\psi\right)}$, as argued in Section \ref{sec:pertRG} .
Hence, the physical meaning of $\theta^{\perp/\parallel}$ is that of the ratio between the effective temperatures of the two fields,  $T_\psi=\tGamma/\Gamma$ and $T_s^{\perp/\parallel}=\tlambda^{\perp/\parallel}/\lambda^{\perp/\parallel}$, namely
\begin{equation}
\theta^{\perp}=\frac{T_\psi}{T_s^\perp}=\frac{\tGamma \lambda^{\perp}}{\Gamma\tlambda^{\perp}} \ , \quad  
\theta^{\parallel}=\frac{T_\psi}{T_s^\parallel}=\frac{\tGamma \lambda^{\parallel}}{\Gamma\tlambda^{\parallel}} \ .
\label{imperia}
\end{equation}
\textcolor{black}{From the dissipative coefficients $\eta$ and $\tilde\eta$, two effective parameters can be obtained
\begin{equation}
\hat{\eta}=
\frac{\eta}{\lambda^{\parallel}\Lambda^{2}+\eta} \ , \quad 
\tildehat{\eta}=
\frac{\tilde{\eta}}{\tlambda^{\parallel}\Lambda^{2}+\tilde{\eta}}
\end{equation}
Here the presence of $\Lambda$ is needed for dimensional reasons. In terms of these {\it reduced} frictions, the conservative dynamics is recovered when $\hat{\eta}=\tildehat{\eta}=0$, while the fully dissipative dynamics is recovered when $\hat{\eta}=\tildehat{\eta}=1$, namely when $\eta\gg\lambda^{\parallel/\perp}\Lambda^{2}$, $\tilde{\eta}\gg\tlambda^{\parallel/\perp}\Lambda^{2}$.
}

The effective coupling constants regulating activity are,
\begin{equation}
c_v=v_0 \frac{{\gamma_v}}{\Gamma}  \sqrt{\frac{\tGamma}{\Gamma}} \sqrt{K_d \Lambda^{d-4}} \quad , \quad  
c_s=v_0\frac{{\gamma_s}}{\lambda^\parallel}   \sqrt{\frac{\tGamma}{\Gamma}} \sqrt{K_d \Lambda^{d-4}} \ ,
 \label{eq:effp1}
\end{equation}
\textcolor{black}{the ferromagnetic coupling $J$ and the mass $m$ have effective couplings given by
\begin{equation}
\tilde u=\frac{\tGamma}{\Gamma}  \, \frac{J}{\Gamma}  K_d \Lambda^{d-4} =\frac{\tGamma}{\Gamma}  \, u\,  K_d \Lambda^{d-4} \quad , \quad 
\tilde r= \frac{\tGamma}{\Gamma} \frac{m}{\Gamma} K_d \Lambda^{d-2} =\frac{\tGamma}{\Gamma} r K_d \Lambda^{d-2}
\ ;
\label{eq:effp2}
\end{equation}
while the mode-coupling and the DYS effective coupling constants respectively are,
\begin{equation}
f=\frac{\tlambda^{\parallel}}{\lambda^{\parallel}}\frac{g^2}{{\lambda^{\parallel}} \Gamma} K_d \Lambda^{d-4}
 \quad , \quad 
\tilde{u}_\kappa = \frac{\tGamma}{\Gamma} \frac{\kappa}{g} K_d \Lambda^{d-4}
\ ;
\label{eq:effp4}
\end{equation}
}
where $\Lambda$ is the cutoff of the theory and $K_d$ is the surface of the unitary sphere in $d$ dimensions (note that in the main text we set $\Lambda=1$ and $K_d=1$ to simplify the notation).
The presence of $\Lambda$ to the power $4-d$ in all effective couplings except for $r$ suggests that the upper critical dimension of the theory is $d_c=4$, which means that all couplings are relevant in $d<4$, while mean-field behaviour is recovered for $d\geq 4$.
\textcolor{black}{The dependance of $r$ on $\Lambda^{d-2}$ indicates instead that the mass is always a relevant perturbation above $d=2$, driving the system away from the critical manifold. Therefore, this requires the mass to be fine-tuned to be near the critical point.}
At equilibrium, where $v_0=0$ and $\tGamma=\Gamma, \tlambda=\lambda, \kappa=u g$, all the effective couplings become identical to their standard equilibrium counterpart \cite{HH1977}, with the equilibrium result $\tilde u=\tilde{u}_\kappa$ \cite{cavagna2021dynamical} being recovered.

\textcolor{black}{Finally, we should add to the list of effective parameters the four adimensional parameters regulting the anomalous mode-coupling and self propulsion non-linearities, namely
\begin{equation}
\Phi_{1}  \quad , \quad \Phi_{2}  \quad , \quad \mu_{1}  \quad , \quad \mu_{2}\, .
\label{eq:effp5}
\end{equation}
}

\subsubsection{RG flow equations and $\beta$-functions}\label{RGflow}
The flow of the effective couplings can be obtained by iterating the RG transformation, thus defining a set of recursive relations. After $l$ iterations, the new parameters will take the form,
\textcolor{black}{
\begin{equation}
    \mathcal{P}_{l+1}=b^{\chi_{\mathcal{P}}}  \mathcal{P}_l
    \label{recup}
\end{equation}}
where $\chi_{\mathcal{P}}$ is evaluated using the values of the parameters at step $l$.
The values $\mathcal{P}^*$ to which the flow of $\mathcal{P}$ approaches when $l\to\infty$ are called fixed points, and play a crucial role in determining the critical behaviour of the theory \cite{goldenfeld_lectures_1992}.
\textcolor{black}{To study the RG flow of our theory, it is convenient to rewrite Eq. \eqref{recup} in the thin shell limit $\ln b\to0$. In this limit, the flow equations become
\begin{equation}
    \dot{\mathcal{P}}=\beta_\mathcal{P}
    \label{paolino}
\end{equation}
where $\beta_\mathcal{P}$ is the derivative of $\mathcal{P}$ with respect to $\ln b$, namely $\beta_\mathcal{P}=\partial\mathcal{P}/\partial( \ln b)$, known as $\beta$-function.
The fixed points $\mathcal{P}^{*}$ of the RG-flow are given by the zeros of $\beta_\mathcal{P}$, since these are the points at which $\dot{\mathcal{P}}= \left. \beta_\mathcal{P}\right|_{\mathcal{P}=\mathcal{P}^{*}}=0$.
Among these fixed points, some are (IR-)stable meaning that the flow is driven towards them, while other may have one or more directions of instability from which the RG flow escapes. Asymptotically IR-stable fixed points are those typically ruling the critical behaviour of systems in the thermodynamic limit.
Note that, as previously discussed, the mass coupling $r$ always represents a source of instability since it drives the system away from criticality. Hence, stability is intended as stability in all directions except $r$.}
 
The $\beta$-function of the generic parameter $\mathcal{P}$ is given by \textcolor{black}{$\beta_\mathcal{P}= \mathcal{P} \chi_\mathcal{P}$}. Hence, for the parameters defined in Eqs. \eqref{sanremo}-\eqref{eq:effp4} the beta-functions take the following form,
\begin{align}
    \beta_w =&\, w \left(\delta\Gamma - \delta\lambda^\parallel\right)\\
    \beta_x =&\, x \left(\delta\lambda^\perp - \delta\lambda^\parallel\right)\\
    \beta_{\theta^\perp} = &\, \theta^\perp \left(\delta\tGamma-\delta\Gamma + \delta\lambda^\perp-\delta \tlambda^\perp +\delta g_s-\delta g_\psi \right)\\
    \beta_{\theta^\parallel} = &\, \theta^\perp \left(\delta\tGamma-\delta\Gamma + \delta\lambda^\parallel-\delta\tlambda^\parallel +\delta g_s-\delta g_\psi\right)\\
    \color{black}{\beta_{\hat{\eta}}} = & \, \color{black}{\hat{\eta}\left(1-\hat{\eta}\right)\left(2-\delta\lambda^\parallel\right)}
    \label{betaeta}\\
    \color{black}{\beta_{\tildehat{\eta}}} = & \, \color{black}{\tildehat{\eta}\left(1-\tildehat{\eta}\right)\left(2-\delta\tlambda^\parallel\right)}
    \label{betatildeeta}
\end{align}
\begin{align}
\color{black}{\beta_{\tilde r}=}&\, \color{black}{\tilde r \left(2+\delta m-\delta \Gamma\right)}\\
    \beta_{c_v}= &\frac{1}{2}c_v \left(\epsilon + \delta\tGamma + 2 \delta\gamma_\psi - 3 \delta\Gamma -{\color{black}\delta\Omega}\right)\\
    \beta_{c_s}= &\frac{1}{2}c_s \left(\epsilon + \delta\tGamma + 2 \delta\gamma_s - \delta\Gamma - 2 \delta\lambda^\parallel -{\color{black}\delta\Omega}\right)\\
    \beta_f =&\, f \left(\epsilon + \delta \tlambda^\parallel + 2 \delta g_\psi - 2 \delta \tlambda^\parallel - \delta \Gamma -{\color{black}\delta\Omega}\right)\\
    \beta_{\tilde{u}} =&\, \tilde{u} \left( \epsilon + \delta \tGamma + \delta J - 2 \delta \Gamma -\delta \omega_{\psi} \right)\\
    \beta_{\tilde{u}_\kappa} =&\, \tilde{u}_\kappa \left( \epsilon + \delta\tGamma + \delta\kappa - \delta\Gamma - \delta g_s -{\color{black}\delta\Omega} \right)
\end{align}
\begin{align}
    \beta_{\mu_1}=&\,\mu_1 \Big(\delta\gamma_{s1}-\delta\gamma_s\Big)\\
    \beta_{\mu_2}=&\,\mu_2 \Big(\delta\gamma_{s2}-\delta\gamma_s\Big)\\
    \beta_{\Phi_1}=&\,\Phi_1 \Big(\delta g_{s1}-\delta g_{s}\Big)\\
    \beta_{\Phi_2}=&\,\Phi_2 \Big(\delta g_{s2}-\delta g_{s}\Big)
\end{align}
where all the perturbative corrections $\delta\mathcal{P}$ are obtained from the Feynman diagram expansion.
The explicit expressions of these beta-functions are given in the Mathematica attached notebook in the ancillary filed ({\it beta\_functions.nb}).

\textcolor{black}{Before we proceed, let us draw the reader's attention to the beta functions of the reduced dissipative coefficients, $\hat\eta$ and $\tildehat\eta$, Eq. \eqref{betaeta} and \eqref{betatildeeta}. Within our perturbative approach, near $4$ dimensions the parameters $\hat{\eta}$ and $\tildehat{\eta}$ may take only two possible fixed point values: $0$ or $1$. This is because the perturbative corrections $\delta\lambda^\parallel$ and $\delta\tlambda^\parallel$ are both of $\mathcal{O}\left(\epsilon\right)$; hence the quantities $2-\delta\lambda^\parallel$ and $2-\delta\tlambda^\parallel$ will always be strictly positive and close to $2$; hence, the only way in which $\beta_{\hat{\eta}}=0$ is by setting $\hat{\eta}=0$ or $\hat{\eta}=1$. Similarly, $\beta_{\tildehat{\eta}}=0$ is achieved only when $\tildehat{\eta}=0$ or $\tildehat{\eta}=1$.
Near the conservative manifold $\hat\eta=\tildehat\eta=0$, the scaling dimensions of both $\hat\eta$ and $\tildehat\eta$ are positive, meaning that the flow drives the system away from it. On the other hand, the fully dissipative manifold $\hat\eta=\tildehat\eta=1$ is attractive. Hence, both $\hat{\eta}$ and $\tildehat{\eta}$ flow from the conservative value of $0$ towards the dissipative value of $1$, regardless of what all the other parameters do.
In terms of the dissipations coefficients $\eta$ and $\tilde\eta$, the stability of the $\hat\eta=\tildehat\eta=1$ manifold reflects into the fact that both parameters grow under the RG flow, eventually diverging: $\eta\to\infty$, $\tilde\eta\to\infty$. This means that the asymptotically stable behaviour of the theory is given by the overdamped limit.}

\textcolor{black}{The power of the RG lies in the fact that critical exponents can be inferred by the study of the RG flow in the neighbourhood of a fixed point. The fixed point at which the critical exponents should be evaluated usually is the stable one, since one expects the RG flow to eventually reach its stable fixed point.
However, unstable fixed points often play a crucial role in determining the behaviour of {\it finite-size} systems, since crossover phenomena may take place. We will show in the following section that this is indeed the case for the field theory we have introduced. Although the IR-stable fixed point in the presence of spin dissipation is the one of the incompressible Toner and Tu theory \cite{chen2015critical}, an RG crossover takes place for small frictions $\hat{\eta}\ll1$, as in \cite{cavagna2019short, cavagna2019long}.
The (IR-unstable) fixed point ruling the behaviour of systems with small frictions can be found by focusing on the RG flow of the conservative theory.
It is worth noting that the conservative subspace is RG-invariant, since $\beta_{\hat{\eta}}=\beta_{\tildehat{\eta}}=0$ when $\hat{\eta}=\tildehat{\eta}=0$. Hence, we shall at first restrict ourself to the sub-space $\hat{\eta}=\tildehat{\eta}=0$, and come back to the dissipative theory later on in Section \ref{sec:diss}.}

\subsection{Properties of the RG solution in absence of dissipation}
\subsubsection{Flow and fixed point}\label{mcenroe}
In the simple cases, fixed points can be found by solving analytically the set of equations $\beta_\mathcal{P}=0$ in the parameters $\mathcal{P}$.
Here this was not possible. Instead, we tackle this problem by integrating numerically the set of partial differential equations defining the RG flow, namely \eqref{paolino}, and looking at what values of the parameters the flow converges.
If the flow does converge, the point to which it will converge is a fixed point of the theory.
This procedure does not allow us to find unstable fixed points, but only the IR-stable ones \textcolor{black}{in the subspace $\hat{\eta}=\tildehat{\eta}=0$.}
To perform this numerical integration, we have to take great care in the choice of initial conditions of the parameters and couplings, as we expect a large portion of this $14$-dimensional space to be just unphysical.
The set of parameters $v_0=0$ (and thus $c_v=c_s=0$), $\theta^{\perp/\parallel}=1$, $\Phi_1=\Phi_2=0$ and $\tilde{u}=\tilde{u}_\kappa\neq0$ identifies Solenoidal Model G \cite{cavagna2021dynamical}, which represents the equilibrium limit of our theory -- Eq. \eqref{eq:psi} \eqref{eq:s}. We expect a system with {\it small}, but non-vanishing activity $v_0$ to belong to the neighbourhood of the equilibrium limit, and thus to be physical in the RG space. If activity is relevant, and indeed we will see it is, the corresponding coupling constants $c_v$ and $c_s$ should grow along the RG flow and drive the system towards an out-of-equilibrium active fixed point, if it exists. We remark that we do not start close to equilibrium because we expect activity to be weak in swarms; in fact, quite the opposite: as we show in the main text and in Section \ref{react}, activity in natural swarms is strong. We start close to the equilibrium fixed point for a mere technical point: we need a safe path through a very dangerous parameter space in which it is far too easy for the RG flow to go bonkers if one deviates too much from a physical initial condition.

\begin{figure}[b]
	\centering
	\includegraphics[width=1\textwidth]{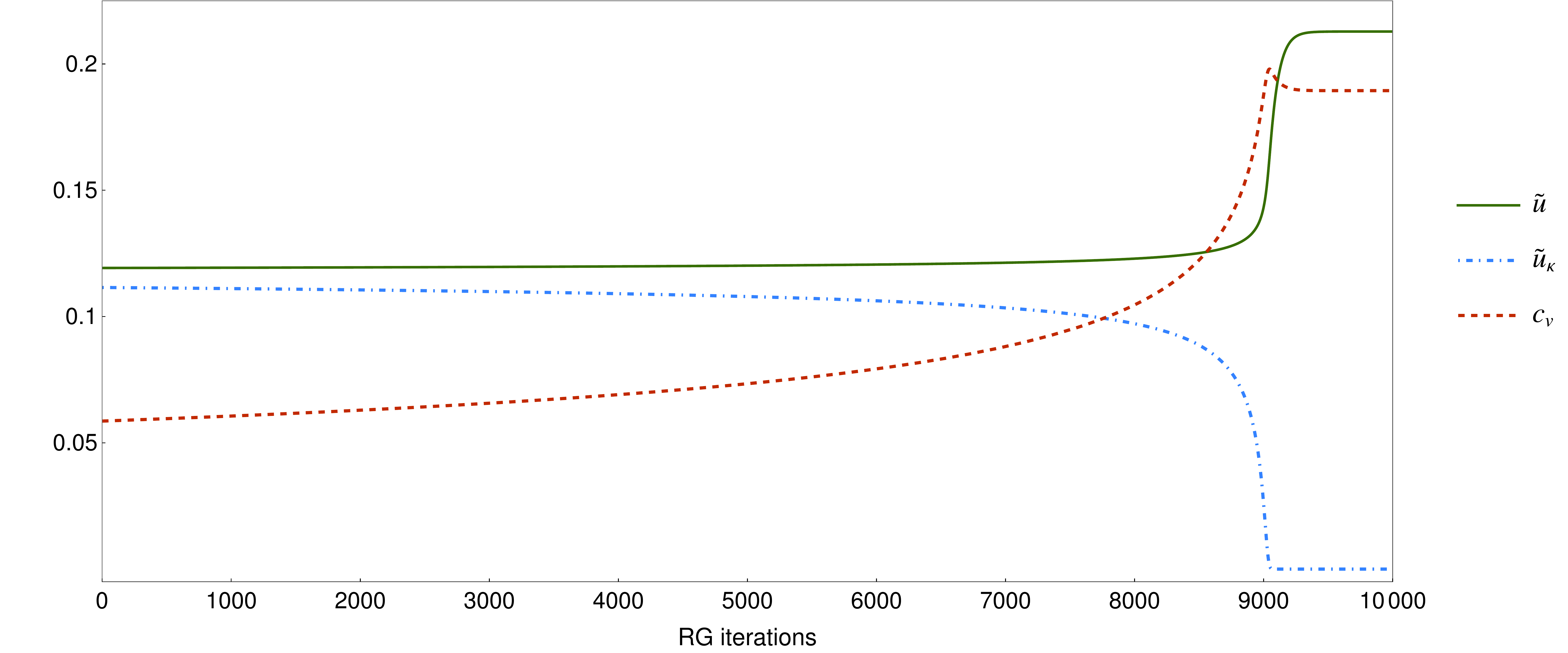}
	\caption{{\bf RG flow of $\tilde{u}$, $\tilde{u}_\kappa$ and $c_v$.} As an example, we show a portion of the RG flow of the couplings $\tilde{u}$, $\tilde{u}_\kappa$ and $c_v$ from the unstable equilibrium fixed point towards the stable off-equilibrium fixed point.
	From the plot we can clearly see that as activity increases, $\tilde{u}$ and $\tilde{u}_\kappa$ become different one from the other. As the fixed point is approached, the value of $\tilde{u}_\kappa$ drops to $0$.
	}
	\label{fig:flowSI}
\end{figure}

The constants $\mu_1$ and $\mu_2$ have undefined values in the equilibrium limit, since they are both multiplied by $v_0 \gamma_s$ in the equations of motion and thus we have no {\it a priori} argument to fix their value near equilibrium.
However, we find that the $\beta$-functions of these two parameters, namely $\beta_{\mu_1}$ and $\beta_{\mu_2}$, vanish when $\mu_1$ and $\mu_2$ take the following values
\begin{align}
	\mu_1=\frac{1}{2} \quad \ ,&\quad \mu_2=-\frac{1}{2}\\
	\mu_1=-1 \quad \ ,&\quad \mu_2=0
\end{align}
independently of the values of all other parameters.
This means that both these combinations of $\mu_{1,2}$ remain constant along with the RG flow, whatever the other couplings and parameters are doing.
The first solution of $\mu_{1,2}$ turns out to be unstable for small perturbations of their values, at least near the equilibrium fixed point, while the second solution is stable.
Hence, we fix from the beginning $\mu_1=-1$ and $\mu_2=0$ and check the consistency of this choice by checking {\it a posteriori} the stability of the RG flow with respect to perturbations in $\mu_1$ and $\mu_2$, thus reducing our problem to $11$ coupled equations instead of $13$.

We simulate the RG flow starting from various initial conditions close to equilibrium by using the built-in {\it NDSolve} function of the software {\it Mathematica}, and we always find the same attractive fixed point, 
\begin{align}
    \tilde{u}^* =& 0.213\epsilon & \tilde{u}_\kappa^* =& 0 & c_v^* =& 0.189\sqrt{\epsilon}\\
    f^* =& 1.68 \epsilon & \Phi_1 =& -0.762 & \Phi_2^* =& -0.137\\
    c_s^* =& 0.882 \sqrt{\epsilon} & \mu_1 =& -1 & \mu_2 =& 0\\
    x^* =& 0.369 & \theta_{\perp}^*=&1.34 & \hat{\eta}^{*}=&0\\
    w^*=&3.95 & \theta_{\parallel}^*= & 0.920 & \tildehat{\eta}^{*}=&0
    \label{unlupomannaroamericanoalondra}
\end{align}
The typical RG flow of a selected number of couplings is represented in Fig. \ref{fig:flowSI}.
Since $c_v$, $c_s$ and $f$ have a finite fixed point value, both activity and mode-coupling are relevant at this fixed point, making it the ideal candidate to describe incompressible inertial active matter. The stability of the novel fixed point is analyzed by looking at the Jacobian matrix of the beta functions
\begin{equation}
	\mathbb{J}_\beta=\pder{\vec{\beta}}{\vec{\mathcal{P}}}
\end{equation}
The eigenvalues of $\mathbb{J}_\beta$ at the new fixed point are given by
\begin{align}
-&2.26\epsilon\, , & -&1.66 \epsilon\, , & -&1.00\epsilon\, , &  (-&0.724+0.107 \, \iu)  \epsilon\, , & (-&0.724-0.107\,  \iu) \epsilon \\
-&0.562  \epsilon\, , & -&0.498  \epsilon\, , & -&0.315  \epsilon\, , & (-&0.247+0.00185\, \iu) \epsilon\, , & (-&0.247-0.00185\, \iu) \epsilon\\
-&0.0584  \epsilon\, , & -&0.0532 \epsilon\, , & -&0.0167\epsilon & 2& - 0.816 \epsilon & 2& - 0.651 \epsilon
\end{align}
A negative (positive) real part of an eigenvalue indicates that the flow near the fixed point converges (diverges) exponentially along the direction of the associated eigenvector. On the other hand, the presence of an imaginary part of the eigenvalues stems from a spiralling convergence or divergence of the flow in the direction of the eigenvalue.

\textcolor{black}{The last two eigenvalues of $\mathbb{J}_\beta$, namely $2 - 0.816 \epsilon$ and $2 - 0.651 \epsilon$, are positive for all values of $\epsilon$ that may be of physical interest ($0\leq\epsilon\leq2$). They indicate the presence of two directions of instability of the fixed point. From an analysis of the eigenvectors, it is possible to see that these instabilities point in the direction of $\hat\eta$ and $\tildehat{\eta}$ respectively, indicating that this fixed point is not stable with respect to friction. 
However, since the real part of all other eigenvalues is negative when $\epsilon>0$, we can conclude that this new fixed point is stable in the subspace $\hat{\eta}=\tildehat{\eta}=0$ for $d<4$ and thus we expect it to rule the long-wavelength behaviour of active conservative systems for $d=3$.}

\subsubsection{The dynamical critical exponent}

The spatio-temporal behaviour of a collective system is described by the two-point connected correlation function of the order parameter, $C(\vx,t)$. In general, the correlation function encodes a very complicated relation between space and time and it depends on the set of parameters $\mathcal{P}$ defining the state of the system. In the case of critical systems, the parameters enter the correlation function only through the correlation length $\xi\left(\mathcal{P}\right)$. This property is known as {\it dynamic scaling} \cite{Ferrell1967,HH1967scaling}, and it states that the correlation function $C$, when expressed as a function of wave-vector and frequency, obeys the following scaling form,
\begin{equation}
C \left  (\vk,\omega ; \mathcal{P}\right)=C_0\left(\vk;\xi\right)\,F\left(\frac{\omega}{\omega_k\left(\mathcal{P}\right)},k\xi\right)
\end{equation}
where $\xi=\xi \left(\mathcal{P}\right)$ is the correlation length and where the static correlation function $C_0$ has in turns the scaling form,
\begin{equation}
C_0(k,\xi)=k^{2-\eta}F_0\left(k\xi\right)
\label{etascaling}
\end{equation}
while the characteristic frequency at scale $k$ is given by,
\begin{equation}
\omega_k \left(\mathcal{P}\right)=k^z\Omega\left(k\xi\left(\mathcal{P}\right)\right)
\end{equation}
In the relations above, $\Omega$, $F_0$ and $F$ are well-behaved scaling functions, whose explicit form is not relevant for our purposes; $\eta$ is the critical exponent for the static correlation function (normally called anomalous dimension of the order parameter \cite{binney_book}), which must {\it not} be confused with the dissipation.
What dynamic scaling asserts is that the divergence of space correlations and time correlations are not independent near the critical point, but they are connected by the dynamic critical exponent $z$.
To find the critical exponent $z$, following a standard procedure \cite{cardy1996scaling,HH1977} we require that the kinetic coefficient of the velocity field is not singular at the RG stable fixed point, namely, 
\begin{equation}
	\Gamma^*=O\left(1\right) \quad \rightarrow \quad \chi_\Gamma=0
	\label{casablanca}
\end{equation}
By plugging $\chi_\Gamma=0$ into Eq. \eqref{eq:scaling_gamma}, we find, 
\begin{equation}
	z=2-\delta\Gamma+{\color{black}\delta\Omega}
	\label{eqz}
\end{equation}
Once we substitute the diagrammatic results for $\delta\Gamma$ and ${\color{black}\delta\Omega}$ \textcolor{black}{at $\hat{\eta}=\tildehat{\eta}=0$} in the equation for $z$, we obtain the following expression for the dynamic critical exponent,
\begin{equation}
\begin{split}
	z=2 & - \frac{3 f (3 w+2 x+1)}{4 (w+1) (w+x)} - \frac{c_v^2}{4} 
	+ \frac{f (\theta^\perp-1) \left(13 w^2+12 w x+5 x^2\right)}{12 \theta^\perp (w+x)^3} - \\
	&-\frac{f (\theta^\parallel-1) \left(13 w^3+w^2 (4 x+75)+w (48 x+51)+24 x+9\right)}{12 (w+1)^3 (w+x)}-\\
	&- \frac{f (\theta^\parallel-1) (\theta^\perp-1) \left(13 w^2+12 w x+5 x^2\right)}{12 \theta^\perp (w+x)^3} -
	\frac{f \theta^\parallel \Phi_1 x (3 w+2 x)}{4 (w+x)^3} -\\
	&-\frac{f \theta^\parallel \Phi_2 \left(9 w^3 x+2 w^2 \left(4 x^2+9 x-1\right)+w x \left(2 x^2+16 x+3\right)+2 x^2 (2 x+1)\right)}{12 (w+1)^2 (w+x)^3}
\end{split}
\label{bruno}
\end{equation}
The value of $z$ is then simply obtained by plugging into \eqref{bruno} the fixed point values of the parameters, which gives,
\begin{equation}
\color{black} z= 2-0.65(2)\epsilon
\label{generica}
\end{equation}
For $d=3$ ($\epsilon=1$), we finally obtain the RG prediction for the dynamic critical exponents of the inertial active theory, 
\begin{equation}
\color{black} z= 1.34(8) \ .
\label{cremedelacreme}
\end{equation}
The correction $\color{black}0.65\epsilon$ with respect to the free value $z=2$ might seem large for a first-order term in a perturbative expansion, \textcolor{black}{in particular when we set $\epsilon=1$ in $d=3$. However, comparing our result with the equilibrium non-inertial theory of Model A might be misleading.} In fact, the new fixed point has been found by adding nonlinear activity to Model G \cite{HH1977}, which has a {\it non-perturbative} dynamic critical exponent $z=\frac{d}{2}=2-\frac{\epsilon}{2}$ \cite{dedominicis1978field}. Thus, our result should be considered as a $\color{black}0.15\epsilon$ departure from Model G's non-perturbative exponent, rather than a $\color{black}0.65\epsilon$ correction to Model A. To compare our result with previous ones in a similar context, the incompressible Toner and Tu theory has a dynamic critical exponent of $\color{black}z=2-0.27\epsilon$ \cite{chen2015critical}, with a perturbative departure of $\color{black}0.27\epsilon$ from equilibrium Model A's exponent, which is its natural expansion point. \textcolor{black}{When setting $\epsilon=1$, the incompressible Toner and Tu theory predicts an exponent $z=1.73$ in $d=3$.} Numerical simulations of Vicsek Model in $d=3$ found a dynamic exponent of $z\simeq1.7$  \cite{cavagna2020equilibrium}, showing that the dynamic exponent found in \cite{chen2015critical} holds with remarkable accuracy also in $d=3$, namely when $\epsilon=1$.

\textcolor{black}{
\subsubsection{Static critical exponents}
In the present section we will derive some of the static critical exponents relative to the active inertial fixed point found in the previous sections.} \textcolor{black}{Before proceeding however, an important caveat is in order. The present RG calculation was performed by imposing a solenoidal constraint on the primary field, $\vnabla\cdot\vpsi=0$, in order to enforce incompressibility. In non-active equilibrium systems, the presence of the solenoidal constraint is known to change the {\it static} universality class \cite{aharony1973critical, fisher1973dipolar}, while leaving unchanged the {\it dynamic} universality class \cite{cavagna2021dynamical}; this means that an equilibrium RG calculation with solenoidal constraint would find the same dynamic critical exponent $z$ as a simulation without solenoidal constraint, but would fail to reproduce the static critical exponents of that same simulation. It is not known how this scenario generalises to the active off-equilibrium case, but some caution is certainly required: although in active systems with mild density fluctuations incompressibility is a reasonable hypothesis to calculate the dynamic critical exponents, one must be careful about the static exponents, as we have no certainty that they must be the same as in system where the solenoidal constraint is not imposed, as in natural swarms or numerical simulations.
}

\textcolor{black}{The first exponent we shall compute is the $\nu$, which characterizes the divergence of the correlation length,
\begin{equation}
\xi\sim \left|T-T_{c}\right|^{-\nu}
\end{equation}
where $T$ is the control parameter (the temperature in the case of equilibrium systems) and $T_{c}$ its value at the critical point.
This exponent can be computed from the runaway exponent of the mass, namely from the mass $\beta$-function. Let us define, 
\begin{equation}
	y_{m}=\left.\pder{\beta_{m}}{m}\right|_{\mathcal{P}=\mathcal{P}^{*}}
\end{equation}
where $\mathcal{P}^{*}$ are the values of the parameters at the fixed point.
The exponent $\nu$ is thus given by
\begin{equation}
\nu=\frac{1}{y_{m}}\simeq\frac{1}{2}+ 0.248 \epsilon
\end{equation}
which in $d=3$ yields,
\begin{equation}
\nu\simeq 0.748 \ .
\end{equation}
}
\textcolor{black}{The second exponent we compute, $\gamma$, characterizes the response of the system to a small external field $\vecc{H}$ coupled to the order parameter. It is crucial here, for the purpose of computing these exponents up to one loop, to notice that no one-loop graphical corrections to such a field term appears in the calculation. Assuming linear response of $\vpsi$ on $\vecc H$, $\langle\vpsi\rangle=\chi_{H}\vecc H$, the exponent $\gamma$ characterizes the divergence of the susceptibility $\chi_{H}$ in the vicinity to the phase transition,
\begin{equation}
\chi_{H}\sim \left|T-T_{c}\right|^{-\gamma}
\end{equation}
where $T$ is the control parameter.
Since $\chi_{H}=\pder{\langle\vpsi\rangle}{H}$, the exponent $\gamma$ will be given by,
\begin{equation}
	\gamma=\nu\left({\color{black}-\chi_{\psi(r,t)}}+y_{H}\right)
\end{equation}
where $\chi_{\psi(r,t)}$ is the scaling dimension of $\vpsi$ in position and time space, and $y_{H}$ is the scaling dimension of the field $\vecc H$. The absence of graphical correction to $\vecc H$ at one loop implies that $y_{H}$ can be determined only by power counting. To determinate $y_{H}$, let us recall how $\vecc{H}$ would enter the equation of motion for $\vpsi$: $\partial_{t}\vpsi\sim\vecc{H}$, hence leading to $y_{H}=z{\color{black}+\chi_{\psi(r,t)}}+\mathcal{O}\left(\epsilon^{2}\right)$. At first order in $\epsilon$, the exponent $\gamma$ is thus given by,
\begin{equation}
	\gamma=\nu z \simeq 1+0.171\epsilon
\end{equation}
which in $d=3$ becomes,
\begin{equation}
	\gamma\simeq1.171
\end{equation}
Let us note that this exponent {\it does not} quantify the divergence of the space integral of the connected correlation function, as it would be the case for equilibrium systems, because the fluctuation-dissipation relation linking the susceptibility $\chi_{H}$ to the correlation function loses its validity when the system is out-of-equilibrium. In previous works \cite{attanasi2014collective,attanasi2014finite,cavagna2017swarm}, we called the space integral of the connected correlation function `generalized susceptibility' and used the symbol $\chi$ for it, so what we are saying here is that, although the two quantities will both diverge at the critical point, in general $\chi\neq\chi_{H}$, so they might have different critical exponents in the off-equilibrium active regime.}

\textcolor{black}{The evaluation of the other static critical exponents requires the knowledge of the scaling dimension of the primary field, $\chi_\psi$, which in turns requires some RG fixing prescription. We had a similar case with the evaluation of $z$ in the previous section, which we resolved through \eqref{casablanca}; however, while the prescription $\Gamma^*=\mathcal{O}(1)$ used to compute $z$ is well-established and well-understood, both in and out of equilibrium \cite{ma1975critical,halperin1976renormalization,HH1977,cavagna2020equilibrium}, and it has been validated by countless simulations and experiments, to our knowledge there is no standard, nor validated prescription to fix $\chi_{\psi}$. Hence, we prefer not to speculate here about this, with the risk of giving wrong predictions. The issue clearly deserves more study, both analytical and numerical.}

\begin{figure}[b]
	\centering
	\includegraphics[width=1\textwidth]{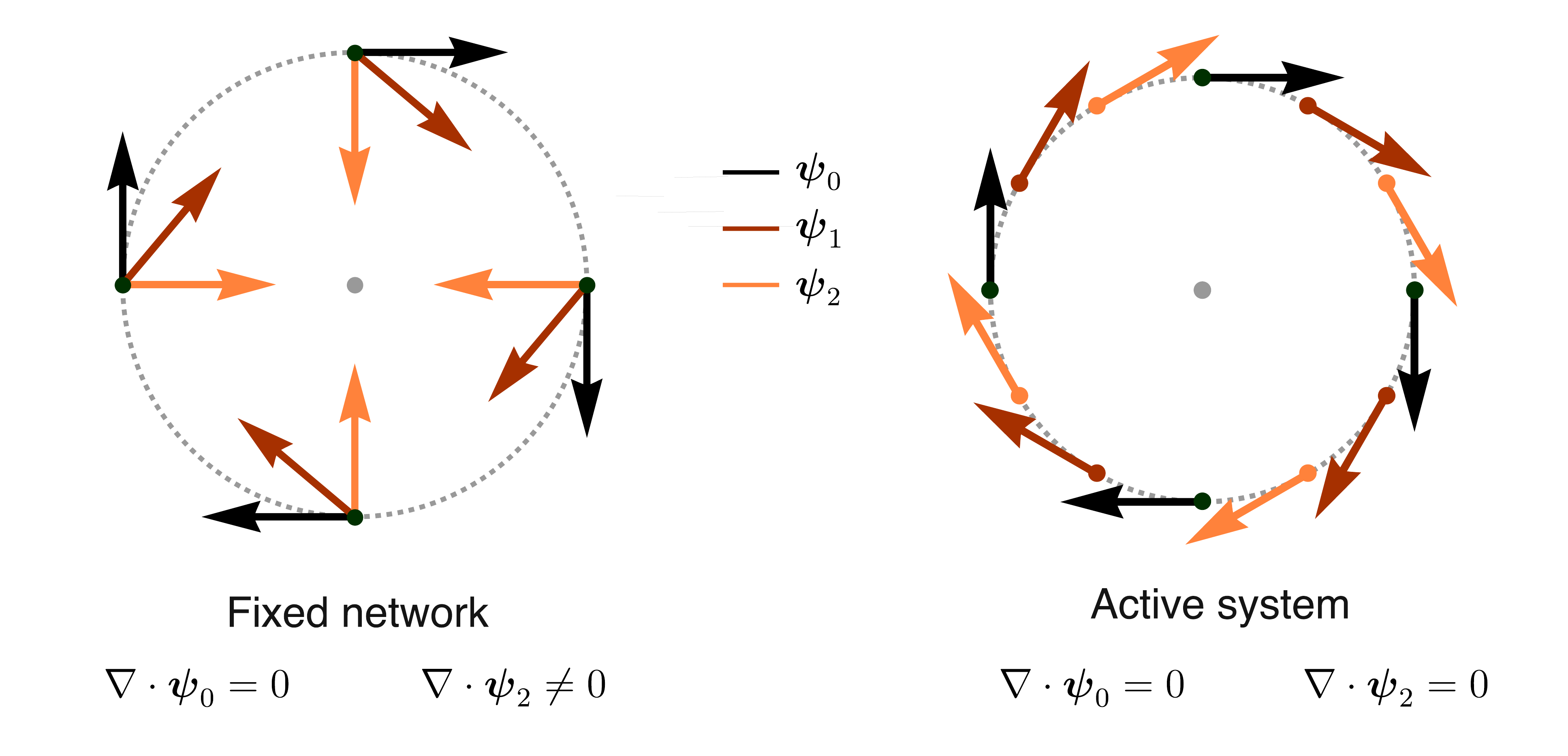}
	\caption{{\bf Incompressibility in the inactive vs active case.} Left: in the inactive case, if we start from a solenoidal configuration of the field ($\psi_0$, black) and we let a constant uniform spin rotate the field, we obtain a non-solenoidal configuration ($\psi_2$, orange). This is equivalent to saying that the solenoidal constraint breaks rotational invariance at equilibrium. Right: in the active case the particles are dragged by the field, hence the rotation generated by the spin rotates both the field {\it and} the position of the field, giving rise to a new configuration that is equally solenoidal (that is incompressible).
	}
	\label{lontra}
\end{figure}

\subsubsection{Off-equilibrium activity restores the conservation of the spin}
\label{deadkatz}

One of the most intriguing features of the novel active inertial fixed point is the vanishing of the DYS vertex effective coupling constant, namely $\tilde{u}_\kappa^*=0$. At equilibrium, the presence of the DYS vertex, with $\tilde{u}_\kappa=\tilde{u}$, had such a crucial role in keeping the dynamics of $\vs$ consistent with the solenoidal constraint on $\vpsi$ \cite{cavagna2021dynamical} that it seems surprising to find that this vertex disappears when the system is the off-equilibrium active phase. Moreover, the couplings $\tilde u$ and $\tilde{u}_\kappa$ both derive from derivatives of the $u \psi^4$ term in the free energy functional $\Ham$, and thus one would expect a deep connection between the two. However, the fact that the DYS vertex vanishes at this novel fixed point avoids the odd scenario of having two different ferromagnetic couplings; both at equilibrium and off-equilibrium, the RG suggests that only {\it one} ferromagnetic coupling should exist, by requiring in the former case that $\tilde{u}=\tilde{u}_\kappa$ and in the latter $\tilde{u}_\kappa=0$.

But by far the most surprising consequence of the off-equilibrium vanishing of the DYS vertex is that when $\tilde{u}_\kappa=0$ (and when the dissipation $\eta=0$) the global spin turns out to be conserved! As mentioned in Section(\ref{sec:nonlin}) and discussed in \cite{cavagna2021dynamical}, the DYS vertex is the only interaction that contributes to the dynamics of the spin at $k=0$, thus violating the spin conservation (although it does not violate it strongly, e.g. through dissipation). Since $\tilde{u}_\kappa$ is the effective coupling associated with the DYS interaction, the fact that it vanishes means that this vertex is irrelevant at the novel fixed point and thus the spin becomes globally conserved. The restored spin conservation in the active off-equilibrium case is surprising, and it can hardly be a mere accident of the RG calculation. In order to understand its meaning we have to go back to the underlying symmetry and conservation law in the theory.

At equilibrium, and in absence of any constraint on the field, the presence of an exact rotational symmetry guarantees that the spin is globally conserved since it is the generator of the symmetry. This is exactly what happens in Model G, described in Section \ref{sec:MG}. Now we add the solenoidal constraint, which is the equilibrium version of incompressibility: as it is clear from Fig.\ref{lontra}-left, the solenoidal constraint breaks the rotational invariance: if we start from a solenoidal configuration of the field and rotate each vector by the same constant amount, we obtain a new field configuration which violates the solenoidal constraint. At the field-theoretical level, this manifests itself by the RG generation of the DYS vertex that indeed breaks the symmetry and conservation of the spin \cite{cavagna2021dynamical}. This is what happens at equilibrium, namely in the non-active case.

According to the RG, if we now turn from the inactive solenoidal case to the active incompressible one, the presence of activity restores the full power of the rotation symmetry - at least on long wavelengths - by making the spin conserved once again. This suggests that - at variance with the inactive case -  activity preserves incompressibility under local rotations. A qualitative cartoon of this mechanism can be seen in Fig.\ref{lontra}-right: in the active case a local rotation generated by the spin has a twofold effect: {\it i)} it rotates the field (as in the inactive case); {\it ii)} it also rotates the positions, through the self-propelled part of the equations, $\dot{\vec x} = \vec v$ (at variance with the inactive case). These two rotations balance each other, giving rise to a new field configuration that is still solenoidal, i.e. incompressible (Fig.\ref{lontra}-right). We stress that this is far from being a general mathematical proof, as it is restricted to the very simple case of a purely rotational field, while one would need to generalize this argument to a generic solution of the dynamical equations. It may be that the full proof of spin conservation in the case of active incompressible dynamics is the very RG calculation that we carried out here; even though one would hope for a simpler and more direct way to prove this result, we could not find it. 

In fact, when we reflect on the whole RG flow, rather than restricting ourselves to the fixed point, the situation becomes even more intriguing. As we wrote above, we found the new active incompressible inertial fixed point by starting close to the inactive (equilibrium) incompressible inertial fixed point. At the starting point, $\tilde{u}_\kappa^*\neq 0$, because - as we said - at equilibrium the DYS vertex is required to enforce incompressibility; along with the flow which goes from equilibrium to off-equilibrium, the coupling  $\tilde{u}_\kappa^*$ decreases weakly (see Fig.\ref{fig:flowSI}) until it abruptly goes to zero right before arriving at the new fixed point. If we tried to start the flow with all parameters close to their equilibrium values, but with $\tilde{u}_\kappa^*= 0$ from the outset, we would {\it not} reach the new off-equilibrium fixed point, and the RG flow would simply go bonkers. Hence, not only the symmetry-breaking coupling $\tilde{u}_\kappa^*$ is necessary {\it at} equilibrium, but it is also necessary to accompany the RG flow to the off-equilibrium fixed point; only there $\tilde{u}_\kappa^*$ is finally allowed to vanish. Clearly, this phenomenon deserves a deeper study.

\subsection{Properties of the RG solution in presence of dissipation}

\subsubsection{Dissipative fixed point}\label{sec:diss}

\textcolor{black}{Up to now, we analyzed the RG flow only in its fully conservative manifold, namely at $\hat{\eta}=\tildehat{\eta}=0$, since we expect finite size systems with small dissipation to be well described by fixed points in this manifold. In the present section we will test the validity of our expectations. First, we will look to the behaviour of the RG flow in the strong-dissipation limit, recovering the results for incompressible Toner and Tu theory. Secondly, we fill focus our attention to the crossover between underdamped (small dissipation) and overdamped (strong dissipation) dynamics by analysing the RG flow in the proximity of the new active, inertial conservative fixed point found in Section \ref{mcenroe}.}

\textcolor{black}{As discussed at the end of Appendix \ref{RGflow}, whenever $\hat{\eta}=\tildehat{\eta}\neq0$ the RG flow makes them grow, approaching the fully dissipative manufold $\hat{\eta}=\tildehat{\eta}=1$. When $\hat{\eta}=\tildehat{\eta}=1$, the $\beta$-functions of the parameters and coupling constants of the theory take a simplified form.
In particular, the $\beta$-functions of $c_{v}$ and $\tilde{u}$ become independent from all the other parameters, while all the critical exponents can be expressed in terms of $c_{v}$ and $\tilde{u}$ alone.
This is a consequence of the fact that, in the presence of friction, the actual mode-coupling effective constant in the perturbative expansion is no longer $f$, but rather
\begin{equation}
    q=\frac{\tlambda^{\parallel}}{\lambda^{\parallel}}\frac{g^2}{\eta \Gamma} K_d \Lambda^{d-2}
\end{equation}
whose scaling dimension in proportional to $2-d$, suggesting that mode-coupling is relevant only below $2$ dimensions, and not $4$, when friction is present. The upper critical dimension of the field theory however remains $d_c=4$, since both $c_{v}$ and $\tilde{u}$ still become relevant below $4$ dimensions. The physical reason behind this change of dimension at which mode-coupling is relevan is the fact that the spin becomes a {\it fast mode}. When $\eta\neq0$, the finite relaxation time for the spin $\eta^{-1}$ will become, close to criticality, much smaller than the velocity relaxation time $\tau=\xi^z$. Hence, the effect of the spin on the order parameter becomes irrelevant, since its effect can be represented as noise.
}

\begin{figure}[b]
	\centering
	\includegraphics[width=0.9\textwidth]{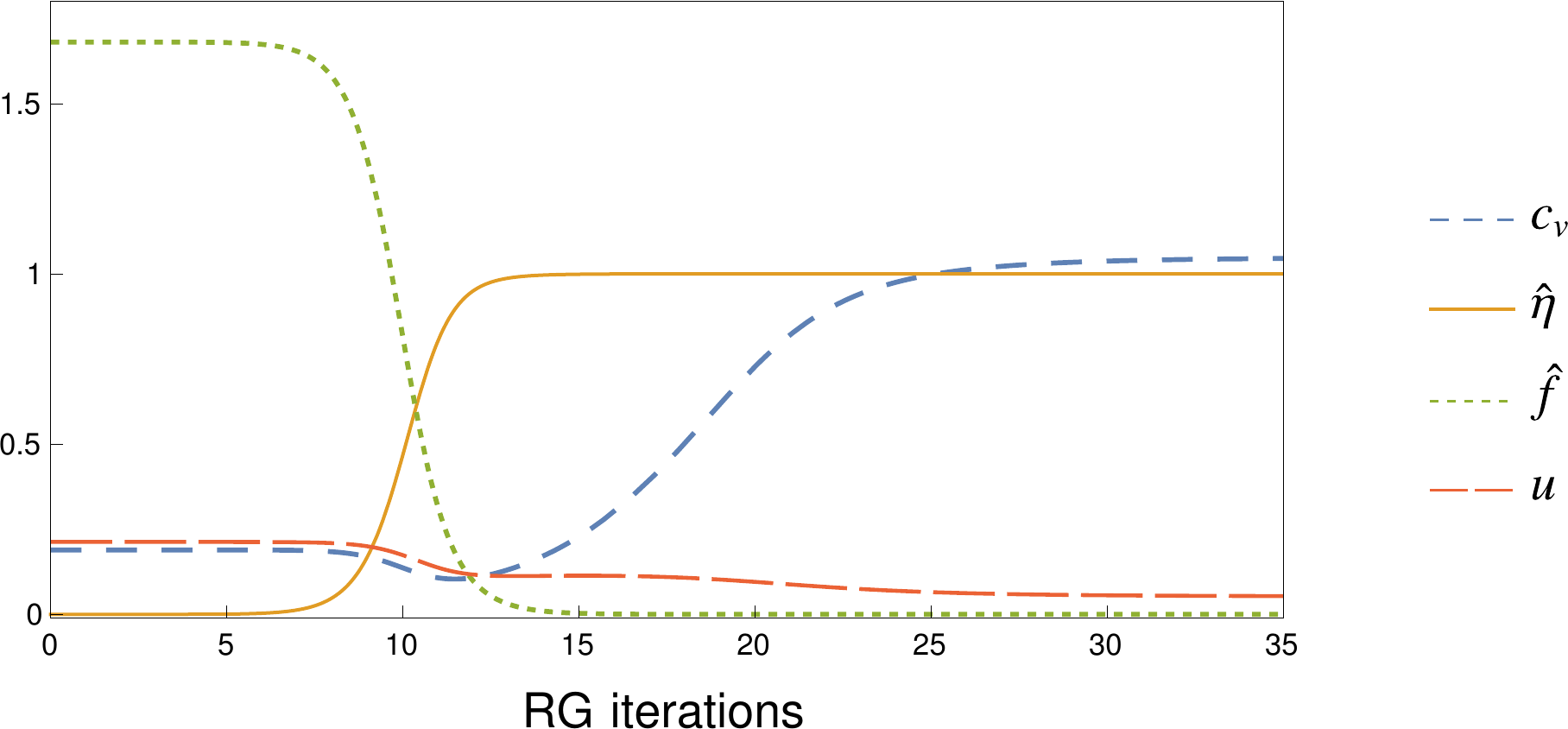}
	\caption{{\bf RG flow in the presence of friction} We show the RG flow of the couplings $c_v$, $\hat\eta$, $\hat f=\left(1-\hat{\eta}\right)f$ and $\tilde{u}$ from the conservative active fixed point towards the dissipative active fixed point. Note that the coupling $\hat f$ interpolates between $f$ and $q$, as the parameter $\eta$ runs from $0$ to $\infty$.
	From the plot we can clearly see that as friction increases, $c_v$ and $\tilde{u}$ flow to the fixed point values of \protect\cite{chen2015critical}.}
	\label{fig:etaflow}
\end{figure}

Therefore, the presence of dissipation strongly modifies the asymptotic critical behaviour, since near criticality only the modes fluctuating on the same scales of the order parameter may affect universal quantities \cite{HH1977}. Therefore, when dissipation is present, the {\it asymptotic} critical dynamics in the {\it thermodynamic limit} is unaffected by the presence of the spin-velocity coupling.

\textcolor{black}{In this limit, $\hat{\eta}=\tildehat{\eta}=1$, our theory should therefore become equivalent to the overdamped incompressible theory of \cite{chen2015critical}. In all perturbative corrections to the equation of motion of $\vv$, the diagrams proportional to $f$ are always also proportional to $\left(1-\hat{\eta}\right)$. Hence, when $\hat{\eta}=1$ these diagrams vanish, giving no contributions to the critical exponents.
The velocity field $\vv$ thus fully decouples from the spin $\vs$ in the overdamped limit, making the behaviour of the fast mode $\vs$ irrelevant in determining the universal quantities, so that for $\hat{\eta}=\tildehat{\eta}=1$, we obtain, 
\begin{align}
	\beta_{c_{v}}&=\frac{1}{2}\,c_{v}\left(\epsilon-\frac{3}{4}c_{v}^2-\frac{10}{3}\tilde{u}\right)\\
	\beta_{\tilde{u}}&=\tilde{u}\left(\epsilon-\frac{1}{2} c_{v}^2-\frac{17}{2}\tilde{u}\right)
\end{align}
which are the same RG flow equations of \cite{chen2015critical}, giving the dynamic critical exponent, 
\begin{equation}
z=1.73  \ .
\end{equation}
This prediction is confirmed also by numerical integration of the full RG flow, as shown in Fig.\ref{fig:etaflow}. As required by physical consistency, the overdamped limit of our theory is the same as the non-inertial theory of \cite{chen2015critical}.
}

\subsubsection{The crossover from underdamped to overdamped dynamics}
\label{sec:crossover}
\textcolor{black}{The asymptotic stable fixed point for our field theory is indeed given by the overdamped, off-equilibrium one already found in \cite{chen2015critical}.
However, as previously mentioned, unstable fixed points can play a crucial role in determining critical exponents for  {\it finite-size} systems \cite{cavagna2019short}.
It often happens that for small enough sizes, the critical properties of a system are determined by the closest fixed point, rather than the fully stable one.
As the size of the system is increased, a crossover between different regime emerges, changing the critical properties of the system.
In the present case, we expect a finite-size system with small enough dissipation to initially behave as if it were inertial, and turn to the overdamped limit only when the thermodynamic limit is approached.}

At the theoretical level, this crossover between a finite-size physics influenced by inertia, and an asymptotic physics completely overdamped manifests itself as a crossover between different RG fixed points. Hence, following \cite{cavagna2019short,cavagna2019long}, we will investigate the crossover between the underdamped dynamic behaviour, ruled by our novel fixed point with $z=1.35$, and the dissipative incompressible theory with $z=1.73$.
The starting point of this analysis is the observation that the ratio between dissipation $\eta$ and conservative kinetic coefficient $\lambda^{\parallel}$ naturally defines, by simple power counting, a length-scale $\mathcal{R}$ given by \cite{cavagna2019short,cavagna2019long},
\begin{equation}
	\mathcal{R}=\sqrt{\frac{\lambda^\parallel}{\eta}}
	\label{fartissimo}
\end{equation}
We could define a second length-scale: $\sqrt{\lambda^\perp/\eta}$. However, it turns out this second length-scale has the same scaling behaviour as $\mathcal{R}$, hence it makes no difference what definition we use.

The parameter $\mathcal{R}$ plays the role of a conservation length-scale, in the sense that fluctuations occurring on length-scales smaller than $\mathcal{R}$ obey a conservative dynamics, while beyond $\mathcal{R}$ fluctuations are insensitive to inertia and conservation laws.
The scaling dimension of $\mathcal{R}$ is that of a length, but only at the naive (non-interacting) level; in fact, we know that the RG coupling between the UV and IR degrees of freedom generates non-trivial modifications of naive scaling dimensions of $\lambda^\parallel$ and in principle also of $\eta$.
The RG flow of $\mathcal{R}$ can be written -- defining its scaling dimension ${\chi_\mathcal{R}}$ -- as (see Eq. \eqref{recup}),
\textcolor{black}{
\begin{equation}
\mathcal{R}_{l+1}= b^{\chi_\mathcal{R}}\  \mathcal{R}_l  \ ,
\label{vendetta}
\end{equation}}
and from equations \eqref{fartissimo},\eqref{eq:scaling_lambda} and \eqref{eq:scaling_eta} we obtain, 
\textcolor{black}{
\begin{equation}
	\chi_\mathcal{R}= \frac{1}{2}(\chi_{\lambda^\parallel}-\chi_\eta) =  -1 +  \frac{1}{2} \delta\lambda^\parallel
	\label{chiR}
\end{equation}}
where \textcolor{black}{$-1$} is the naive dimension of $\mathcal R$, which allowed us to naively identify it as a length-scale, while $\delta\lambda^\parallel$ the correction to the kinetic coefficient $\lambda^\parallel$. We recall that $\eta$ has no perturbative correction due to the symmetry of the problem.
The physical origin of this fact is that all diagrams contributing to the self energy $\Pi$ (which contains corrections to both $\eta$ and $\lambda^\parallel$) vanish at $k=0$ as a consequence of the form of the mode-coupling and the self-propulsion vertices for $\vs$. This has been verified in our calculation at first order in perturbation theory, where higher order effects of diagrams containing the DYS vertex on $\Pi$ are not taken into account; however, if the result $\kappa^*=0$ is exact to all orders, the DYS vertex is always irrelevant and thus no diagrams could ever generate dissipation. This is what we expect, as $\kappa^*=0$ implies a global conservation of the spin (at $\eta=0$), which does not seem a perturbative accident (see Section \ref{deadkatz}). But even in the unlikely case in which the result $\kappa^*=0$ were an accident of first order perturbation theory, we prove here that the dissipation of the spin still can {\it never} receive any perturbative corrections. In general, the self-energy $\Pi$ takes contributions only from the following two distinct classes of diagrams,
\begin{equation}
\Pi=
	\begin{tikzpicture}[baseline=-\the\dimexpr\fontdimen22\textfont2\relax]
		\begin{feynman}[small]
		    \vertex (a) at (0,0);
			\vertex (v1) [blob] at (1,0) {};
			\vertex (v2) [dot] at (2,0) {};
		    \vertex (b) at (3,0);
			\diagram* {
				(a) -- [charged boson] (v1) -- [half left, fermion] (v2) -- [boson] (b),
				(v2) -- [half left] (v1),
			};
		\end{feynman}
	\end{tikzpicture}
	+
	\begin{tikzpicture}[baseline=-\the\dimexpr\fontdimen22\textfont2\relax]
		\begin{feynman}[small]
		    \vertex (a) at (0,0);
			\vertex (v1) [blob] at (1,0) {};
			\vertex (v2) [dot] at (2,0) {};
		    \vertex (b) at (3,0);
			\diagram* {
				(a) -- [charged boson] (v1) -- [half left, charged boson] (v2) -- [boson] (b),
				(v2) -- [half left] (v1),
			};
		\end{feynman}
	\end{tikzpicture}
\end{equation}
where the blobs represent the renormalized mode-coupling and self-propulsion vertices respectively, in which all the possible diagrammatic corrections (at all orders in $\vk$ and $\vq$) are taken into account.
Because the renormalized spin mode-coupling and self-propulsion vertices vanish at zero external momentum $\vk$, then these diagrams are zero at $k = 0$, implying that no dissipation is generated. Therefore, as long as the structure of the equations of motion \eqref{eq:psi} and \eqref{eq:s} is preserved under the RG, no spin dissipation can be generated.

\begin{figure}[b]
	\centering
	\includegraphics[width=0.6\textwidth]{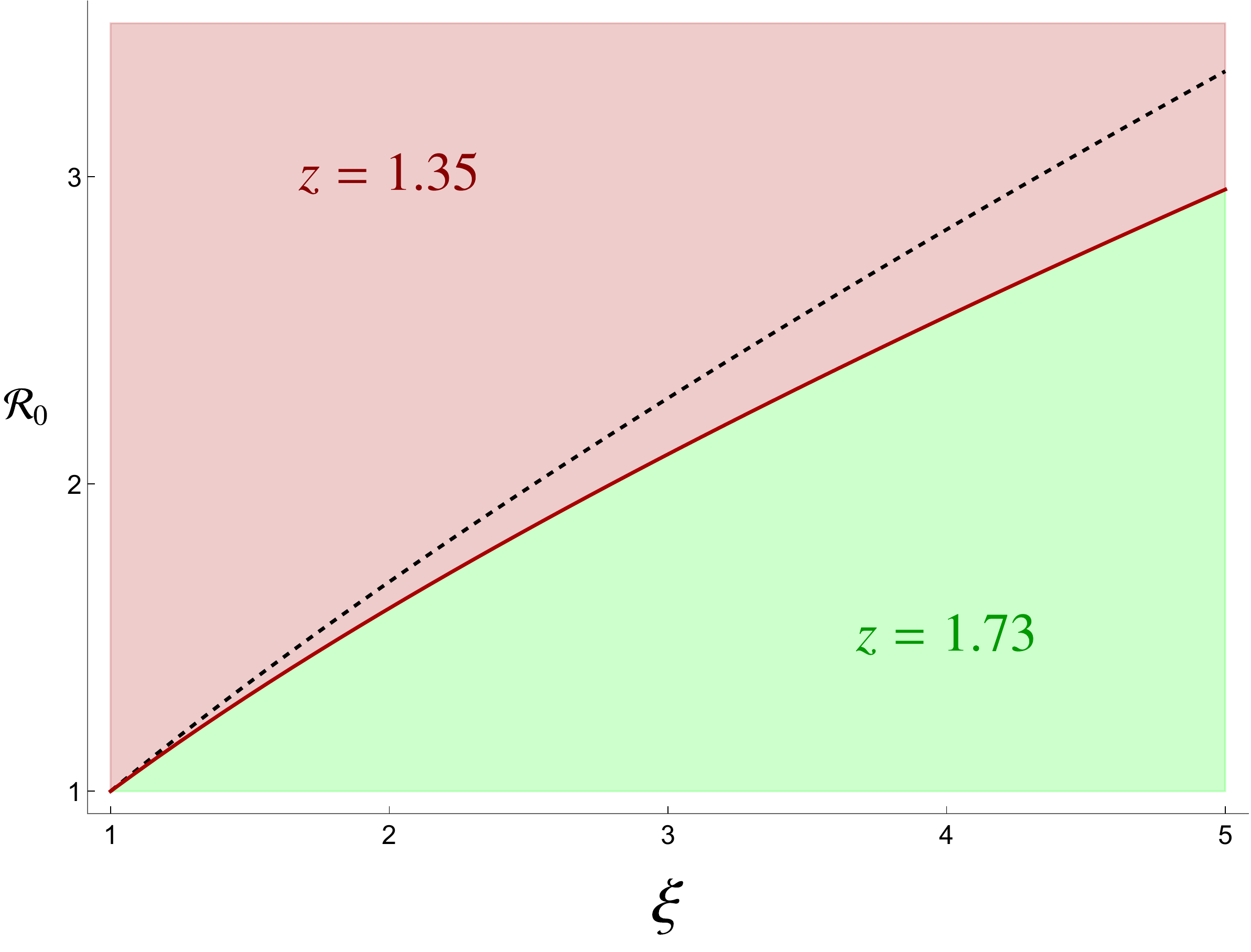}
	\caption{{\bf Crossover between different critical regions.} Different values of $\mathcal{R}_0$ lead to different critical behaviour, depending on the correlation length $\xi$.
	We restrict ourself to the physical range $\xi_0>\Lambda^{-1}$, and measure lengths in units of $\Lambda^{-1}$. The figure refers to the $d =3$ case.
	The dark red region is where dissipation is weak enough that conservative dynamics still holds, thus leading to $z=1.35$: this is the underdamped regime. In the light green dissipation overcomes and overdamped regime is achived, where the dynamic behaviour is controlled by $z=1.73$.
	The red line represents $L_c$, the threshold between the two behaviour in the active case.
	The black dashed line represents instead the threshold between underdamped and overdamped dynamics at equilibrium: since the undedamped region is lower here, we conclude that activity protects underdamped behaviour up to high scales.
	}
	\label{fig:crossover}
\end{figure}

To work out the correction $\delta\lambda^\parallel$ we need a different argument. As we have seen, the ratio $w= \Gamma/\lambda^\parallel$ is finite at the novel fixed point, $w^*=3.95$ (equation \eqref{unlupomannaroamericanoalondra}); moreover, the kinetic coefficient of the primary field is also finite at the fixed point, $\Gamma^*=O(1)$ (this is how one works out the dynamic exponent $z$), implying that also the kinetic coefficient of the spin is finite, ${\lambda^\parallel}^*=O(1)$, and thus that
$\chi_{\lambda^\parallel}=0$. From \eqref{eq:scaling_gamma} we conclude, 
\begin{equation}
\delta\lambda^\parallel=2-z \ .
\label{shapiro}
\end{equation}
From equations \eqref{chiR} 
and \eqref{shapiro}, we finally obtain the scaling dimension of $\mathcal{R}$ near the active conservative fixed point, 
\textcolor{black}{
\begin{equation}
	\chi_\mathcal{R} = - \frac{z}{2} \ .
	\label{interstellar}
\end{equation}}
\textcolor{black}{where $z$ is the dynamic exponent of the conservative inertial fixed point we are interested in.}
The fact that the scaling dimension of $\mathcal R$ is \textcolor{black}{negative} implies (through \eqref{vendetta}) that it {\it decreases} along the RG flow, making the length-scale within which critical dynamics is underdamped shorter and shorter; this is another way to see that dissipation eventually takes over in the hydrodynamic limit.
However, thanks to \eqref{interstellar}, we can now quantitatively describe the RG crossover from the conservative active fixed point to the dissipative one, or -- more precisely -- describe the departure of the RG flow from the conservative active fixed point when we start close to it.
Close to criticality, the correlation length-scales as,
\begin{equation}
\xi_l=b^{-1} \xi_{l-1} = b^{-l} \xi_0 \ ,
\end{equation}
where $\xi_0$ is its physical value in the original system under study. The RG flow stops when the system is far from the critical manifold, namely when the correlation length becomes of the same size of the microscopic scale $\Lambda^{-1}$, giving $b^{l_\mathrm{stop}}=\xi_0 \Lambda$ \cite{HH1977, cardy1996scaling}.
Let us consider a system with small bare dissipation, $\eta_0$, and therefore with a large conservation length-scale $\mathcal{R}_0$ (the subscript zero indicates the bare values of the parameter, namely the starting values of the RG flow or the parameters that are those of the equations of motion).
The RG flow will rapidly approach the conservative fixed point, remaining in its neighborhood for a large number of RG iterations and eventually flowing towards the dissipative fixed point \cite{cavagna2019long}. Whether the system is ruled by the conservative or dissipative fixed point depends on how large $\mathcal{R}_{l_\mathrm{stop}}$ is when the RG flow leaves the critical region: if when the flow stops $\mathcal{R}_{l_\mathrm{stop}} \gg \Lambda^{-1}$, the critical behaviour is ruled by the conservative fixed point; this is equivalent to the condition, 
\textcolor{black}{
\begin{equation}
\mathcal{R}_0 (b^{\chi_\mathcal{R}})^{l_\mathrm{stop}} =   \mathcal{R}_0 (\xi_0\Lambda)^{\chi_\mathcal{R}} \gg \Lambda^{-1}
\end{equation} }
that is, 
\begin{equation}
\xi_0 \ll (\mathcal{R}_0 \Lambda)^{-\frac{1}{\chi_\mathcal{R}}} \Lambda^{-1} \ .
\end{equation}
By using equation \eqref{interstellar} and by measuring all lengths in units of the microscopic scale $\Lambda^{-1}$ (which is equivalent to simply set $\Lambda=1$ in all equations), we obtain the following condition for active underdamped dynamics,  
\begin{equation}
	\xi_0 \ll \left(\mathcal{R}_0\right)^{2/z} \ ,
	\label{merdu}
\end{equation}
where $\kappa=2/z$ is the so-called {\it crossover exponent}, which determines how slowly the dissipative dynamics becomes relevant. Beyond this regime the overdamped fixed point with dynamic exponent $1.73$ takes over. In Fig. \ref{fig:crossover} we show the regions corresponding to the two dynamical behaviours. In the case of finite-size systems as natural swarms, the size of the system $L$ is a physical upper bound to the correlation length $\xi_0$. Therefore, if the dissipation $\eta_0$ is small enough to have $L\ll\left(\mathcal{R}_0\right)^{2/z}$, the ruling fixed point is the  active inertial one. By recalling the definition of $\mathcal{R}$, equation \eqref{fartissimo}, we get,
\begin{equation}
L\ll\left(\frac{\lambda_0^\parallel}{\eta_0}\right)^{1/z}  \longrightarrow \quad z=1.35 \ .
\label{calipso}
\end{equation}
Because $\lambda_0^\parallel$ is always finite, the condition above can be finally rewritten as, 
\begin{equation}
\eta_0 L^z \ll 1 \longrightarrow \quad z=1.35 \ ,
\end{equation}
which clearly shows that the regime of the system (underdamped vs overdamped) depends essentially on the balance between system size and spin dissipation. 

\textcolor{black}{A final remark is in order. From \eqref{calipso} we clearly see that the smaller is $z$, the larger the system can be before crossing over to the dissipative fixed point.
Since the new active inertial critical exponent, $z=1.35$, is smaller than its equilibrium inertial counterpart, $z=1.5$, this means that in the active off-equilibrium case 
 conservative dynamics rules the collective behaviour of the system up to larger scales compared to the equilibrium case (see Fig. (\ref{fig:crossover})); this means that activity {\it protects} the conservative structure against dissipation, rather than thwarting it, which is quite remarkable.}

\newpage


\section{Feynman Diagrams}\label{sec:diagrams}

\subsection{Self Energies}\label{sec:selfen}
\subsubsection{Self Energy $\Sigma$}
The self energy $\Sigma_{\alpha\beta}$ corrects the inverse bare propagator of $\vpsi$, and is given by
\begin{equation}
	\begin{tikzpicture}[baseline=(a.base)]
		\begin{feynman}
			\vertex at (0,0) (a) {\(\hpsi_\alpha(-\tvk)\)};
			\vertex at (2.2,0) [blob] (b){};
			\vertex at (4,0) (c) {\(\psi_\beta(\tvk)\)};
			\diagram* {
				(a) -- [fermion] (b) -- (c),
			};
		\end{feynman}
	\end{tikzpicture}
\end{equation}
with the following non-vanishing diagrams contributing to it
\begin{figure}[h!]
	\centering
	\begin{subfigure}[c]{0.24\textwidth}
		\centering
		\includegraphics[page=1,width=0.75\linewidth]{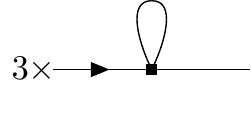}
	\end{subfigure}
	\begin{subfigure}[c]{0.24\textwidth}
		\centering
		\includegraphics[page=2, width=1\linewidth]{figure_SI/Sigma.pdf}
	\end{subfigure}
	\begin{subfigure}[c]{0.24\textwidth}
		\centering
		\includegraphics[page=3, width=1\linewidth]{figure_SI/Sigma.pdf}
	\end{subfigure}
	\begin{subfigure}[c]{0.24\textwidth}
		\centering
		\includegraphics[page=4, width=1\linewidth]{figure_SI/Sigma.pdf}
	\end{subfigure}
	\caption{Diagrams contributing to leading order of $\Sigma$}
	\label{fig:Sig}
\end{figure}

\subsubsection{Self Energy $\Pi$}
The self energy $\Pi_{\alpha\beta\gamma\nu}$ corrects the inverse bare propagator of $\vs$, and is given by
\begin{equation}
	\begin{tikzpicture}[baseline=(a.base)]
		\begin{feynman}
			\vertex at (0,0) (a) {\(\hs_{\alpha\beta}(-\tvk)\)};
			\vertex at (2.2,0) [blob] (b){};
			\vertex at (4,0) (c) {\(s_{\gamma\nu}(\tvk)\)};
			\diagram* {
				(a) -- [charged boson] (b) --  [boson] (c),
			};
		\end{feynman}
	\end{tikzpicture}
\end{equation}
with the following non-vanishing diagrams contributing to it
\begin{figure}[h!]
	\begin{subfigure}[c]{0.24\textwidth}
		\centering
		\includegraphics[page=1,width=1\linewidth]{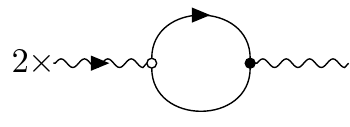}
	\end{subfigure}
	\begin{subfigure}[c]{0.24\textwidth}
		\centering
		\includegraphics[page=2, width=1\linewidth]{figure_SI/Pi.pdf}
	\end{subfigure}
	\caption{Diagrams contributing to leading order of $\Pi$}
	\label{fig:Pi}
\end{figure}

\subsubsection{Self Energy $\tSigma$}
The self energy $\tSigma_{\alpha\beta}$ corrects the noise variance of $\vpsi$, and is given by
\begin{equation}
	\begin{tikzpicture}[baseline=(a.base)]
		\begin{feynman}
			\vertex at (0,0) (a) {\(\hpsi_\alpha(-\tvk)\)};
			\vertex at (2.2,0) [blob] (b){};
			\vertex at (4,0) (c) {\(\hpsi_\beta(\tvk)\)};
			\diagram* {
				(a) -- [fermion] (b) -- [anti fermion] (c),
			};
		\end{feynman}
	\end{tikzpicture}
\end{equation}
with the following non-vanishing diagram contributing to it
\begin{figure}[h!]
	\centering
	\includegraphics[page=1,width=0.24\linewidth]{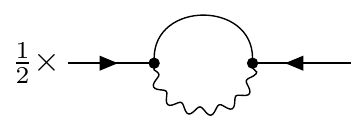}
	\caption{Diagram contributing to leading order of $\tSigma$}
	\label{fig:tSig}
\end{figure}

\subsubsection{Self Energy $\tPi$}
The self energy $\tPi_{\alpha\beta\gamma\nu}$ corrects the noise variance of $\vs$, and is given by
\begin{equation}
	\begin{tikzpicture}[baseline=(a.base)]
		\begin{feynman}
			\vertex at (0,0) (a) {\(\hs_{\alpha\beta}(-\tvk)\)};
			\vertex at (2.2,0) [blob] (b){};
			\vertex at (4,0) (c) {\(\hs_{\gamma\nu}(\tvk)\)};
			\diagram* {
				(a) -- [charged boson] (b) --  [anti charged boson] (c),
			};
		\end{feynman}
	\end{tikzpicture}
\end{equation}
with the following non-vanishing diagram contributing to it
\begin{figure}[h!]
	\begin{subfigure}[c]{0.24\textwidth}
		\centering
		\includegraphics[page=1,width=1\linewidth]{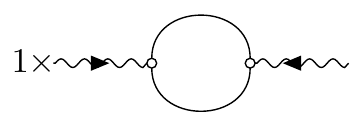}
	\end{subfigure}
	\begin{subfigure}[c]{0.24\textwidth}
		\centering
		\includegraphics[page=2, width=1\linewidth]{figure_SI/tPi.pdf}
	\end{subfigure}
	\caption{Diagrams contributing to leading order of $\tPi$}
	\label{fig:tPi}
\end{figure}

\newpage
\subsection{Vertex functions}\label{sec:vertexfun}

\subsubsection{Vertex function of the Mode-Coupling vertex of $\psi$}
The vertex function $V^{\hpsi\psi s}$ represents the corrections to the mode-coupling vertex in the equation of $\vpsi$ and is given by
\begin{equation}
	V_{\alpha\beta\gamma\nu}^{\hpsi \psi s}(\tvk,\tvq)=
	\begin{tikzpicture}[baseline=(a.base)]
		\begin{feynman}
			\vertex (a) at (0,0) {\(\hpsi_\alpha(-\tvk)\)};
			\vertex (b) [blob] at (2,0) {};
			\vertex (c) at (2.87,1.5) {\(\psi_\beta(\tvk-\tvq)\)};
			\vertex (d) at (2.87,-1.5) {\(s_{\gamma\nu}(\tvq)\)};
			\diagram* {
				(a) -- [fermion] (b) -- (c),
				(b) -- [boson] (d),
			};
		\end{feynman}
	\end{tikzpicture}
\end{equation}
with the following non-vanishing diagram contributing to it
\begin{figure}[h!]
	\begin{subfigure}[c]{0.15\textwidth}
		\centering
		\includegraphics[page=1,width=1\linewidth]{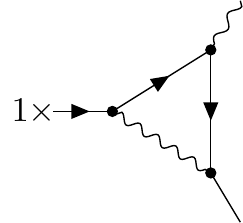}
	\end{subfigure}
	\begin{subfigure}[c]{0.15\textwidth}
		\centering
		\includegraphics[page=2, width=1\linewidth]{figure_SI/Vvvs.pdf}
	\end{subfigure}
	\begin{subfigure}[c]{0.15\textwidth}
		\centering
		\includegraphics[page=3, width=1\linewidth]{figure_SI/Vvvs.pdf}
	\end{subfigure}
	\caption{Diagrams contributing to leading order of $V^{\hpsi\psi s}$}
	\label{fig:Vvvs}
\end{figure}

\newpage

\subsubsection{Vertex function of the Mode-Coupling vertex of $s$}
The vertex function $V^{\hs\psi\psi}$ represents the corrections to the mode-coupling vertex in the equation of $\vs$, and is given by
\begin{equation}
	V_{\alpha\beta\gamma\nu}^{\hs \psi\psi}(\tvk,\tvq)=
	\begin{tikzpicture}[baseline=(a.base)]
		\begin{feynman}
			\vertex (a) at (0,0) {\(\hs_{\alpha\beta}(-\tvk)\)};
			\vertex (b)[blob] at (2,0) {};
			\vertex (c) at (2.87,1.5) {\(\psi_\gamma(\tvk/2-\tvq)\)};
			\vertex (d) at (2.87,-1.5) {\(\psi_\nu(\tvk/2+\tvq)\)};
			\diagram* {
				(a) -- [charged boson] (b) -- (c),
				(b) -- (d),
			};
		\end{feynman}
	\end{tikzpicture}
\end{equation}
with the following non-vanishing diagram contributing to it
\begin{figure}[h!]
	\begin{subfigure}[c]{0.15\textwidth}
		\includegraphics[page=1,width=1\linewidth]{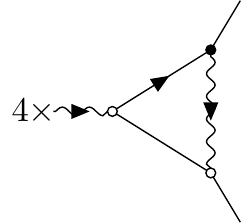}
	\end{subfigure}
	\begin{subfigure}[c]{0.15\textwidth}
		\centering
		\includegraphics[page=2, width=1\linewidth]{figure_SI/Vsvv.pdf}
	\end{subfigure}
	\begin{subfigure}[c]{0.15\textwidth}
		\centering
		\includegraphics[page=3, width=1\linewidth]{figure_SI/Vsvv.pdf}
	\end{subfigure}
	\begin{subfigure}[c]{0.15\textwidth}
		\centering
		\includegraphics[page=4, width=1\linewidth]{figure_SI/Vsvv.pdf}
	\end{subfigure}
	\begin{subfigure}[c]{0.15\textwidth}
		\centering
		\includegraphics[page=5, width=1\linewidth]{figure_SI/Vsvv.pdf}
	\end{subfigure}
	\begin{subfigure}[c]{0.15\textwidth}
		\centering
		\includegraphics[page=6, width=1\linewidth]{figure_SI/Vsvv.pdf}
	\end{subfigure}
	\begin{subfigure}[c]{0.15\textwidth}
		\centering
		\includegraphics[page=7, width=1\linewidth]{figure_SI/Vsvv.pdf}
	\end{subfigure}
	\begin{subfigure}[c]{0.15\textwidth}
		\centering
		\includegraphics[page=8, width=1\linewidth]{figure_SI/Vsvv.pdf}
	\end{subfigure}
	\begin{subfigure}[c]{0.15\textwidth}
		\centering
		\includegraphics[page=9, width=1\linewidth]{figure_SI/Vsvv.pdf}
	\end{subfigure}
	\begin{subfigure}[c]{0.15\textwidth}
		\centering
		\includegraphics[page=10, width=1\linewidth]{figure_SI/Vsvv.pdf}
	\end{subfigure}\\
	\begin{subfigure}[c]{0.17\textwidth}
		\centering
		\includegraphics[page=11, width=1\linewidth]{figure_SI/Vsvv.pdf}
	\end{subfigure}
	\begin{subfigure}[c]{0.15\textwidth}
		\centering
		\includegraphics[page=12, width=1\linewidth]{figure_SI/Vsvv.pdf}
	\end{subfigure}
	\caption{Diagrams contributing to leading order of $V^{\hs\psi\psi}$}
	\label{fig:Vsvv}
\end{figure}

\newpage

\subsubsection{Vertex function of the advection vertex of $\psi$}
The vertex function $V^{\hpsi\psi\psi}$ represents the corrections to the advection vertex in the equation of $\vpsi$, and is given by
\begin{equation}
	V_{\alpha\beta\gamma}^{\hpsi \psi \psi}(\tvk,\tvq)=
	\begin{tikzpicture}[baseline=(a.base)]
		\begin{feynman}
			\vertex (a) at (0,0) {\(\hpsi_\alpha(-\tvk)\)};
			\vertex (b) [blob] at (2,0) {};
			\vertex (c) at (2.87,1.5) {\(\psi_\beta(\tvq)\)};
			\vertex (d) at (2.87,-1.5) {\(\psi_\gamma(\tvk-\tvq)\)};
			\diagram* {
				(a) -- [fermion] (b) -- (c),
				(b) -- (d),
			};
		\end{feynman}
	\end{tikzpicture}
\end{equation}
with the following non-vanishing diagram contributing to it
\begin{figure}[h!]
	\begin{subfigure}[c]{0.15\textwidth}
		\includegraphics[page=1,width=1\linewidth]{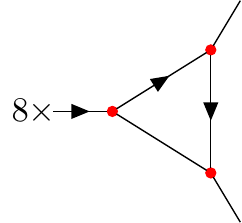}
	\end{subfigure}
	\begin{subfigure}[c]{0.15\textwidth}
		\centering
		\includegraphics[page=2, width=1\linewidth]{figure_SI/Vvvv.pdf}
	\end{subfigure}
	\begin{subfigure}[c]{0.15\textwidth}
		\centering
		\includegraphics[page=3, width=1\linewidth]{figure_SI/Vvvv.pdf}
	\end{subfigure}
	\begin{subfigure}[c]{0.15\textwidth}
		\centering
		\includegraphics[page=4, width=1\linewidth]{figure_SI/Vvvv.pdf}
	\end{subfigure}
	\begin{subfigure}[c]{0.15\textwidth}
		\centering
		\includegraphics[page=5, width=1\linewidth]{figure_SI/Vvvv.pdf}
	\end{subfigure}
	\begin{subfigure}[c]{0.15\textwidth}
		\centering
		\includegraphics[page=6, width=1\linewidth]{figure_SI/Vvvv.pdf}
	\end{subfigure}
	\begin{subfigure}[c]{0.15\textwidth}
		\centering
		\includegraphics[page=7, width=1\linewidth]{figure_SI/Vvvv.pdf}
	\end{subfigure}
	\begin{subfigure}[c]{0.15\textwidth}
		\centering
		\includegraphics[page=8, width=1\linewidth]{figure_SI/Vvvv.pdf}
	\end{subfigure}
	\begin{subfigure}[c]{0.15\textwidth}
		\centering
		\includegraphics[page=9, width=1\linewidth]{figure_SI/Vvvv.pdf}
	\end{subfigure}
	\begin{subfigure}[c]{0.15\textwidth}
		\centering
		\includegraphics[page=10, width=1\linewidth]{figure_SI/Vvvv.pdf}
	\end{subfigure}\\
	\begin{subfigure}[c]{0.17\textwidth}
		\centering
		\includegraphics[page=11, width=1\linewidth]{figure_SI/Vvvv.pdf}
	\end{subfigure}
	\begin{subfigure}[c]{0.17\textwidth}
		\centering
		\includegraphics[page=12, width=1\linewidth]{figure_SI/Vvvv.pdf}
	\end{subfigure}
	\caption{Diagrams contributing to leading order of $V^{\hpsi\psi\psi}$}
	\label{fig:Vvvv}
\end{figure}

\newpage

\subsubsection{Vertex function of the advection vertex of $s$}
The vertex function $V^{\hs s\psi}$ represents the corrections to the advection vertex in the equation of $\vs$, and is given by
\begin{equation}
	V_{\alpha\beta\gamma\nu\eta}^{\hs s \psi}(\tvk,\tvq)=
	\begin{tikzpicture}[baseline=(a.base)]
		\begin{feynman}
			\vertex (a) at (0,0) {\(\hs_{\alpha\beta}(-\tvk)\)};
			\vertex (b) [blob] at (2,0) {};
			\vertex (c) at (2.87,1.5) {\(s_{\gamma\nu}(\tvq)\)};
			\vertex (d) at (2.87,-1.5) {\(\psi_\eta(\tvk-\tvq)\)};
			\diagram* {
				(a) -- [charged boson] (b) -- [boson] (c),
				(b) -- (d),
			};
		\end{feynman}
	\end{tikzpicture}
\end{equation}
with the following non-vanishing diagram contributing to it
\begin{figure}[h!]
	\begin{subfigure}[c]{0.15\textwidth}
		\centering
		\includegraphics[page=1,width=1\linewidth]{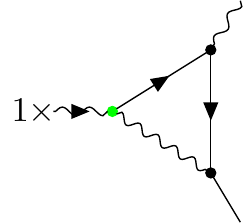}
	\end{subfigure}
	\begin{subfigure}[c]{0.15\textwidth}
		\centering
		\includegraphics[page=2, width=1\linewidth]{figure_SI/Vssv.pdf}
	\end{subfigure}
	\begin{subfigure}[c]{0.15\textwidth}
		\centering
		\includegraphics[page=3, width=1\linewidth]{figure_SI/Vssv.pdf}
	\end{subfigure}
	\caption{Diagrams contributing to leading order of $V^{\hs s \psi}$}
	\label{fig:Vssv}
\end{figure}

\newpage

\subsubsection{Vertex function of the ferromagnetic vertex of $\psi$}
The vertex function $V^{\hpsi\psi\psi\psi}$ represents the corrections to the ferromagnetic vertex in the equation of $\vpsi$, and is given by
\begin{equation}
	V_{\alpha\beta\gamma\nu}^{\hpsi \psi \psi \psi}(\tvk,\tvq,\tvh)=
	\begin{tikzpicture}[baseline=(a.base)]
		\begin{feynman}
			\vertex (a) at (0,0) {\(\hpsi_\alpha(-\tvk)\)};
			\vertex (b)[blob] at (2,0) {};
			\vertex (c) at (2.87,1.7) {\(\psi_\beta(\tvq)\)};
			\vertex (d) at (3.5,0) {\(\psi_{\gamma}(\tvh)\)};
			\vertex (e) at (2.87,-1.7) {\(\psi_{\gamma}(\tvk-\tvq-\tvh)\)};
			\diagram* {
				(a) -- [fermion] (b) -- (c),
				(b) -- (d),
				(b) -- (e),
			};
		\end{feynman}
	\end{tikzpicture}
\end{equation}
with the following non-vanishing diagram contributing to it
\begin{figure}[h!]
	\begin{subfigure}[c]{0.15\textwidth}
		\centering
		\includegraphics[page=3, width=1\linewidth]{figure_SI/Vvvvv.pdf}
	\end{subfigure}
	\begin{subfigure}[c]{0.15\textwidth}
		\centering
		\includegraphics[page=4, width=1\linewidth]{figure_SI/Vvvvv.pdf}
	\end{subfigure}
	\begin{subfigure}[c]{0.15\textwidth}
		\centering
		\includegraphics[page=5, width=1\linewidth]{figure_SI/Vvvvv.pdf}
	\end{subfigure}
	\begin{subfigure}[c]{0.15\textwidth}
		\centering
		\includegraphics[page=6, width=1\linewidth]{figure_SI/Vvvvv.pdf}
	\end{subfigure}
	\begin{subfigure}[c]{0.15\textwidth}
		\centering
		\includegraphics[page=7, width=1\linewidth]{figure_SI/Vvvvv.pdf}
	\end{subfigure}\\
	\begin{subfigure}[c]{0.15\textwidth}
		\centering
		\includegraphics[page=8, width=1\linewidth]{figure_SI/Vvvvv.pdf}
	\end{subfigure}
	\begin{subfigure}[c]{0.15\textwidth}
		\centering
		\includegraphics[page=9, width=1\linewidth]{figure_SI/Vvvvv.pdf}
	\end{subfigure}
	\begin{subfigure}[c]{0.15\textwidth}
		\centering
		\includegraphics[page=10, width=1\linewidth]{figure_SI/Vvvvv.pdf}
	\end{subfigure}
	\begin{subfigure}[c]{0.15\textwidth}
		\centering
		\includegraphics[page=11, width=1\linewidth]{figure_SI/Vvvvv.pdf}
	\end{subfigure}\\
	\begin{subfigure}[c]{0.15\textwidth}
		\centering
		\includegraphics[page=12, width=1\linewidth]{figure_SI/Vvvvv.pdf}
	\end{subfigure}
	\begin{subfigure}[c]{0.15\textwidth}
		\centering
		\includegraphics[page=13, width=1\linewidth]{figure_SI/Vvvvv.pdf}
	\end{subfigure}
	\begin{subfigure}[c]{0.15\textwidth}
		\centering
		\includegraphics[page=14, width=1\linewidth]{figure_SI/Vvvvv.pdf}
	\end{subfigure}
	\begin{subfigure}[c]{0.15\textwidth}
		\centering
		\includegraphics[page=15, width=1\linewidth]{figure_SI/Vvvvv.pdf}
	\end{subfigure}\\
	\begin{subfigure}[c]{0.15\textwidth}
		\centering
		\includegraphics[page=16, width=1\linewidth]{figure_SI/Vvvvv.pdf}
	\end{subfigure}
	\begin{subfigure}[c]{0.15\textwidth}
		\centering
		\includegraphics[page=17, width=1\linewidth]{figure_SI/Vvvvv.pdf}
	\end{subfigure}
	\begin{subfigure}[c]{0.15\textwidth}
		\centering
		\includegraphics[page=18, width=1\linewidth]{figure_SI/Vvvvv.pdf}
	\end{subfigure}
	\begin{subfigure}[c]{0.15\textwidth}
		\centering
		\includegraphics[page=19, width=1\linewidth]{figure_SI/Vvvvv.pdf}
	\end{subfigure}\\
	\begin{subfigure}[c]{0.15\textwidth}
		\centering
		\includegraphics[page=20, width=1\linewidth]{figure_SI/Vvvvv.pdf}
	\end{subfigure}
	\begin{subfigure}[c]{0.15\textwidth}
		\centering
		\includegraphics[page=21, width=1\linewidth]{figure_SI/Vvvvv.pdf}
	\end{subfigure}
	\begin{subfigure}[c]{0.15\textwidth}
		\centering
		\includegraphics[page=22, width=1\linewidth]{figure_SI/Vvvvv.pdf}
	\end{subfigure}
	\begin{subfigure}[c]{0.15\textwidth}
		\centering
		\includegraphics[page=23, width=1\linewidth]{figure_SI/Vvvvv.pdf}
	\end{subfigure}\\
	\begin{subfigure}[c]{0.19\textwidth}
		\centering
		\includegraphics[page=1,width=1\linewidth]{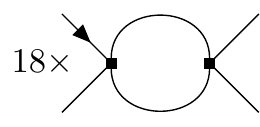}
	\end{subfigure}
	\begin{subfigure}[c]{0.22\textwidth}
		\centering
		\includegraphics[page=2, width=1\linewidth]{figure_SI/Vvvvv.pdf}
	\end{subfigure}
	\caption{Diagrams contributing to leading order of $V^{\hpsi\psi\psi\psi}$}
	\label{fig:Vvvvv}
\end{figure}

\newpage

\subsubsection{Vertex function of the DYS vertex of $s$}
The vertex function $V^{\hs\psi\psi\psi\psi}$ represents the corrections to the DYS vertex in the equation of $s$, and is given by
\begin{equation}
	V_{\alpha\beta\gamma\nu\sigma\tau}^{\hs \psi\psi\psi\psi} (\tvk,\tvq,\tvh,\tvp)=
	\begin{tikzpicture}[baseline=(a.base)]
		\begin{feynman}
			\vertex (a) at (0,0) {\(\hs_{\alpha\beta}(-\tvk)\)};
			\vertex (b) [blob] at (2,0) {};
			\vertex (c) at (1.44,1.71) {\(\psi_\gamma(\tvq)\)};
			\vertex (d) at (3.53,1.12) {\(\psi_\nu(\tvh)\)};
			\vertex (e) at (3.53,-1.12) {\(\psi_\sigma(\tvp)\)};
			\vertex (f) at (1.44,-1.71) {\(\psi_\tau(\tvk-\tvq-\tvh-\tvp)\)};
			\diagram* {
				(a) -- [charged boson] (b) -- (c),
				(b) -- (d),
				(b) -- (e),
				(b) -- (f),
			};
		\end{feynman}
	\end{tikzpicture}
\end{equation}
with the following non-vanishing diagram contributing to it
\begin{figure}[h!]
	\centering
	\begin{subfigure}[c]{0.25\textwidth}
		\centering
		\includegraphics[page=1,width=1\linewidth]{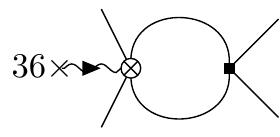}
	\end{subfigure}
	\begin{subfigure}[c]{0.2\textwidth}
		\centering
		\includegraphics[page=2, width=1\linewidth]{figure_SI/Vsvvvv.pdf}
	\end{subfigure}
	\begin{subfigure}[c]{0.2\textwidth}
		\centering
		\includegraphics[page=3, width=1\linewidth]{figure_SI/Vsvvvv.pdf}
	\end{subfigure}
	\caption{Diagrams contributing to leading order of $V^{\hs \psi\psi\psi \psi}$}
	\label{fig:Vsvvvv}
\end{figure}

\newpage

\section{Assessing the relevance of activity in natural swarms} \label{react}

In systems of active individuals interacting via local alignment there are two distinct mechanisms allowing propagation of directional information through the group. The first consists in passing the information from neighbour to neighbour via local interaction links: this mechanism would be present even at equilibrium, i.e. for a Heisenberg model on a fixed network. The second is due to the individuals' movement, which carry the information along with their motion causing a rearrangement of the interaction network in time: this second (activity-related) mechanism would be absent in standard ferromagnets and it is an out-of-equilibrium feature. A quantitative way to assess the role of activity in determining the relaxation properties of the system is therefore to compare the typical timescales related to the rearrangement of the interaction network and the relaxation of directional observables. In order to do so, we consider the following two correlation functions.

To characterize the dynamical evolution of the network we define the network overlap function $C_{\rm net}(r,t)$ as,
\begin{equation}
C_{\rm net}(r,t)=\frac{1}{N}\sum_i \bigg\langle \frac{1}{n_i(r,t_0)}\sum_j n_{ij}(r,t_0)n_{ij}(r,t_0+t) \bigg\rangle_{t_0} \ ,
\label{cnet}
\end{equation}
where the matrix $n_{ij}(r,t)$ is equal to $1$ if the two individuals $i$ and $j$ are at mutual distance $r_{ij}<r$ at time $t$ and zero otherwise, $n_i(r, t_0)=\sum_j n_{ij}(r,t_0)$ represents the number of neighbors in a region of size $r$ around individual $i$ at time $t_0$, \textcolor{black}{while $N$ is the total number of individuals} and $\langle\cdots\rangle_{t_0}$ represents a time average over $t_0$. The function $C_{\rm net}(r,t)$ measures how much on average a neighborhood  of size $r$ of a given individual remains the same in a time interval $t$. For systems with a metric interaction - like swarms are \cite{attanasi2014collective} - if we set $r$ equal to the interaction range $r_c$, then $C_{\rm net}(r_c,t)$ measures the reshuffling of the interaction neighborhood (i.e. how many individuals interacting at time $t_0$ are still interacting at time $t_0+t$).

To characterize the directional relaxation we consider the space-time correlation function of velocity fluctuations. In the context of the field theory described in the main text, this function would be given by the connected correlations of the field $ \vpsi(\vx,t)$. However, if we want to measure such quantity on real data, it is convenient to define it in terms of the individual velocities:
\textcolor{black}{
\begin{equation}
	C(r,t) = \left\langle  
	\frac{
		\sum_{i,j}  \delta\hat{\bf v}_i(t_0) \cdot \delta\hat{\bf v}_j(t_0+t) \, \delta[r - r_{ij}(t_0,t)] 
	}{
		\sum_{i,j}  \, \delta[r - r_{ij}(t_0,t)] 
	}
	\right\rangle_{t_0} \ ,
	\label{cdir}
\end{equation}
where $r_{ij}(t_0,t) = |{\bf r}_i(t_0)-{\bf r}_j(t_0+t)|$ is the mutual distance.The fluctuations $\delta\hat{\bf v}_{i}$ are the dimensionless velocity fluctuations, defined as
\begin{equation}
	\delta \hat{\bf v}_i \equiv \frac{\delta{\bf v}_i}{\sqrt{\frac{1}{N} \sum_k \delta{\bf v}_k \cdot \delta{\bf v}_k}}  \ ,
	\nonumber
\end{equation}
where $\delta{\bf v}_i = {\bf v}_i - {\bf V}$ indicates the individual fluctuation with respect to the collective velocity of the swarm ${\bf V}$ that takes into account global translation, rotation and dilation modes. Note that in Fig.3b of the main, we plot the normalized correlation function
\begin{equation}
\hat C(r,t)=\frac{C(r,t)}{C(r,t=0)}
\end{equation}.}

\begin{figure}[t]
	\centering
	\includegraphics[width=0.5\textwidth]{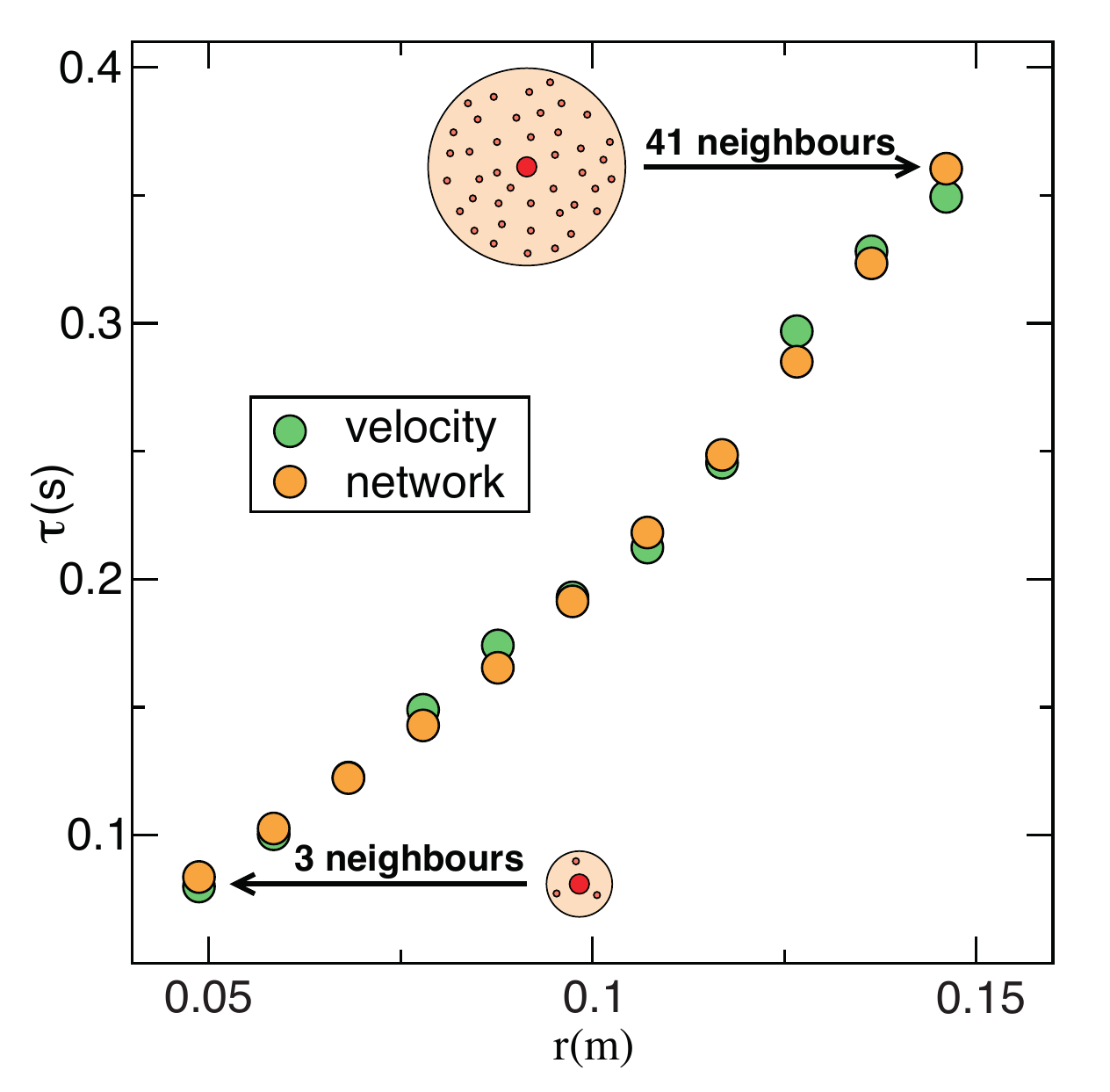}
	\caption{\textcolor{black}{{\bf Natural swarms are strongly active.} We report here the network relaxation time and the velocity relaxation time, calculated at various spatial scales $r$, in natural swarms. The two relaxational processes have the same time scale at all values of $r$, thus indicating that the reshuffling of the interaction network occurs on the same time scale as velocity relaxation. We conclude that natural swarms are highly active systems. }}
	\label{fig:manazza}
\end{figure}

\textcolor{black}{These two correlation functions can be easily computed from the experimental and the numerical data, and compared with each other; their characteristic timescales (or relaxation times) can be computed through the formula introduced by Halperin and Hohenberg in \cite{HH1969scaling} (see equation \eqref{taudata}), which provides a reliable evaluation of $\tau$ independently of the functional form of the correlation function. 
The question then is: over what spatial scale should we compare network vs velocity relaxation?
Ideally, one would like to compare these two curves over the scale of the interaction range, to check whether relaxation is affected by activity even locally. 
Previous experiments \cite{attanasi2014collective} indicate that interaction in natural swarms is metric, with an interaction range $r_c$ of a few centimetres. However, since that estimate is far from precise, here we compare network and velocity relaxation at several spatial scales, from $r=5$cm up to $r=15$cm. To have a better intuition of what these scales entail, we note that $r=5$cm corresponds to a neighbourhood of approximately 3 individuals, while $r=15$cm to one of 41 individuals.
In Fig.\ref{fig:manazza} we show the behaviour of $\tau$ for the network and the \textcolor{black}{normalized} velocity correlations as a function of the spatial scale $r$: 
we can see that in the entire considered range the characteristic timescale is the same in the two cases. Given that the correlation functions quantify, respectively, how quickly the network changes on scale $r$, and how quickly directional correlations decay on the same scale, we conclude that activity is always strong in natural swarms, certainly affecting  the relaxation properties of the system. }

\textcolor{black}{The experimental information that network and velocity relaxation times are the same in natural swarms, was used to fix the level of activity in numerical simulations
of the Inertial Spin Model (ISM); in the model, the interaction range $r_c$ (and therefore the mean number of interacting neighbours $n_c$) is a parameter set by us, hence we can  calculate the two correlation functions in ISM simulations over the exact scale of the interaction range, and increase the particles speed $v_0$ until the two timescales are approximately the same, to be sure that numerical simulations -- as natural swarms --  are in the active regime. In Fig.3b of the main text we display the behaviour of both $C_\mathrm{net}(r,t)$ and $\hat C(r,t)$ in simulation, at $r=r_c$, corresponding to $n_c = 18$ interacting neighbours; for a comparison, in the same figure we report the two correlation functions in natural swarms, at the same value of $n_c$; however, as we have seen from Fig.\ref{fig:manazza}, in natural swarms the two timescales are the same over a very wide range of scales.
}

\newpage
\textcolor{black}{
\section{Determination of the dynamic exponent from the experimental data}
To compare our theoretical RG prediction of the dynamic critical exponent $z$ with experiments on swarms in the field we must provide a robust experimental evaluation of $z$. To understand what is the best way to experimentally infer $z$, let us go back to the dynamic scaling hypothesis, which links the relaxation time of the mode at wave-vector $k$, namely $\tau_k$, to the correlation length of the system $\xi$,
\begin{equation}
	\tau_k=\xi^z\Omega(k\xi)\ .
	\label{scaling}
\end{equation}
To fit $z$ from experimental data, we proceed as follows. For each swarm, we compute the relaxation time at $k=\xi^{-1}$ \cite{cavagna2017swarm}, so that the scaling law \eqref{scaling} becomes,
\begin{equation}
	\tau=\Omega(1) \xi^z\propto \xi^z\ .
\end{equation}
where $\tau\equiv\tau_{k=\xi^{-1}}$.
In log-log scale, this becomes a linear relation between $\log\xi$ vs $\log\tau$. The exponent can be thus obtained as the slope of the regression line fitted on $\log\tau$ vs $\log\xi$:
\begin{equation}
	\log\tau=z\log\xi + c
	\label{giggino}
\end{equation}
Hence, to fit $z$ we need a robust estimate of $\xi$ and $\tau$.
}

\textcolor{black}{
\subsection{Evaluation of $\xi$ and $\tau$}\label{errors}
In this Section we describe how the correlation function is computed from the trajectories of midges, and then how we can extract the correlation length and the relaxation time from the data. This is the same procedure used in \cite{cavagna2017swarm}. For each swarming event, the best estimate of $\xi$, $\tau$ and of their experimental errors can be found in Table \ref{data}.
}

\begin{figure}[t]
	\centering
	\includegraphics[width=1\textwidth]{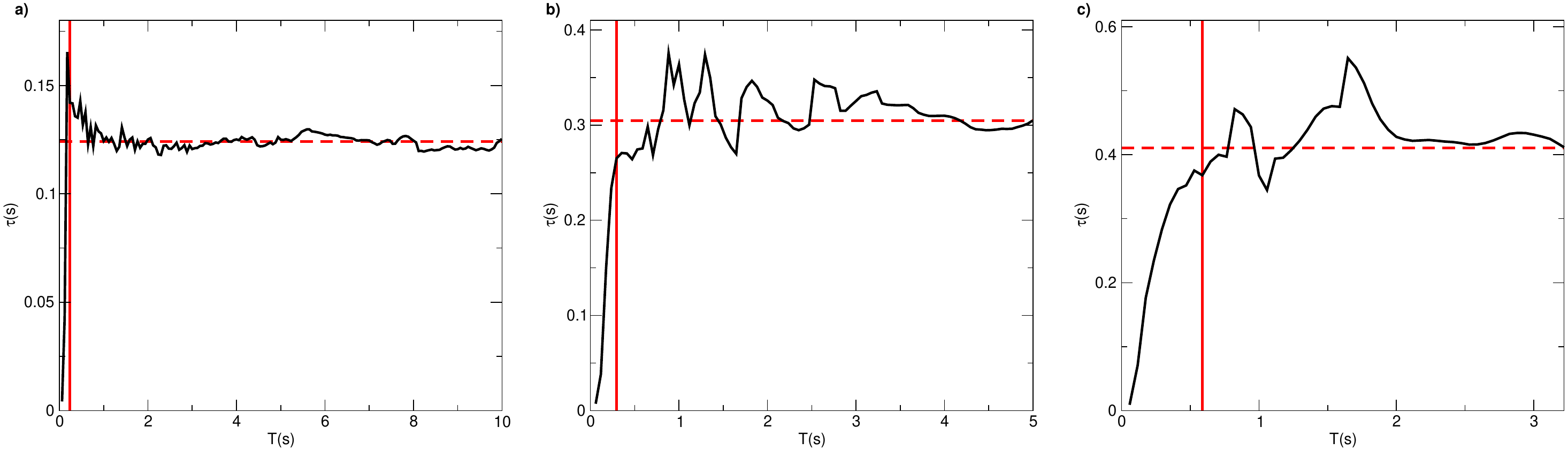}
	\caption{\textcolor{black}{{\bf Time series of $\tau$ vs $T$ for three different swarming events.} The estimation of $\tau$ as a function the time $T$ over which temporal averages are computed is plotted for three different swarming events. The horizontal dashed line represents our best estimation of $\tau$, obtained from Eq. \eqref{taudata} using the full temporal range. This corresponds to the last point of the series, namely $\tau\left(T=t_\mathrm{max}\right)$. The vertical red line indicates the value of $T^*$ starting from which data is used to compute the variance, and therefore the error $\delta\tau_\mathrm{TS}$. This value is chosen by eye as the value at which $\tau\left(T\right)$ reaches its plateau. {\bf a)} Data from swarming event 20120918\_ACQ3; {\bf b)} Data from swarming event 20150924\_ACQ3; {\bf c)} Data from swarming event 20120704\_ACQ1.}}
	\label{fig:tau}
\end{figure}

\textcolor{black}{
Let us here remind that we are dealing with experimental data in which time is intrinsically discrete, where the temporal resolution $\Delta t$ is fixed by the experimental setup. Data presented in this work are collected with cameras shooting at $170\,\mathrm{fps}$ (see Methods), thus $\Delta t=170^{-1}\,\mathrm{s}$. In the following data analysis, time is therefore treated as {\it discrete}, namely measured in units of $\Delta t$. Hence, it will typically range from $1$ to the length of tha data acquisition $t_{\mathrm{max}}$. The only exception to this convention are plots in Fig.\ref{fig:tau}, where we use time measured in seconds.}

\textcolor{black}{
\subsubsection{Correlation functions}
As a starting point, we define the dimensionless velocity fluctuations as,
\begin{equation}
	\delta \hat{\bf v}_i \equiv \frac{\delta{\bf v}_i}{\sqrt{\frac{1}{N} \sum_k \delta{\bf v}_k \cdot \delta{\bf v}_k}}  \ ,
	\label{fluctu}
\end{equation}
where,
$\delta {\bf v}_i \equiv {\bf v}_i - {\bf V} $ and ${\bf V}$ is the collective velocity of the swarm which takes into account global translation, rotation and dilation modes, see \cite{attanasi2014finite}.
The spatio-temporal correlation function is the time generalization of the static space correlation function previously studied in \cite{cavagna2010scalefree, attanasi2014finite,attanasi2014collective},
\begin{equation}
	C(r,t) = \left\langle  
	\frac{
		\sum_{i,j}^N  \delta\hat{\bf v}_i(t_0) \cdot \delta\hat{\bf v}_j(t_0+t) \, \delta[r - r_{ij}(t_0,t)] 
	}{
		\sum_{i,j}^N  \, \delta[r - r_{ij}(t_0,t)] 
	}
	\right\rangle_{t_0} \ ,
	\nonumber
	\label{mingus}
\end{equation}
where $r_{ij}(t_0,t) = |{\bf r}_i(t_0)-{\bf r}_j(t_0+t)|$ and the positions are calculated with respect to the center of mass of the swarm, that is ${\bf r}_i(t_0) = {\bf R}_i(t_0) -{\bf R}_{\mathrm{CM}}(t_0)$, and $N$ is the total number of individuals.}
\textcolor{black}{
The brackets $\langle \dots \rangle_{t_{0}}$ indicate an average over time, 
\begin{equation}
	\langle f(t_0, t)\rangle_{t_0}= \frac{1}{t_\mathrm{max}-t} \sum_{t_{0}=1}^{t_\mathrm{max}-t} f(t_{0}, t) \ ,
\end{equation}
where $t_\mathrm{max}$ is the total available time in the simulation or in the experiment and $\Delta t$ is the time interval between two frames.
}
\textcolor{black}{
The purpose of $C(r,t)$ is to measure how much a  change of velocity of an individual at time $t_0$ influences a change of velocity of another individual at distance $r$ at a later time $t_0+t$. The (dimensionless) correlation function in Fourier space is given by,
\begin{equation}
	C(k, t) = \rho \int d{\bf r} \ e^{i {\bf k}\cdot{\bf r}} C(r,t) \ .
\end{equation}
}\textcolor{black}{
with $\rho$ being the average density.}
\textcolor{black}{
By using the definition of $C(r,t)$ and the approximation $\sum_{i,j}^N  \, \delta[r - r_{ij}(t_0,t)] \sim 4\pi r^2 \rho N$ in the integral, we obtain,
\begin{equation}
	\begin{split}
	C(k,t) 
	&= \left\langle \frac{1}{N} \sum_{i,j}^N \;  \int_{-1}^{+1} d(\cos \theta) e^{ikr_{ij} \cos(\theta)}   \ \delta\hat{\bf v}_i(t_0) \cdot \delta\hat{\bf v}_j(t_0+t) \, \right\rangle_{t_0} =
	\\
	&= \left\langle \frac{1}{N} \sum_{i,j}^N \;  \frac{\sin(k\,r_{ij}(t_0,t))}{k\,r_{ij}(t_0,t)} \ \delta\hat{\bf v}_i(t_0) \cdot \delta\hat{\bf v}_j(t_0+t) \, \right\rangle_{t_0} \ ,
	\end{split}
	\label{cicciobello}
\end{equation}
which is the correlation function that we compute experimentally in the present work. Notice that, by definition, $\sum_i \delta \hat{\bf v}_i =0$; due to this sum rule we obtain $C(k=0,t)=0$. Hence, $k=2\pi/L$  is the smallest non-trivial value of the momentum at which we can evaluate the correlation.
}

\textcolor{black}{
\subsubsection{Correlation length}\label{xi}
To compute the correlation length, $\xi$, we can directly work in $k$ space. The static correlation function, $C_0(k)\equiv C(k,t=0)$, is,
\begin{equation}
	C_0(k)= \left\langle \frac{1}{N} \sum_{i,j}^N \;  \frac{\sin(k\,r_{ij})}{k\,r_{ij}} \ \delta\hat{\bf v}_i(t_0) \cdot \delta\hat{\bf v}_j(t_0) \, \right\rangle_{t_0} \ .
	\label{galore}
\end{equation}
where now both $i$ and $j$ are evaluated at equal time, $t_0$. }\textcolor{black}{For $k\to\infty$, only the self-correlations contribute to $C_{0}$ since $\sin(k\,r_{ij}) / (k\,r_{ij})\to \delta_{ij}$, and hence $C_{0}\left(k\right)\to 1$. }\textcolor{black}{
By decreasing $k$ we are averaging over larger length scales, therefore adding to \eqref{galore} more
correlated pairs, making $C_0(k)$ increase. 
When the momentum arrives at $k\sim 1/\xi$, we start adding uncorrelated pairs, hence, $C_0(k)$ must level. If we further decrease $k$ and reach $1/L$ (where $L$ is the system's size) we start to be affected by the sum rule, $C_0(k=0)=0$, hence the static correlation $C_0(k)$ decreases, until eventually it vanishes for $k=0$ \cite{cavagna2016spatio}.
In a system where $\xi \ll L$ the static correlation therefore has -- in log scale -- a broad plateau between $k\sim1/\xi$ and $k\sim1/L$. However, natural swarms are scale-free systems, where $\xi\sim L$ \cite{attanasi2014finite}; in this case, $C_0(k)$ has a well-defined maximum at $k_\mathrm{max} \sim 1/\xi \sim 1/L$. This is a very practical way to evaluate $\xi$ if one is already working in $k$ space and it is the one we used in \cite{cavagna2017swarm}. Alternatively, in the scale-free regime, one could define $\xi$ as the point where the static correlation in $r$ space $C_0(r)=C(r,t=0)$
reaches zero, $C_0(r =\xi )=0$. These two definitions of $\xi$ are consistent with each other, as shown in \cite{cavagna2017swarm}.
}

\textcolor{black}{
The experimental evaluation of the correlation function is not exempted from experimental errors, which in turn affect our determination of $\xi$.
} \textcolor{black}{
Our best estimate of $\xi$ is obtained by looking for the value of $k$, $k_\mathrm{max}$, correspondent to the maximum of $C_0(k)$, which is computed by averaging Eq. \eqref{galore} over $t_{0}$ using the full time interval $t_{\mathrm{max}}$ experimentally available, namely by using all the $t_{\mathrm{max}}$ frames. To associate an experimental error $\delta k_\mathrm{max}$ to this evaluation of $k_\mathrm{max}$, we use a resampling method. Among the $t_{\mathrm{max}}$ different frames, we randomly sort half of them and compute $k_\mathrm{max}$ from the sub-sample.
} \textcolor{black}{
We repeat this $1000$ times, thus working out the experimental distribution of $k_\mathrm{max}$. The standard deviation of this distribution is our estimate of the error $\delta k_\mathrm{max}$. The experimental error associated to $\log\xi$ is given by $\delta\xi/\xi=\delta k_\mathrm{max}/k_\mathrm{max}$, since $\xi=k_\mathrm{max}^{-1}$. The relative error $\delta\xi/\xi$ evaluated in this way is given in Table \ref{data}. As we shall see in the next section, assigning an experimental error to the relaxation time is far less straightforward. 
}

\textcolor{black}{
\subsubsection{Relaxation time}\label{errtime}
To calculate the relaxation time at wavevector $k$, $\tau_k$, we use the formula \cite{HH1969scaling},
\begin{equation}
	\int^{\infty}_{0} \frac{dt}{t} \, \sin(t/\tau_k)  \hat C(k,t)= \pi/4 \ ,
	\label{tau}
\end{equation}
where $\hat C(k,t)=C(k,t)/C(k,t=0)$ is the normalized correlation function. Notice that this is nothing else than the time domain version of the classic definition of the characteristic frequency as the half-width of the correlation function in the $\omega$ domain \cite{HH1969scaling}.
For a purely exponential correlation, $\tau_k$ coincides with the exponential decay time, while for more complex functional forms, $\tau_k$ is the most relevant time scale of the system. Relation \eqref{tau} gives an estimate of $\tau_k$ that is more robust than simply crossing $\hat C(k,t)$ with a constant and more reliable than a fit, as it does not require a priori knowledge of the functional form of $\hat C(k,t)$.
Dealing with real data, to estimate $\tau=\tau_{k=\xi^{-1}}=\tau_{k_\mathrm{max}}$ we discretize the integral \eqref{tau} approximating it with trapezoids and hence numerically solve:
} \textcolor{black}{
\begin{equation}
	\frac{1}{2}\frac{\Delta t}{\tau}+\sum_{t=1}^{t_{\mathrm{max}}} \frac{\Delta t}{t} \, \sin(t/\tau)  \hat C(k_\mathrm{max},t)= \pi/4 \ ,
	\label{taudata}
\end{equation}
where we remind that $t_\mathrm{max}$ is the time duration of the event of interest while $\Delta t$ is the time resolution.
}

\textcolor{black}{
How can we be sure that the value of $\tau$ so determined is a fair estimate of the {\it actual} relaxation time? Whenever we estimate a relaxation time, we must be very careful about the balance between $\tau$ and the total length of the time series, $t_\mathrm{max}$. Ideally, one would like to have an infinitely long time series, which would give a perfect determination of the correlation function; let us call the relaxation time relative to this ideal correlation function, $\tau_\infty$. In the real cases, $t_\mathrm{max}<\infty$, so that in general $\tau(t_\mathrm{max}) \neq \tau_\infty$. Although the way in which $\tau(t_\mathrm{max})$ converges to  $\tau_\infty$ for $t_\mathrm{max}\to\infty$ depends on the specific form of the correlation function, there are some general traits: }\textcolor{black}{when $t_\mathrm{max} < \tau_\infty$ the estimated relaxation time $\tau$ is strongy biased by the shortness of the interval, 
while it saturates to $\tau_\infty$ when $t_\mathrm{max} \gg \tau_\infty$}\textcolor{black}{; hence, this saturation can (and must) be used as a check that $\tau$ is a fair estimate of $\tau_\infty$. This is what we have done in our data.
Given that $t_\mathrm{max}$ is fixed by the experiment, to do this test we calculated the relaxation time from Eq. \eqref{taudata} with $t_\mathrm{max}$ replaced by various observation times $T\leq t_\mathrm{max}$, with increment }\textcolor{black}{$\Delta T=10$ frames. }\textcolor{black}{Whenever Eq.\eqref{taudata} has no solution, we take the boundary value, $\tau\left(T\right)=T$.
For $T<t_\mathrm{max}/2$, the relaxation time can be computed in several non-overlapping intervals, hence we use their average as an estimate of $\tau(T)$.  In Fig.\ref{fig:tau} we report $\tau\left(T\right)$ for three different swarming events: because we see a rather clear saturation of $\tau\left(T\right)$ for larger $T$, we trust the value of the relaxation time.
}

\textcolor{black}{
Given the procedure to calculate $\tau$, it should be clear that the estimate of its experimental uncertainty, $\delta\tau$, is way more problematic than for $\xi$. The main point is that, because we need the whole time series to evaluate $\tau$, we cannot use temporal sampling to estimate $\delta\tau$, as we instead did for $\delta\xi$. Moreover, unlike what one would do in a numerical simulation, we do not have different independent samples of the same swarm to make statistics. Admittedly, the procedure to assign an error on $\tau$ is rather arbitrary, as it strongly depends on our prior about what is the main source of experimental variability of $\tau$; unfortunately, while different methods to estimate $\tau$ give very similar results, strongly correlated to each other, different procedures to estimate $\delta\tau$ have far greater variability and weaker correlation. We propose in the following two different methods to estimate the experimental error on $\tau$; the two methods will give results that, although not outlandishly so, {\it are} in fact quite different from each other. This will hopefully convince the reader of the difficulties in the determination of the experimental error on the relaxation time, and of the reason why we ultimately decided to use a fitting method for $z$ that does {\it not} require the experimental errors as an input.
}


\textcolor{black}{
\vskip 0.5 truecm
{\it *** Experimental error from the Time Series - $\delta\tau_\mathrm{TS}$}
\vskip 0.2 truecm
As we explained above, when the duration of the time series increases, $\tau$ eventually reaches a plateau (Fig.\ref{fig:tau}); but of course, there are fluctuations around this saturation value, fluctuations that should become smaller and smaller the larger $t_\mathrm{max}$ is. We use these fluctuations from the time series around the plateau to estimate the error, $\delta\tau_\mathrm{TS}$; more precisely, we calculate the variance of the fluctuations of $\tau$ around the plateau. There is however a significant problem with this definition of the error: to apply the procedure we need to define a time
$T^*$ at which $\tau(T)$ has reached the plateau, so we can sample the fluctuations of $\tau(T)$ only for $T\geq T^*$; but deciding where to locate $T^*$ is not straightforward  and the reason can be understood from Fig. \ref{fig:tau}: the relatively large fluctuations on $\tau$ and the different shapes of $\tau\left(T\right)$ make it not possible to find a safe universal criterion to decide when the plateau is actually reached. But unfortunately, the final value of $\delta\tau_\mathrm{TS}$ depends rather strongly on $T^*$. This problem is particularly severe because -- even though we only consider swarms that do reach a plateau -- in some cases we do not have a very large interval of time on the plateau. These problems notwithstanding, we have estimated to the best of our judgement $\delta\tau_\mathrm{TS}$ in all swarms, and the results are reported in Table \ref{data}.
}

\textcolor{black}{
\vskip 0.5 truecm
{\it *** Experimental error from the Correlation Functions - $\delta\tau_\mathrm{CF}$}
\vskip 0.2 truecm
The second method consists in propagating the error in the determination of the correlation function $C(t)$, by using Eq. \eqref{taudata}. To do so, let us call
}\textcolor{black}{
\begin{equation}
	I\left(\tau\right)=\left[\frac{1}{2\tau}+\frac{1}{C_{0}} \sum_{t=1}^{t_\mathrm{max}}\frac{1}{t} \, \sin(t/\tau)\,C_{t}\right]\Delta t
\end{equation}
}\textcolor{black}{
the quantity on the l.h.s. of Eq. \eqref{taudata}, where }\textcolor{black}{$C_{t}\equiv C(k_\mathrm{max}, t)$. }\textcolor{black}{A variation $\delta C$ and $\delta\tau$ would affect the quantity $I$ as
}\textcolor{black}{
\begin{equation}
	\delta I=\frac{\partial I}{\partial \tau}\delta\tau + \frac{\partial I}{\partial C_{0}}\delta C_{0}+\sum_{t=1}^{t_\mathrm{max}} \frac{\partial I}{\partial C_{t}}\delta C_{t}
\end{equation}
}\textcolor{black}{
However, from Eq. \eqref{taudata} we know that $I\left(\tau\right)=\pi/4$, meaning that $\delta I= 0$ by definition. The variations $\delta C$ and $\delta\tau$ cannot be independent from each other, but they must keep $\delta I=0$. We can therefore express $\delta\tau$ as a function of $\delta C$ at different times. After some algebraic manipulation, we obtain that
}\textcolor{black}{
\begin{equation}
	\left| \frac{\delta\tau_\mathrm{CF}}{\tau}\right| \leq 
	2 \frac{ \sum_{t=1}^{t_\mathrm{max}}
	\left| \frac{\sin(t/\tau)}{t/\tau}  \delta C_{t}\right|
	+\left| \delta C_{0}
	\sum_{t=1}^{t_\mathrm{max}}
	\frac{\sin(t/\tau)}{t/\tau}\frac{C_{t}}{C_{0}} \right|}{ \left| C_{0} +
	2 \sum_{t=1}^{t_\mathrm{max}} \cos(t/\tau)  C_{t} \right|}
	\label{erroraccio}
\end{equation}
}\textcolor{black}{
To avoid an underestimation of the uncertainty on $\tau$, we approximate $\delta\tau_\mathrm{CF}$ with the upper bound given by Eq. \eqref{erroraccio}. 
Errors of the correlation function, $\delta C_{n}$ are evaluated as follows: reminding that $C(k,t)$ is computed as a temporal average - see Eq. \eqref{mingus}, \eqref{cicciobello} - we define the experimental error $\delta C_{n}$ as the standard error associated to this temporal average.
}

\textcolor{black}{
Even though this method to assess $\delta\tau$ may seem quite promising, as it does not require any arbitrary definition (in contrast with the choice of $T^*$ in the previous method), it has its own shortcomings. The correlation function $C(t)$ is the more unreliable the closer $t$ is to $t_\mathrm{max}$, as the number of temporal pairs of frames used to compute $C(k,t)$ in Eq. \eqref{mingus} is given by $t_\mathrm{max}-t$; hence, one would expect the actual uncertainty on the correlation to be larger at larger $t$. But in fact, this is not necessarily what happens to the standard error: $\delta C_{t}$ is obtained by sampling the distribution of the correlation at that time, but this sampling is very imprecise if the number of pairs is very limited. Moreover, in the determination of the standard error on $C(k,t)$ one is implicitly assuming that all the temporal pairs over which we are averaging are uncorrelated, which is clearly blatantly false. While this does not impact on the me an, that is on $C(k,t)$ itself, it certainly impacts -- and a lot -- on the determination of the variance, namely on $\delta C_{t}$, and therefore in turns on $\delta\tau_\mathrm{CF}$.}

\textcolor{black}
{In conclusion, even though both definitions of the error on the relaxation time seem reasonable, they both have some critical issues; their values are reported in Table \ref{data}, where one can see that -- despite some correlation -- they are in fact quite different from each other. If the value of the experimental errors, $\delta\xi$ and $\delta\tau$ only affected the error on the critical exponent, $\delta z$, we would not be too worried; however, with certain fitting methods the experimental errors affect the value of $z$ itself, which is quite another story. Because of this, we ultimately decided to rely on a method (RMA) that does {\it not} need the errors as an input, hence giving a value of $z$ that is not affected by our estimate of $\delta\xi$ and $\delta\tau$. Nevertheless, we describe in the next Sections also a method that does use $\delta\xi$ and $\delta\tau$ to estimate $z$ (EV) and see how it performs with the two different definitions of $\delta\tau$ given above. Fortunately, we will see that all three estimates (RMA, EV with $\delta\tau_\mathrm{TS}$ and EV with $\delta\tau_\mathrm{CF}$) are all quite compatible with each other.
}

\textcolor{black}{
\subsection{Different fitting methods}
}

\textcolor{black}{
\subsubsection{The problem with Least Squares}
The Least Squares (LS) regression method is perhaps the most common and used regression method to fit experimental data \cite{taylor1997introduction}. This has also been the method with which previous determinations of the dynamic critical exponent $z$ in natural swarms were made \cite{cavagna2017swarm}. Its wide use is due to the simplicity of its assumptions and to the fact that the equations to solve in order to find the coefficients of the fitting line are linear. LS works under the hypothesis of zero experimental errors on one of the two variables, which then takes the role of independent variable $x$, and uniform variance $\sigma_y^2$ for the second variable, which takes the role of dependent variable $y$. Note that, although with LS the variable $y$ can be affected by experimental uncertainty, the actual errors are not needed as an input, hence results given by LS do not depend on the protocol used to evaluate the errors. The best fit line $y=\alpha x+\beta$ is obtained by minimizing the average square scatter $\Delta y_i^2$ of the points from the line in the direction of $y$ only, where
\begin{equation}
	\Delta y_i=y_i-\alpha x_i -\beta
\end{equation}
This translates in the minimization of the following quantity
\begin{equation}
	\Sigma\left(\alpha,\beta\right)=\frac{1}{N}\sum_{i=1}^N \Delta y_i^2=\frac{1}{N}\sum_{i=1}^N \left(y_i-\alpha x_i -\beta\right)^2
\end{equation}
which can be done explicitly by solving $\partial_\alpha\Sigma=\partial_\beta\Sigma=0$. The system that has to be solved to find the minimum point is given by
\begin{align}
	\partial_\alpha \Sigma&=\frac{2}{N}\sum_{i=1}^N\alpha x_i^2+\beta x_i-x_iy_i=0\\
	\partial_\beta \Sigma &=-\frac{2}{N}\sum_{i=1}^N \left(y_i-\alpha x_i -\beta\right)=0
\end{align}
The linear system in $\alpha,\,\beta$ can be easily solved, leading to
\begin{align}
	\alpha&=\frac{\E{xy}-\E{x}\E{y}}{\E{x^2}-\E{x}^2}\label{LSa}\\
	\beta&=\E{y}-\alpha\E{x}
	\label{LSb}
\end{align}
where
\begin{equation}
	\E{g(x,y)}=\frac{1}{N}\sum_{i=1}^N g(x_i,y_i)
\end{equation}
When LS is applied to the data of Table \ref{data}, we obtain
\begin{equation}
	z_\mathrm{LS}=1.16\pm0.12 \ ,
\end{equation}
where the error is obtained through a resampling method, namely by performing many times the fit over half of the data and taking the square root of the variance of $z$ (see main text). 
}

\textcolor{black}{
The LS value of $z$  is in line with what was obtained with fewer data in \cite{cavagna2017swarm}, which also used LS. But unfortunately, there is a serious problem.
Despite its simplicity, the main limit of LS is the very assumption of the absence of experimental uncertainty on the $x$ variable; in our case this means assuming that the correlation length is perfectly determined, which is definitely not true in the case of our experiments. In fact, errors on both $\xi$ and $\tau$ are significant, as it can be seen from Table \ref{data}. When the hypothesis of exact determination of $x$ is violated, LS is known to underestimate the slope, and to severely do so if the uncertainty in $x$ is significant. A consequence of the LS assumption is that the method is {\it not} symmetric under exchange of $x$ and $y$.
In our case, when the dependent and independent variables are swapped in LS, namely when we fit the relation $\log \xi = z^{-1} \log \tau + c'$, we obtain a completely different result for the dynamic exponent, $z_\mathrm{LS}=1.62\pm0.20$; in this case LS is assuming that there is no experimental uncertainty in the relaxation time, which is even worse than assuming no uncertainty in $\xi$, given that errors on $\log\tau$ are usually even larger than those on $\log\xi$. 
Summarizing, the fact that experimental errors on $\log\tau$ and $\log\xi$ are present, no matter how difficult they are to determine, indicates that one critical assumption of LS fails in our case. This is the reason why the exponent $z$ was unfortunately significantly underestimated in \cite{cavagna2017swarm} as well as in \cite{cavagna2021natural}.
}

\textcolor{black}{
\subsubsection{Reduced Major Axis} 
The shortcomings of LS can be easily overcome by treating the two variables $x$ and $y$ in a statistically symmetric way, which is what the Reduced Major Axis (RMA) method does.
We will not repeat here the detailed description of RMA, as this is already provided in the main text, Methods Section. We simply recall here that RMA - as LS - does {\it not} require the experimental errors in order to work, but - unlike LS - RMA works under the hypothesis that both $x$ and $y$ are affected by experimental uncertainty and therefore gives the same fitted line under exchange of $x$ and $y$. The fit of $z$ with RMA gives,
\begin{equation}
	z_\mathrm{RMA}=1.37\pm0.11 \ .
\end{equation}
}

\begin{figure}[b]
	\centering
	\includegraphics[width=1\textwidth]{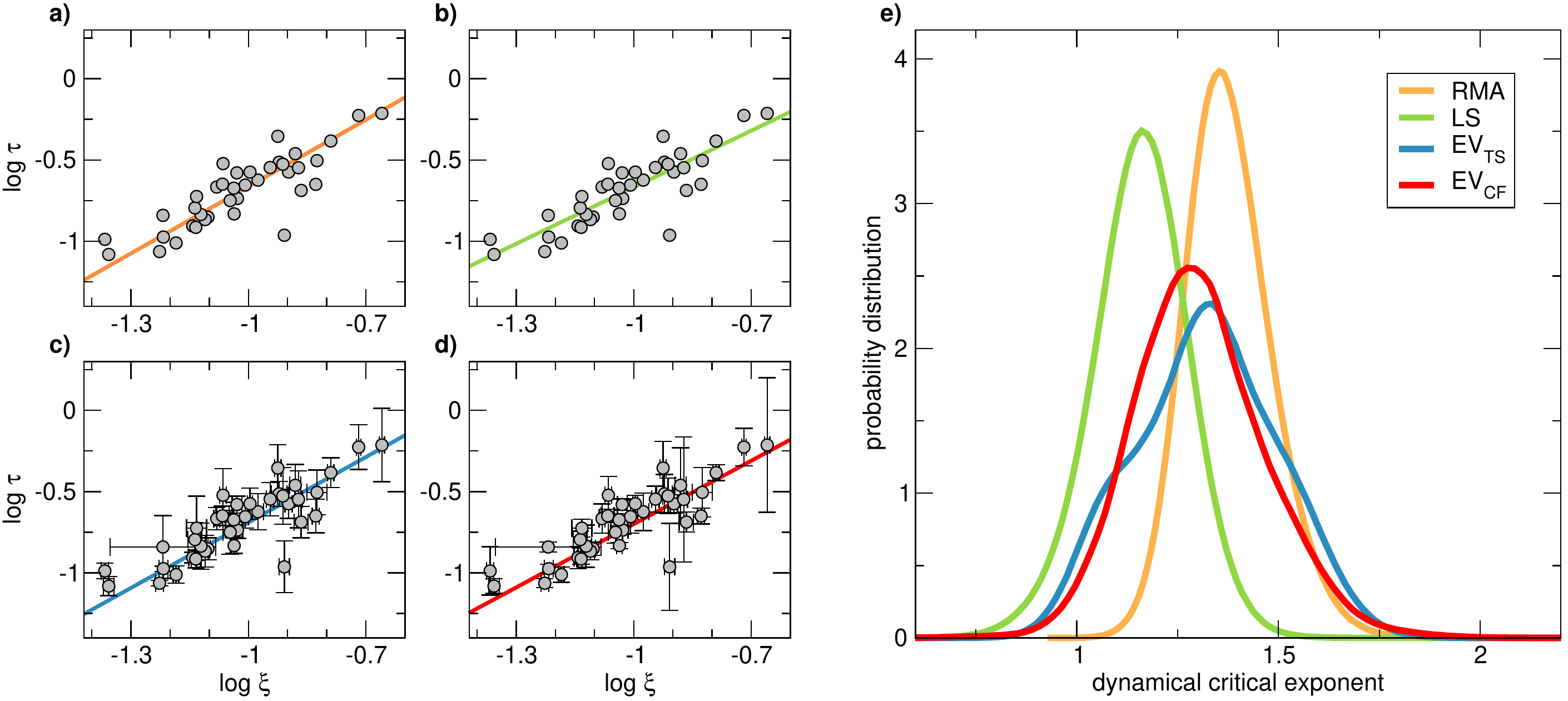}
	\caption{\textcolor{black}{{\bf Different fitting methods and experimental distribution of $z$.} We show here the results of all the four different regression methods discussed in the present work when applied to experimental data of Table \ref{data}. Best fit obtained by using {\bf a)} Reduced Major Axis (RMA) method, {\bf b)} Least Square (LS) method, {\bf c)} Effective Variance ($\text{EV}_\text{TS}$) method with error-bars on $\log\tau$ determined through the temporal series method, {\bf d)} Effective Variance ($\text{EV}_\text{CF}$) method with error-bars on $\log\tau$ determined from the correlation function errors. {\bf e)} Comparison between the four experimental distributions of $z$, one for each method, obtained through resampling.}}
	\label{fig:exp_plot}
\end{figure}

\textcolor{black}{
\subsubsection{Effective Variance}
The Effective Variance (EV) regression method -- unlike LS and RMA -- makes explicit use of the errors on the experimental data-points. 
To derive the EV method, let us start by making some improvement on LS by taking into account the errors experimentally measured for different swarms. To do this, we minimize the average square scatter from the fitting line divided by the experimental error on $y$, namely
\begin{equation}
	\Sigma\left(\alpha,\beta\right)=\frac{1}{N}\sum_{i=1}^N \Delta y_i^2=\frac{1}{N}\sum_{i=1}^N \left(\frac{y_i-\alpha x_i -\beta}{\delta_i}\right)^2
	\label{SEV}
\end{equation}
where $\delta_i^2=\sigma_{y_i}^2$ is the variance of the measurement of $y_i$.
This version of LS does resolve the critical issues raised in the previous section and it only slightly changes the determination of the fitting parameters, which are the same as Eq. \eqref{LSa},\eqref{LSb}, but with an average which weights the different points by their inverse variance
\begin{equation}
	\E{g(x,y)}=\frac{\sum_{i=1}^N \sigma_i^{-2}\,g(x_i,y_i)}{\sum_{i=1}^N \sigma_i^{-2}}
\end{equation}
To fully take into account also the presence of experimental uncertainty also on the variable $x$ we must replace the variance $\sigma_{y_i}^2$ with an {\it effective} variance that takes into account the errors on both variables \cite{orear1982least}.
Hence, the EV regression method consists in minimizing the quantity defined in Eq. \eqref{SEV} with $\delta_i$ taken as,
\begin{equation}
	\delta_i^2=\sigma_{y_i}^2 + \left(\left.\frac{\partial f}{\partial x}\right|_{x=x_i}\right)^2 \sigma_{x_i}^2=\sigma_{y_i}^2 + \alpha^2 \sigma_{x_i}^2
\end{equation}
No explicit solutions exists for EV method in the general case in which $\sigma_{x_i}$ and $\sigma_{y_i}$ are different from point to point. Minimization must be thus performed numerically with a Monte Carlo algorithm, where as starting value we choose the solution given by LS.
We applied the EV method to our data using the experimental errors on $\log\xi$ and $\log\tau$ estimated above. When the error on the relaxation time is estimated from the fluctuations around the plateau, $\delta\tau_\mathrm{TS}$, we obtain,
\begin{equation}
	z_\mathrm{TS}=1.34\pm0.18\,
\end{equation}
On the other hand, when the error on the relaxation time is estimated from the fluctuations of the correlation function, $\delta\tau_\mathrm{CF}$, we obtain,
\begin{equation}
	z_\mathrm{CF}=1.30\pm0.17\,
\end{equation}
Errors on the determination of $z$ are computed through the usual resampling method, as for RMA and LS. 
}

\textcolor{black}{
In Fig. \ref{fig:exp_plot} we compare the different experimental distributions obtained through all the fitting methods discussed, namely LS, RMA, EV$_\mathrm{TS}$ and EV$_\mathrm{CF}$. The only clear outlier is LS, which is expected because of the shortcomings we have discussed above; note also that LS is the method that gives the largest fraction of exponents smaller than $1$, which is unphysical. On the other hand, RMA and EV are quite compatible with each other, although only RMA has virtually zero fraction of exponents smaller than $1$ ($0.002\%$ of the sub-samples). Moreover, EV quite strongly depends on the determination of the errors, which -- as we explained above -- has its own criticalities. This is why we ultimately decided to estimate $z$ through a method that does not depend on the determination of the experimental errors (RMA), although we believe that the consistency between RMA and EV is a sanity check of our experimental determination of $z$.
}




\newpage
\textcolor{black}{ \section{Numerical simulations}
We perform numerical simulations in three dimensions of the Inertial Spin Model (ISM). Although the ISM was introduced in \cite{cavagna2015flocking} to study information propagation in birds flocks, recent experimental data on natural swarms of insects \cite{cavagna2017swarm}, and preliminary theoretical investigations \cite{cavagna2019short}, strongly suggest that it is also a useful model to describe natural system. The ISM is a model of self-propelled particles moving at constant speed $v_0=|\boldsymbol v_i(t)|$ $\forall \ i, t$ and interacting with a short-range metric alignment force. The main difference with similar models of polar active particles, for instance the Vicsek model \cite{vicsek1995novel}, is that the dynamics is inertial in the velocities $\boldsymbol v_i$, that is mediated by the generator of internal rotational symmetry, i.e. the spin $\boldsymbol s_i$. For completeness we report also here the equations already explained in the main text,
\begin{equation}
\label{microISMsupp}
\begin{split}
\frac{d  \boldsymbol v_i }{dt} &= \frac 1 {\chi} \boldsymbol s_i \times \boldsymbol v_i \\
\frac{d \boldsymbol s_i}{dt} &=  
\boldsymbol {v}_i \times 
\frac{J}{n_i} \sum_j n_{ij}(t) \boldsymbol {v}_j
 - \frac{\eta}{\chi} \boldsymbol {s_i} + \boldsymbol  {v}_i \times \boldsymbol{\zeta}_i   \\
\frac{d \boldsymbol r_i}{dt} &= \boldsymbol v_i  
\end{split}
\end{equation}
where $\boldsymbol \zeta_i$ is a gaussian white noise with variance,
\begin{equation}
\langle \boldsymbol{\zeta}_i (t) \cdot \boldsymbol{\zeta}_j(t')\rangle = 2dT\; \eta\; \delta_{ij}\delta(t-t') 
\end{equation}
We will study the ISM in the near-ordering phase.  Among all the parameters entering the equations, two are the most important for our purposes: the first is the microscopic friction $\eta$, which represents the spin dissipation and is directly connected to the mesoscopic parameter discussed in section A.3. In \cite{cavagna2019short} we studied the equilibrium ($v_0=0$) version of ISM and we unveiled the role of this parameter, showing that for $\eta=1.0$ an underdamped value of the equilibrium dynamic critical exponent $z$ is found. To be sure we are in the underdamped phase also in the active regime, we decide to use the same value of spin friction in the present simulations. Accordingly, in Fig.3a of the main text we report the velocity correlation functions computed with this value of $\eta$ that clearly show an inertial decay.
The second important parameter is what makes the system active, namely the particle speed $v_0$. The study we carried out in \cite{cavagna2020equilibrium} shows that in order to observe the active critical exponents, $v_0$ has to be large enough to yield a substantial reshuffling of the interaction network. We chose $v_0=2.0$ since, as it is shown in Fig.3b of the main text, at this speed the relaxation time of the network becomes comparable to that of the velocity.}

\begin{figure}[t]
	\centering
	\includegraphics[width=1.0 \textwidth]{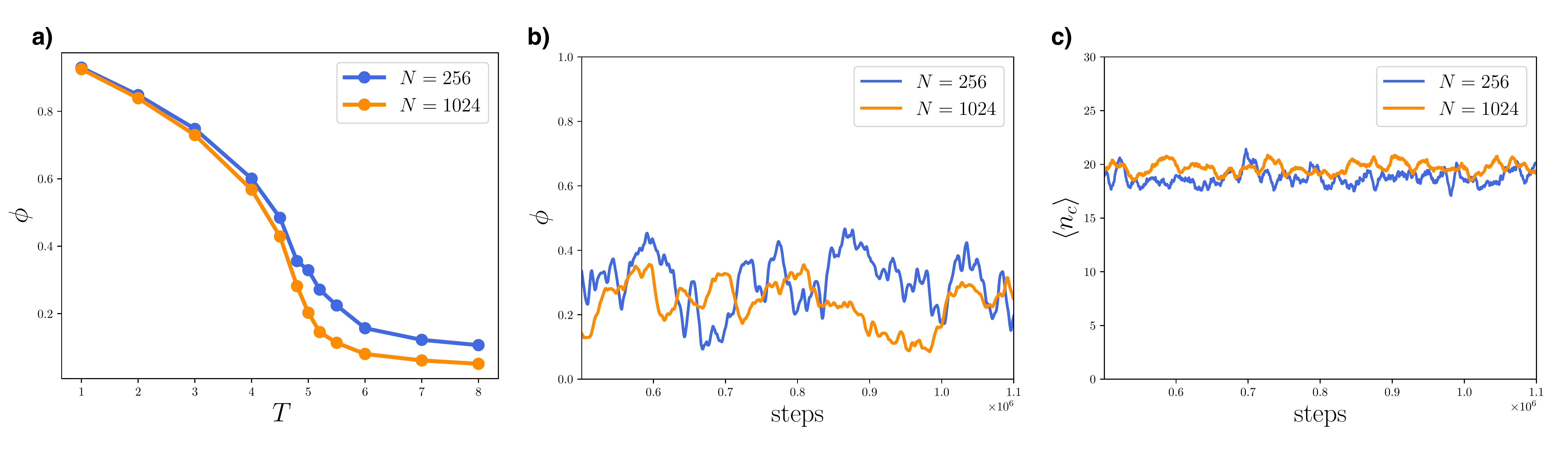}
	\caption{\textcolor{black}{{\bf  Phase transition of ISM.} Panel a): time averaged polarization $\phi =|\sum_i \boldsymbol v_i|/(N v_0)$ vs $T$ for sizes $N=256$ and $N=1024$, the phase transition between order and disorder shows a second order phenomenology. Panel b): time series of polarization after the termalization of two samples of the same size of panel a. No trend in time is recognizable, the time scale of decorrelation increases with the size as expected. Panel c) time series of interacting neighbours's number $\langle n_c\rangle$ averaged on system's particles for the same samples of panel b). Fluctuations are limited and stationary, confirming the absence of polarized clusters. The parameters used in the simulations are $\chi=1, J=18, r_c=1.6, \rho=1, \eta=1, v_0=2$. In perfectly homogenous systems we expect $\langle n_c\rangle \sim 18$.  }}
	\label{fig:ncphi}
\end{figure}

\textcolor{black}{\subsection{Thermalization}
To perform numerical simulations, eqs \eqref{microISMsupp} are discretized in time, and the constraint of constant speed is imposed using the RATTLE algorithm known from molecular dynamics simulation schemes \cite{andersen1983rattle,cavagna2016spatio}. An order-disorder phase transition is observed when the amplitude of the noise $T$ is varied in the range $[1:8]$. For the set of parameters and sizes chosen, the phase transition is smooth and it does not show a jump usually due to the presence of heterogeneities.
This is also due to the fact that the alignment strength is rescaled with $J \to J/n_i$ where $n_i=\sum_j n_{ij}(t)$ are the neighbours of particle $i$ at time $t$. Doing this corresponds to implementing a non-additive interaction rule, which is known to suppress density fluctuations and to prevent aggregation events typical of the critical phase of  polar active systems \cite{chepizhko2021revisiting}. Without this normalization, the system is very unstable and the heterogeneities affect the behaviour of the order parameter appearing non-stationary, thus quite far from a scaling regime. Since our ultimate purpose is the computation of a critical exponent, we check that every sample used for the analysis has reached a stationary state and it does not show abrupt fluctuations. In panel a of Fig \ref{fig:ncphi} we show the curve of  time averaged polarization $  \phi =|\sum_i \boldsymbol v_i|/(N v_0)$ vs $T$ for two of the sizes simulated: the smooth transition is traceable in a well-behaved time series of this quantity (panel b). We also monitor density fluctuations looking at the trend of  the average number of interacting neighbours $\langle n_c\rangle$, where the average is done on the particles of the system at a given instant of time. This quantity shows a stationary trend and limited fluctuations reflecting the absence of polarized clusters (panel c). Because of these evidences, we conclude that at these sizes and for this set of parameters it is reasonable to apply a second-order paradigm under the assumption of incompressibility in agreement with the analytical calculation.
}

\begin{figure}[b]
	\centering
	\includegraphics[width=1.0 \textwidth]{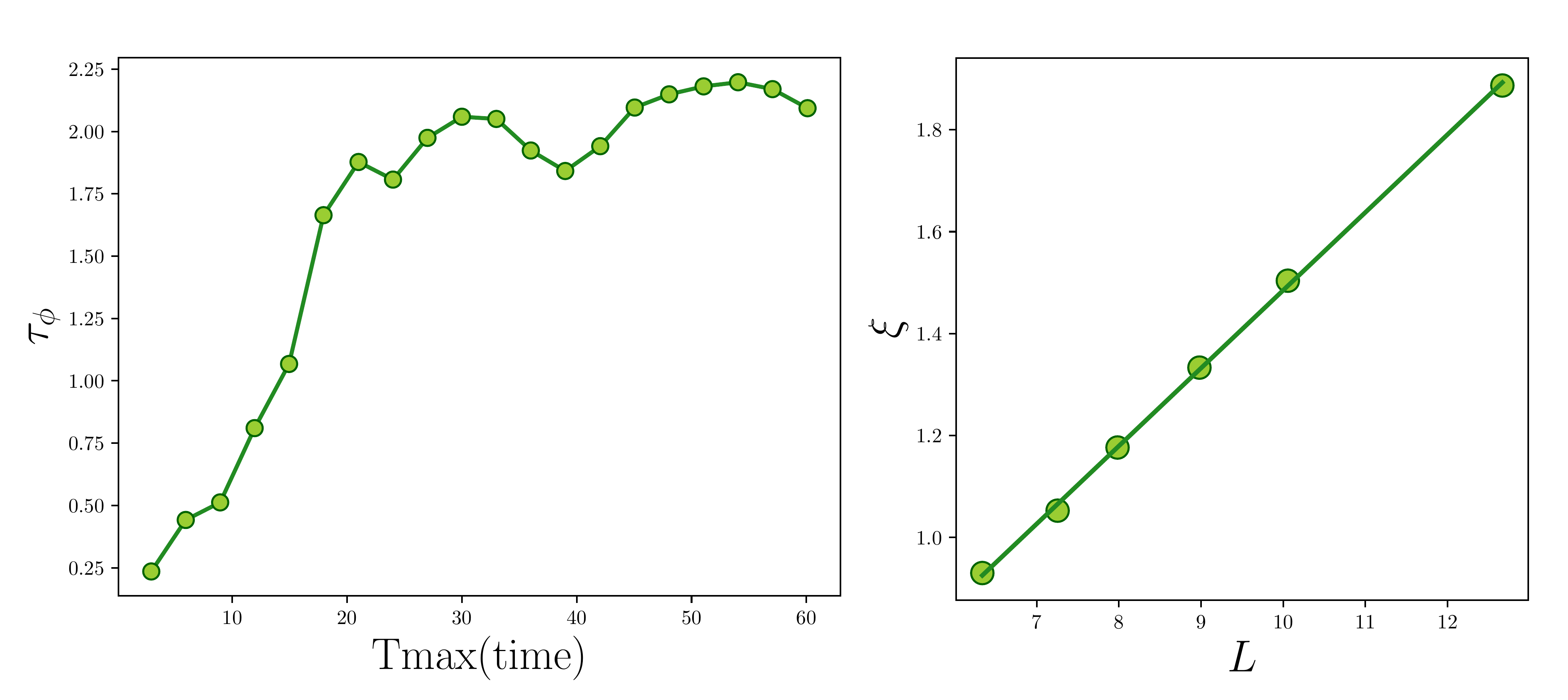}
	\caption{\textcolor{black}{{\bf Scaling regime} Left: characteristic time scale of polarization vs $t_{\mathrm{max}}$ for $N=256$ and $T=5.1$; time is computed as $t=N_{steps} dt$ with $dt=0.0001$. Choosing $t_{\mathrm{max}} \geq 20 \tau_\phi$ ensures that the system is well termalized and the correlation functions are correctly computed as well as the characteristic time scales. Right: correlation length $\xi=1/k_\mathrm{max}$ vs $L=N^{1/3}$. $k_\mathrm{max}$ is the wave-number at which the static correlation function at $T_c(N)$ reaches a maximum. The linear trend confirms the scale-free behaviour of the system close to criticality.  }}
	\label{fig:csiL}
\end{figure}

\textcolor{black}{To extrapolate the dynamical critical exponent we use a finite-size scaling analysis simulating systems of sizes $N=256, 384, 512, 729, 1024, 2048$ at constant average density $\rho=1$. The other parameters of the model are fixed at: $\chi=1, J=18, r_c=1.6, \eta=1, v_0=2$.  For each size we simulated 5 statistical independent samples for every value of temperature $T$ explored. We initialize each simulation in a perfectly ordered state in a random spatial direction, and we let it evolve for  $\sim 6 \times 10^5$ simulation steps to reach the termalization. After it, the sample is collected as a simulation of length $t_{\mathrm{max}}$  depending on the system's size. To check that the choice of the latter provides a reliable calculation of correlation functions, we compute the auto-correlation of polarization as,
\begin{equation}
C_\phi(t) = \frac{1}{t_{\mathrm{max}}-t} \sum_{t_0=1}^{t_{\mathrm{max}}-t} \langle \delta \phi(t_0) \delta \phi(t_0+t) \rangle, \qquad \text{with} \qquad 
\delta \phi(t) = \phi(t) - \langle \phi \rangle_{t_{\mathrm{max}}} 
\end{equation}
and we compute its characteristic time scale $\tau_\phi$ using the definition \eqref{tau} for different choices of $t_{\mathrm{max}}$. The left panel of Fig \ref{fig:csiL} shows this quantity averaged on 5 simulations of size $N=256$ and $T=5.1$. The decorrelation time reaches a plateau around $30\,\mathrm{s}$, where $\text{time}=N_{steps} dt$ and $dt=0.0001$. The asymptotic $\tau_\phi$ is $\sim 2.1\,\mathrm{s}$, therefore choosing $t_{\mathrm{max}} \sim 20 \tau_\phi$ seems already a safe choice to compute correct correlation functions. Moreover, the $C_\phi$ takes into account the global collective dynamics, thus it is expected to relax on time-scales larger than when involving modes with higher wave-number. To perform an analysis similar to that of the experiments, we are going to compute spatio-temporal correlation functions only at $k \neq 0$, therefore this ensures that using $t_{\mathrm{max}}=60\,\mathrm{s}$ (or greater) always provides reliable calculations of correlation functions, and consequently of the corresponding characteristic time-scales. We chose $t_{\mathrm{max}}=60\,\mathrm{s}$ for $N=256$, up to $t_{\mathrm{max}}=80\,\mathrm{s}$ at $N=2048$.
}

\textcolor{black}{\subsection{Scaling and critical exponent} 
For every set of simulation at given $T$ and $N$ we compute the static correlation functions in $k$-space, following definition \eqref{galore} and using the simplified fluctuations $\delta \boldsymbol v_i(t) = \boldsymbol v_i(t) - \sum_k \boldsymbol v_k(t) /N$. Because of the sum rule $\sum_i \delta \boldsymbol v_i =0$, this quantity is null at $k=0$, it reaches a saturation value depending on $v_0$ and it shows a maximum at a point $k_\mathrm{max}$. The value of correlation at this point $C_0(k_\mathrm{max})$ represents a measure of the generalized integral susceptibility $\chi$ of the system (note that, because the system is out of equilibrium, the integral susceptibility $\chi$ that we are using here needs not be equal to the field-susceptibility, $\chi_H$, defined as the static response to an external field \cite{cavagna2018correlations}). Computing $\chi$ for all the values of $T$ considered, we obtain a curve which peaks at $T_c(N)$, namely the size-dependent critical point that we identify for every $N$ simulated. At the corresponding temperature we compute the dynamical correlation functions $C(k,t)$ following \eqref{cicciobello} at $k=k_\mathrm{max}$. The latter is the wave-number at which the system is maximally correlated, it can thus be interpreted as the inverse of the correlation length $k_\mathrm{max}=1/\xi$. When a finite-size system is close to criticality it shows a scale-free behaviour where $\xi$ scales linearly with $L$: in the right panel of Fig \ref{fig:csiL} is shown that this is exactly what we observe in our simulations. This evidence ensures we are in the correct scaling regime where the dynamic scaling hypothesis can be tested. }

\begin{figure}[b]
	\centering
	\includegraphics[width=1.0 \textwidth]{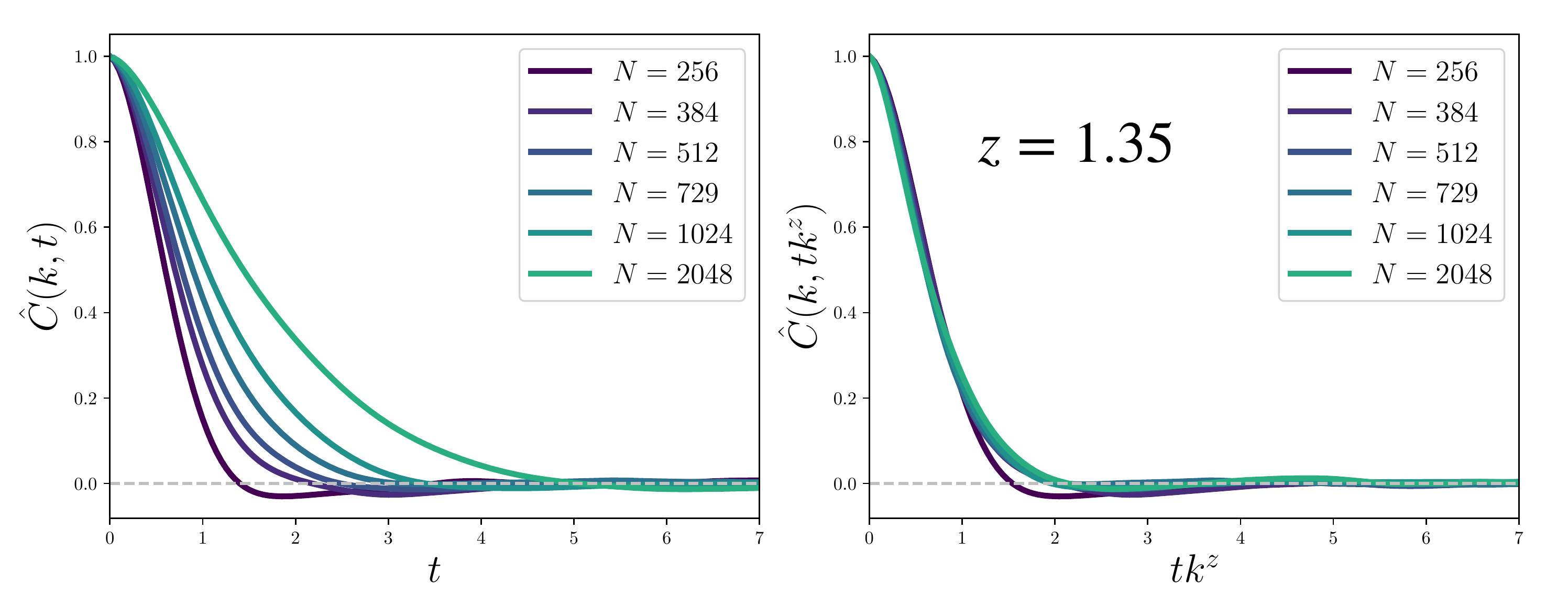}
	\makeatletter\long\def\@ifdim#1#2#3{#2}\makeatother
	\captionof{figure}{\textcolor{black}{{\bf Dynamical correlation functions.} Left: dynamical correlation functions $\hat C(k,t)= C(k,t)/C(k,0)$ at different sizes $N$ and at $T_c(N)$, the wave-number is $k=k_\mathrm{max}=1/\xi$. Right: same correlation functions when time is rescaled $t \to tk^z$, using the analytical RG prediction $z=1.35$. The curves collapse one onto each other, fully confirming the validity of scaling hypothesis and of the analytical calculation.}}
	\label{fig:simckt}
\end{figure}

\textcolor{black}{
For the six sizes simulated we compute the $C(k,t)$ at $T_c(N)$ on the 5 samples, the average curves are shown in the left panel of Fig \ref{fig:simckt}. From them we extrapolate the characteristic time-scales using definition \eqref{tau}; in the main text is reported the linear regression of $\ln(\tau)$ vs $\ln (\xi)$ that provides a numerical estimate of the dynamical critical exponent,
\begin{equation}
z_\mathrm{sim}= 1.35\pm 0.04. \ ,
\end{equation}
which is in perfect agreement with the RG predictions and in good agreement with the experimental data.
An additional confirmation of the validity of the dynamical scaling hypothesis with the exponent just found is the collapse of the dynamical correlation functions when time is properly rescaled $t \to tk^z$ \cite{HH1969scaling}. This theory indeed predicts that the correlation function and the characteristic time scale of the system are homogenous functions of the the product $k \xi$, namely $\hat{C}(k,t)= f(k \xi, t/\tau_c(k))$ with $\tau_c(k) = k^{-z} \Omega(k\xi)$. Therefore, at fixed $ k \xi =1$, we should observe $\hat{C}(k,t)=g(tk^z)$ with the correct rescaling of time.
Using the analytical prediction $z=1.35$, the numerical curves  almost perfectly collapse one onto each other in a single shape function, asserting the validity of the scaling hypothesis and fully confirming the RG perturbative calculation. }

\newpage
\section{Experimental data}

\begin{table*}[h!]
	\vskip 0.1 in
	\begin{tabular}{|c|c|c|c|c|c|c|c|c|}
		\hline
		\hline
		{\sc Event label} \hspace{0.2cm} &
		$N$ \hspace{0.4cm} &
		$t_\mathrm{max} (s)$\hspace{0.2cm}&
		$\tau (s)$\hspace{0.4cm}&
		$\xi (cm)$\hspace{0.2cm}&
		$r_1 (cm)$\hspace{0.4cm}&
		$\delta\xi/\xi$\hspace{0.2cm}&
		$\delta\tau_{TS}/\tau$\hspace{0.2cm} &
		$\delta\tau_{CF}/\tau$\hspace{0.2cm}\\
		\hline
		\hline
		20110511\_A2 & 279 & 0.88 & 0.12 & 12.3 & 5.33 & 0.01 & 0.16 & 0.27\\
		20110906\_A3 & 138 & 2.05 & 0.09 & 4.40 & 2.94 & 0.01 & 0.06 & 0.05\\
		20110908\_A1 & 119 & 4.41 & 0.11 & 4.30 & 3.59 & 0.01 & 0.05 & 0.15\\
		20110909\_A3 & 312 & 2.73 & 0.10 & 6.53 & 2.59 & 0.01 & 0.05 & 0.05\\
		20110930\_A1 & 173 & 5.88 & 0.47 & 11.9 & 5.72 & 0.01 & 0.14 & 0.16\\
		20110930\_A2 & 99  & 5.88 & 0.27 & 12.7 & 6.32 & 0.01 & 0.09 & 0.06\\
		20111003\_A1 & 89 & 5.58 & 0.14 & 6.05 & 7.74 & 0.14 & 0.19 & 0.03\\
		20111006\_A1 & 88 & 11.76 & 0.18 & 9.37 & 8.28 & 0.01 & 0.05 & 0.02\\
		20111011\_A1 & 131  & 5.88 & 0.23 & 14.9 & 7.52 & 0.01 & 0.10 & 0.05\\
		20120702\_A1 & 98 & 2.14 & 0.22 & 8.30 & 6.16 & 0.01 & 0.07 & 0.09\\
		20120702\_A2 & 111 & 7.29 & 0.14 & 7.88 & 5.57 & 0.01 & 0.07 & 0.04\\
		20120702\_A3 & 80 & 9.99 & 0.11 & 6.06 & 5.97 & 0.01 & 0.02 & 0.03\\
		20120703\_A2 & 167 & 4.41 & 0.09 & 5.93 & 4.65 & 0.01 & 0.02 & 0.03\\
		20120704\_A1 & 152 & 9.99 & 0.13 & 7.21 & 4.98 & 0.01 & 0.03 & 0.07\\
		20120704\_A2 & 154 & 5.29 & 0.13 & 7.34 & 5.32 & 0.01 & 0.06 & 0.03\\
		20120705\_A1 & 188 & 5.88 & 0.15 & 9.19 & 5.54 & 0.01 & 0.05 & 0.02\\
		20120828\_A1 & 89 & 6.29 & 0.11 & 7.75 & 6.18 & 0.01 & 0.10 & 0.05\\
		20120907\_A1 & 169 & 3.23 & 0.62 & 21.9 & 6.21 & 0.01 & 0.23 & 0.41\\
		20120910\_A1 & 219  & 1.76 & 0.24  & 10.6 & 4.68 & 0.01 & 0.11 & 0.12\\
		20120917\_A2 & 212 & 5.88 & 0.35 & 13.2 & 5.15 & 0.01 & 0.13 & 0.23\\
		20120917\_A3 & 554 & 4.12 & 0.28 & 13.4 & 4.40 & 0.01 & 0.18 & 0.38\\
		20120918\_A2 & 69 & 15.8 & 0.22 & 8.58 & 6.06 & 0.02 & 0.06 & 0.04\\
		20120918\_A3 & 215 & 5.00 & 0.31 & 12.0 & 5.04 & 0.02 & 0.08 & 0.12\\
		20150729\_A1 & 110 & 5.87 & 0.32 & 8.61 & 4.63 & 0.01 & 0.16 & 0.11\\
		20150910\_A2 & 99 & 2.99 & 0.15 & 7.56 & 4.61 & 0.01 & 0.14 & 0.13\\
		20150921\_A1 & 201 & 4.11 & 0.23 & 9.81 & 4.21 & 0.01 & 0.07 & 0.10\\
		20150922\_A1 & 94 & 5.87 & 0.19 & 8.98 & 6.04 & 0.02 & 0.13 & 0.07\\
		20150922\_A2 & 126 & 5.87 & 0.29 & 11.4 & 5.29 & 0.01 & 0.10 & 0.08\\
		20150924\_A1 & 115 & 5.87 & 0.30 & 12.2 & 4.81 & 0.01 & 0.17 & 0.11\\
		20150924\_A2 & 781 & 2.94 & 0.59 & 19.1 & 4.08 & 0.01 & 0.14 & 0.12\\
		20150924\_A3 & 232 & 3.23 & 0.41 & 16.2 & 5.55 & 0.01 & 0.09 & 0.05\\
		20150924\_A4 & 107 & 4.38 & 0.32 & 15.0 & 6.38 & 0.03 & 0.14 & 0.15\\
		20151008\_A2 & 92 & 3.51 & 0.27 & 10.1 & 5.33 & 0.01 & 0.10 & 0.05\\
		20151008\_A3 & 91 & 5.87 & 0.16 & 7.30 & 4.41 & 0.02 & 0.10 & 0.05\\
		20151008\_A4 & 87 & 3.29 & 0.21 & 13.6 & 8.63 & 0.02 & 0.10 & 0.06\\
		20151026\_A1 & 85 & 5.87 & 0.19 & 7.37 & 6.67 & 0.03 & 0.20 & 0.06\\
		20151030\_A1 & 274 & 5.87 & 0.27 & 9.34 & 3.96 & 0.01 & 0.05 & 0.04\\
		20151030\_A2 & 123 & 5.81 & 0.21 & 9.18 & 4.96 & 0.02 & 0.11 & 0.04\\
		\hline
		\hline
	\end{tabular}
	\caption{Summary of experimental data. Swarming events are labelled according to experimental date and acquisition number. $N$ indicates the number of insects (and reconstructed trajectories) in the swarm, $\tau$ is the relaxation time and $\xi$ is the correlation length. The average nearest neighbour distance $r_1$ is calculated by averaging over all individuals in the swarm, and over the event duration.}
	\label{data}
\end{table*}

\putbib
\end{bibunit}

\end{document}